\documentclass[prl,aps,twocolumn,superscriptaddress,10pt]{revtex4-2}
\usepackage{amsmath}
\usepackage{amssymb}
\usepackage{amsfonts}
\usepackage{graphicx}
\usepackage[colorlinks=true,citecolor=blue,linkcolor=black]{hyperref}
\usepackage[nameinlink,capitalize]{cleveref}
\usepackage{wrapfig}
\usepackage{caption}
\usepackage{comment}
\usepackage{color}

\newcommand*{\Tr}{\ensuremath{\mathbf{Tr}}}
\newcommand*{\sgn}{\ensuremath{\mathrm{sgn}}}

\begin{document}

\title{Framing Anomaly in Lattice Chern-Simons-Maxwell Theory}
\author{Ze-An Xu}
\affiliation{Institute for Advanced Study, Tsinghua University, Beijing, 100084, China}
\author{Jing-Yuan Chen}
\affiliation{Institute for Advanced Study, Tsinghua University, Beijing, 100084, China}
\begin{abstract}
Framing anomaly is a key property of $(2+1)d$ chiral topological orders, for it reveals that the chirality is an intrinsic bulk property of the system, rather than a property {that requires a} boundary between two systems. Understanding framing anomaly in lattice models is particularly interesting, as concrete, solvable lattice models of chiral topological orders are rare. In a previous work \cite{Xu:2024hyo}, we defined and solved the $U(1)$ Chern-Simons-Maxwell theory on spacetime lattice, showing its chiral edge mode and the associated gravitational anomaly on boundary. In this Letter, we show its framing anomaly in the absence of boundary by computing the expectation of a lattice version of the modular $T$ operator in the ground subspace on a spatial torus, from which we extract that $\langle T \rangle$ has a universal phase of $-2\pi/12$ as expected: $-2\pi/8$ from the Gauss-Milgram sum of the topological spins of the ground states, and $2\pi/24$ from the framing anomaly. We can also extract the $2\pi/24$ framing anomaly phase alone from the full spectrum of $T$ in the ground subspace by computing $\langle T^m \rangle$. This pins down the last and most crucial property required for a valid lattice definition of $U(1)$ Chern-Simons theory.
\end{abstract}
\maketitle

\noindent\emph{\bf Introduction--- } Chern-Simons theory is a landmark in the formal development of quantum field theory \cite{Schwarz:1978cn, Witten:1988hf}, and is the effective theory that describes the quantum Hall effect \cite{Wen:1995qn}. Among its fascinating topological properties, {one} key characterization of a Chern-Simons theory is its \emph{chirality}. 

Chirality usually manifests in the form of gapless chiral edge modes, as has been observed in thermal Hall measurements \cite{Banerjee:2017}. {Nonetheless, edge modes and thermal Hall conductivity are only definable as \emph{relative} notions between two systems on the two sides of a boundary} \cite{Kapustin:2019net}{. Meanwhile, it is conceptually important that chirality is an intrinsic bulk property of a system. (For comparison, electric Hall conductivity is an intrinsic bulk property, and since the early years, this understanding has played an important role in the development of the theory of the quantum Hall effect, e.g.,} \cite{Niu:1984ks}{.)} In the absence of boundary, chirality manifests itself as a \emph{framing anomaly} \cite{Witten:1988hf}, which suggests the phase of the partition function has a dependence, in units of $2\pi c_-/24$, on the global choice of a basis of tangent vectors (a frame) over the spacetime, where $c_-$ is the \emph{chiral central charge} of the theory.

\begin{figure}[t]
    \centering
    \includegraphics[width=0.3\linewidth]{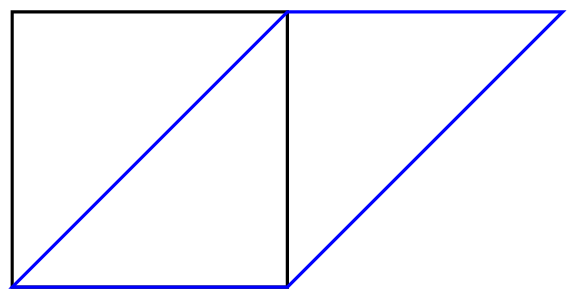}
    \caption{Our convention of modular $T$ operator \emph{actively} brings the values of the fields at $(x, y)$ to $(x+yL_x/L_y, y)$. (In the literature the convention is often passive.)}
    \label{fig:modular-T}
\end{figure}

Associated to this, the modular $T$ operator (see \cref{fig:modular-T}) acts on the ground subspace of a spatial torus as \cite{Verlinde:1988sn}
\begin{equation}
T=e^{-i2\pi (h- c_-/24)}
\label{eqn:modular_T}
\end{equation}
where $h$ is the operator measuring the conformal weight of a ground state, and $2\pi c_-/24$ is the framing anomaly. $h$ and $c_-$ are not entirely independent. For a bosonic topological order, the Gauss-Milgram formula reads \cite{Turaev:1994xb, Kitaev:2005hzj}
\begin{equation}
    \sum_{\mathrm{gnd \: states}\: n=1}^N e^{i2\pi h_n} = \sqrt{N} \, e^{i2\pi c_-/8} \ .
    \label{eqn:Gauss_Milgram}
\end{equation}
In particular, for bosonic $U(1)$ Chern-Simons of even level-$k$, the ground states are labeled by $n\in\mathbb{Z}_{N=|k|}$, with purely imaginary $i2\pi h_n=i\pi n^2/k$ being the topological spin, and $c_-=\sgn(k)$. For fermionic systems, it is easy to see the modular $T$ operator (\cref{fig:modular-T}) is only defined when the fermions obey \emph{periodic} boundary conditions across the $x$ direction, and in particular for $U(1)$ Chern-Simons of odd level $k$, the periodic boundary condition makes the ground state topological spins $i2\pi h_n=i\pi (n+1/2)^2/k$ \cite{Belov:2005ze,Lapa:2019uim,Cheng:2025ube} in \cref{eqn:Gauss_Milgram} (see Supplemental Material \cite{SupplementalMaterial} for explanation), yielding the same right-hand side.

Solvable lattice models have played an important role in the development of the subject of topological order. However, most of the systematically constructed lattice models were for theories that are gappable on the boundary, hence nonchiral \cite{Kitaev:2011dxc}. Concrete, solvable models for chiral topological orders are very rare (an important example is \cite{Kitaev:2005hzj}). Particularly, the lattice realization of chiral $U(1)$ Chern-Simons has been a problem of interest for decades. Recently, by carefully putting together existing but scattered ideas, $U(1)$ Chern-Simons-Maxwell theory has been defined on a lattice (with the Maxwell term needed for fundamental reasons) \cite{Peng:2025nfa, Xu:2024hyo} and analytically solved. [A lattice model for $SU(N)$ Chern-Simons-Yang-Mills has also been constructed by generalizing the $U(1)$ case via higher category theory \cite{Chen:2024ddr, Zhang:2024sgm}, though the $SU(N)$ case is not analytically solvable.] The chiral edge mode and the associated gravitational anomaly on the boundary have been solved for in \cite{Xu:2024hyo}, showing the lattice model is indeed chiral. But it will be much more interesting if we can show the celebrated framing anomaly on the lattice \emph{in the absence of boundary}. This will not only pin down the last key requirement from a successful lattice realization of $U(1)$ Chern-Simons, but also provide an important example of a UV-complete study of the framing anomaly.

{More broadly speaking, the general motivation of our program to realize Chern-Simons, the framing anomaly, and other formal aspects of quantum field theories on the lattice }\cite{Chen:2024ddr} {is to make everything regularized in an explicit manner (and it is nontrivial to show that this is indeed possible) and hence more assessable.}

In this Letter, we will define a spacetime lattice realization of the modular $T$ operator acting on a spatial torus, and evaluate its expectation in the ground subspace of the lattice Chern-Simons-Maxwell theory. We will find, as expected, that
\begin{align}
\langle T \rangle =\frac{ e^{-i2\pi\sgn(k)(1/8-1/24)}
}{\sqrt{|k|}} 
\ e^{-\alpha L^2 + \, \cdots}
\label{eqn:expected_result}
\end{align}
where the first fraction is the universal contribution from
\cref{eqn:modular_T} averaged over the ground states using \cref{eqn:Gauss_Milgram} (or its fermionic counterpart), while in the second factor $L^2$ is the size of the spatial torus with $\alpha$ a nonuniversal complex number that can be removed by adding local counterterm to the definition of $T$ (the $L^2$ scaling is because $T$ is an extensive operator defined over the spatial torus), and ``$\cdots$'' represents finite size effects that vanish as $L\rightarrow \infty$.

We can also separate the $1/24$ framing anomaly contribution from the $-1/8$ Gauss-Milgram sum contribution by evaluating a lattice realization of $\langle T^m \rangle$ for $m\in\mathbb{Z}_+$---which can be recognized as a \emph{higher central charge} calculation \cite{Ng:2018ddj,Kaidi:2021gbs}---in order to extract the full spectrum of $T$ in the ground subspace. The result indeed agrees with \cref{eqn:modular_T} for both bosonic and fermionic $U(1)$ Chern-Simons.

\

\noindent\emph{\bf Lattice Chern-Simons-Maxwell theory--- \hspace{.0cm}} We first review the Chern-Simons-Maxwell theory on a spacetime cubic lattice of Euclidean signature. The path integral reads \cite{Peng:2025nfa, Xu:2024hyo}
\begin{align}
    Z=&\left[\prod_{\text{link} \: l} \int_{-\pi}^\pi \frac{dA_l}{2\pi}\right] \left[\prod_{\text{plaq} \: p} \sum_{s_p\in\mathbb{Z}} \right]
    \left[\prod_{\text{cube} \: c} \int_{-\pi}^\pi \frac{d\lambda_c}{2\pi} \: e^{i\lambda_c ds_c}\right] \nonumber\\
    &\exp \left\{\frac{ik}{4\pi}\sum_c \left[(A\cup dA)_c - (A\cup 2\pi s)_c - (2\pi s\cup A)_c\right] \right. \nonumber\\
    & \left. \hspace{.7cm} -\frac{1}{2e^2}\sum_p F_p^2 \right\} \ .
    \label{eqn:model}
\end{align}
The second line is the lattice Chern-Simons term of level $k$ \cite{Chen:2019mjw,Jacobson:2023cmr} which we will explain soon. The last line is the Maxwell term, which is needed for \emph{fundamental} reasons as explained in \cite{Chen:2019mjw, Peng:2025nfa, Xu:2024hyo}, with $e^2$ the Maxwell coupling in units where the lattice length is set to $1$.

The dynamical $U(1)$ gauge connection is $A_l\in (-\pi, \pi]$ on each lattice link $l$. The gauge flux around each plaquette $p$ takes the ``Villainized form \cite{Einhorn:1977qv, Sulejmanpasic:2019ytl, Chen:2019mjw} (see \cite{Xu:2024hyo} for an intuitive interpretation and \cite{Chen:2024ddr} for the mathematical exposition)
\begin{equation}
    F_p:=dA_p - 2\pi s_p \in \mathbb{R},
\end{equation}
where $dA_p$ is the lattice curl of $A_l$, and $s_p\in \mathbb{Z}$ is an independent dynamical variable, the ``Dirac string'' threading through $p$. Note $A_l$ is not apparently $2\pi$-periodic in $F_p$; rather, $F_p$ has a one-form $\mathbb{Z}$ gauge invariance
\begin{equation}
    A_l \mapsto A_l+2\pi m_l, \ \ \ s_p\mapsto s_p+dm_p, \ \ \ m_l\in\mathbb{Z} ,
    \label{eqn:1-form_Z}
\end{equation}
{and our restriction $A_l\in (-\pi, \pi]$ fixes this gauge.} The Dirac monopole number inside a cube $c$ is given by the lattice divergence $dF_c/2\pi=-ds_c\in\mathbb{Z}$, satisfying Dirac quantization. We have forbidden the monopoles by a Lagrange multiplier field $e^{i\lambda_c}\in U(1)$ \cite{Sulejmanpasic:2019ytl, Chen:2019mjw}. But over a noncontractible surface, $\sum_p F_p=-2\pi\sum_p s_p\in 2\pi\mathbb{Z}$ can still be a Dirac quantized nonzero total flux, as is desired for a $U(1)$ gauge theory. 

\begin{figure}[t]
    \centering
        \includegraphics[width=0.15\linewidth]{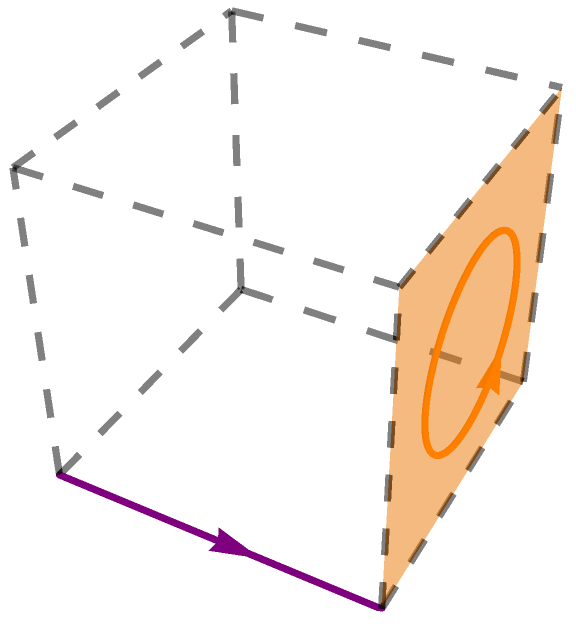} \hspace{.15cm}
        \includegraphics[width=0.15\linewidth]{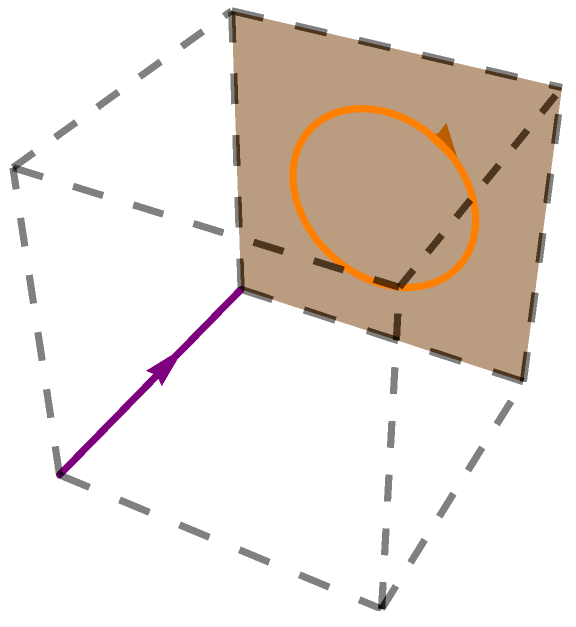} \hspace{.15cm}
        \includegraphics[width=0.15\linewidth]{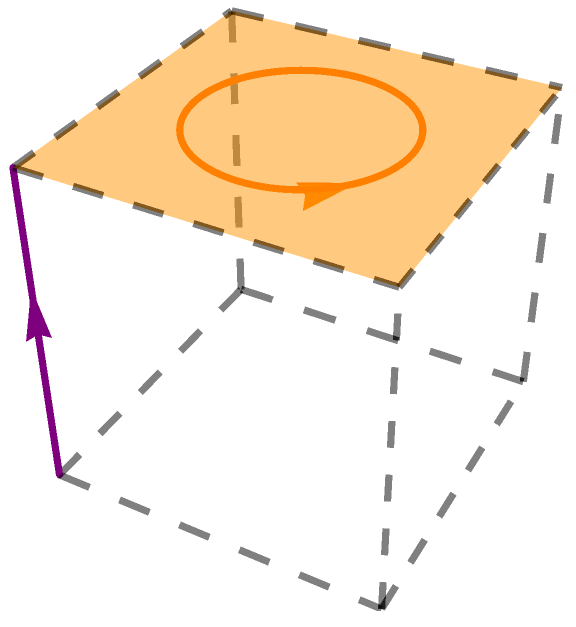}
        \\[.2cm]
        \includegraphics[width=0.15\linewidth]{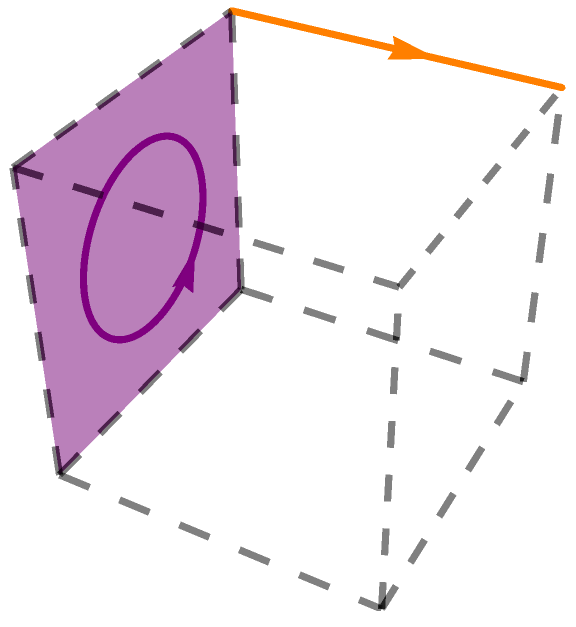} \hspace{.15cm}
        \includegraphics[width=0.15\linewidth]{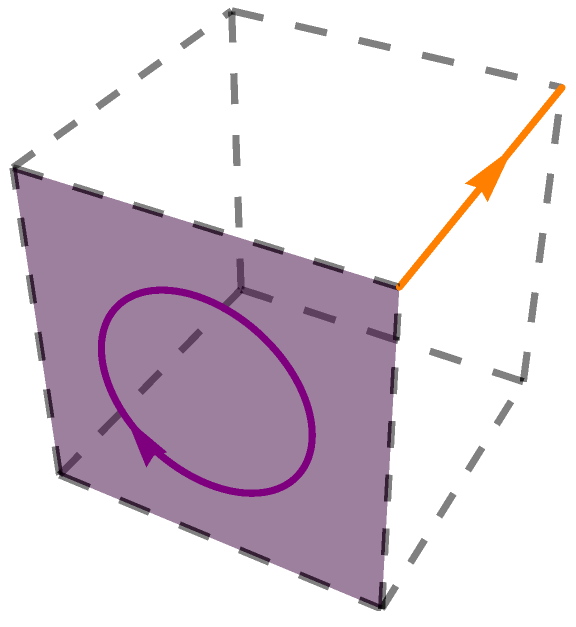} \hspace{.15cm}
        \includegraphics[width=0.15\linewidth]{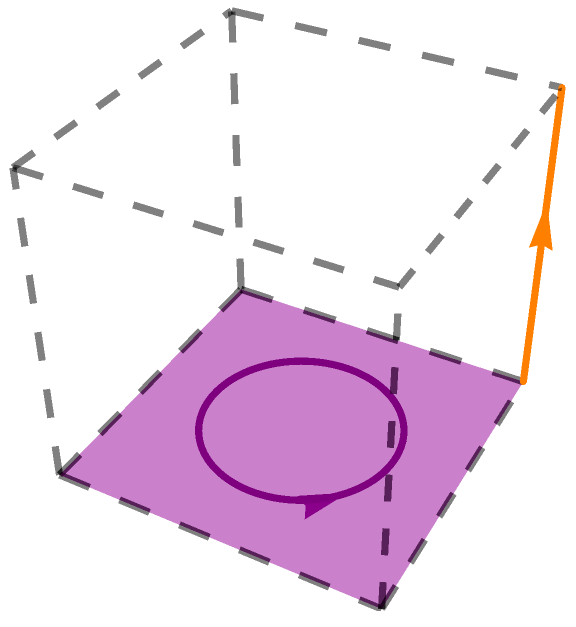}
    \caption{Top: The cup product $(X\cup Y)_c$ for a lattice 1-form $X$ and 2-form $Y$ on a cube $c$ is a sum of three terms, each with the $X$ value on a purple link multiplied to the $Y$ value on an associated orange plaquette. \hspace{1.5cm} Bottom: $(Y\cup X)_c$ is a different sum of three terms.}
    \label{fig:cup}
\end{figure}

The cup product in the lattice Chern-Simons term is defined as \cref{fig:cup}, which satisfies the Leibniz rule under lattice exterior derivative $d$. Thanks to this lattice Leibniz rule, in the absence of spacetime boundary, the model has gauge invariance under $A_l\mapsto A_l+d\phi_l$ (accompanied by suitable transformation of $\lambda_c$) for arbitrary $\phi_v\in\mathbb{R}$ on each vertex $v$. Furthermore, $A_l$ (and hence $\phi_v$) should be effectively $2\pi$-periodic in the sense of \cref{eqn:1-form_Z}. We can check the path integral is invariant under \cref{eqn:1-form_Z} if and only if $k\in 2\mathbb{Z}$---this is the lattice origin of the level quantization of the bosonic Chern-Simons term \cite{Chen:2019mjw,Jacobson:2023cmr}. In this Letter we will not consider spacetime boundary, but a careful treatment of the boundary can be found in \cite{Xu:2024hyo}. Also, the Chern-Simons term has some global symmetries and associated anomalies (see \cite{Jacobson:2023cmr}) that are not altered by the inclusion of the Maxwell term.

When $k$ is odd, under \cref{eqn:1-form_Z}, the partition function has a sign ambiguity. This ambiguity can be canceled by including in the path integral an extra fermionic path integral $z_\chi[s\ \mathrm{mod}\ 2]$, which is a functional of $s_p\ \mathrm{mod}\ 2$ and takes value $\pm 1$ \cite{Gu:2012ib}, that depends on spin structure (fermion boundary condition) data \cite{Gaiotto:2015zta}. The construction of $z_\chi$ on a cubic lattice \cite{Chen:2019mjw} is reviewed in Supplemental Material \cite{SupplementalMaterial}. This is the level quantization of fermionic Chern-Simons.

\

\noindent\emph{\bf Lattice modular $T$ operator--- \hspace{.0cm}} Consider a spatial torus with $L\times L$ lattice vertices. $\Tr\, e^{-\beta H}$ is constructed by the path integral $Z_{\mathrm{T}^3}$ over a three-torus $\mathrm{T}^3=S^1\times S^1\times S^1$ with $L\times L\times \beta$ vertices. In \cite{Xu:2024hyo} we have shown $Z_{\mathrm{T}^3}=|k| \exp(-\beta L^2 \epsilon_0+\cdots)$ where $|k|$ is the desired ground state degeneracy on the spatial torus, $\epsilon_0$ is some non-universal real-valued ground state energy density that can be removed by local counterterm, and ``$\cdots$'' are finite size effects that vanish exponentially in $\beta L^2$. Now, our task is to define the modular $T$ operator on the spacetime lattice and evaluate $\Tr\left( T\, e^{-\beta H}\right)$ as a lattice path integral.

\begin{figure}[t]
    \centering
    \includegraphics[width=0.6\linewidth]{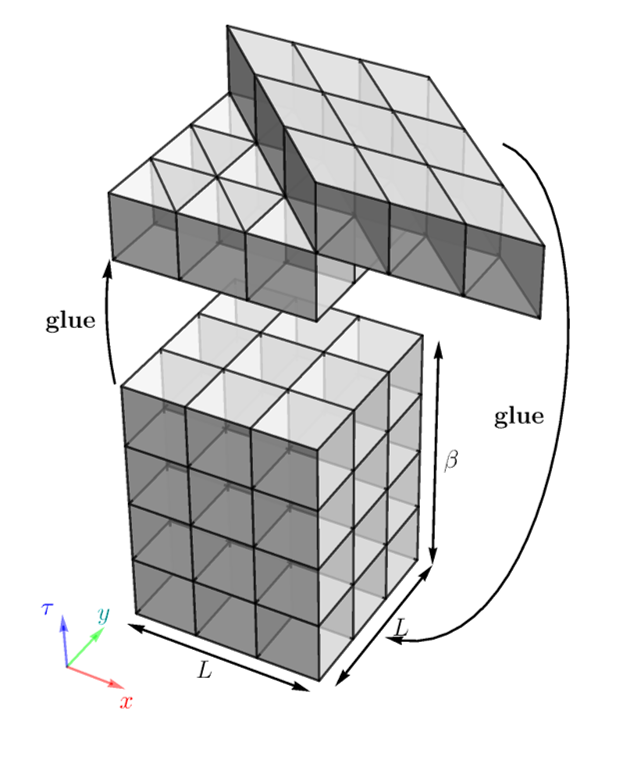}
    \caption{The lower part of the path integral constructs $e^{-\beta H}$, while the upper part defines the lattice modular $T$ operator, and they glue as indicated, to form $Z_\mathcal{T}=\Tr(T e^{-\beta H})$. {(Periodicity is understood in the $x,y$-directions, so the exposed triangles and the dangling ones in the lattice modular $T$ operator are identified.)}}
    \label{fig:T_path_int}
\end{figure}

First, $e^{-\beta H}$ before taking the trace is constructed by the path integral over $S^1\times S^1\times I$ of size $L\times L \times \beta$, as a function of the $A$ and $s$ fields on the $\tau=0$ and the $\tau=\beta$ boundaries, see the lower part of \cref{fig:T_path_int}. Then, the modular $T$ operator is constructed as in the upper part of \cref{fig:T_path_int}, with two layers of cubes in the $\tau$ direction glued in a twisted manner as shown, creating some triangular shaped plaquettes in between. To take the trace, as shown in \cref{fig:T_path_int}, the square grid at the bottom of the lattice $T$ operator is glued to that at the $\tau=\beta$ top layer of $e^{-\beta H}$, while the square grid at the top of the $T$ operator is glued to that at the $\tau=0$ bottom layer of $e^{-\beta H}$. (By ``glue'' we mean the $A_l$ and $s_p$ on the grids being identified are integrated or summed over.) The closed manifold thus created, which we denote as $\mathcal{T}$, is no longer a three-torus because a loop running across the $x$ direction becomes contractible, see Supplemental Material \cite{SupplementalMaterial}.

\begin{figure}[t]
    \centering
    \includegraphics[width=0.3\linewidth]{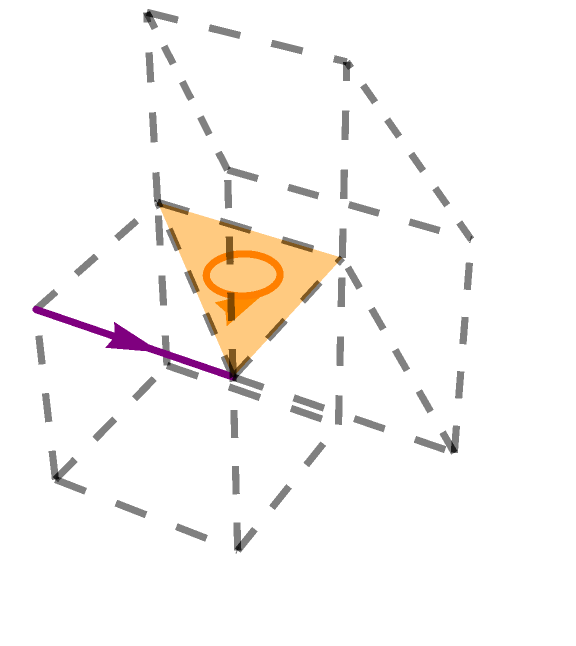} \hspace{.6cm}
    \includegraphics[width=0.3\linewidth]{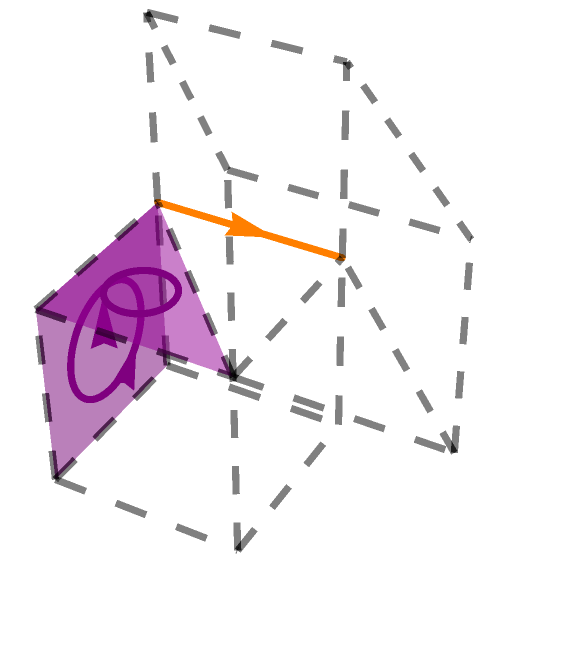}
    \caption{Left: $(X \cup Y)_c$ for cube $c$ at the lower layer of the lattice $T$ operator receives an extra term in addition to the usual ones in \cref{fig:cup}, given by the product of $X$ on the purple link and $Y$ on the orange triangular plaquette. \hspace{1.5cm} Right: $(Y \cup X)_c$ for cube $c$ at the lower layer of the lattice $T$ operator also receives an extra term: now $X$ on the orange link multiplies not only to $Y$ on the purple square plaquette on the side as usual, but to the sum of $Y$ on both the square and the triangular purple plaquettes.}
    \label{fig:special_cup}
\end{figure}

How are the lattice Chern-Simons and Maxwell terms defined on those special layers of lattice cells in the modular $T$ operator? It turns out, we can modify the cup product on the two special layers of cubes in $\mathcal{T}$ as \cref{fig:special_cup}, and find the lattice Leibniz rule and hence the gauge invariances remain valid over $\mathcal{T}$. When $k$ is odd, i.e., in fermionic theory, we also need to define $z_\chi$ on the special layers. A detailed, systematic approach toward these problems, inspired by \cite{Thorngren:2018ziu}, can be found in Supplemental Material \cite{SupplementalMaterial}. And regarding the Maxwell term, we still have it on each plaquette, including the triangular ones. In principle, the Maxwell coupling to be used on those triangular plaquettes is unimportant because it is only going to affect $\alpha$ in \cref{eqn:expected_result}. For concreteness, we can make a ``uniform'' choice with $e^2/2$ in replacement of $e^2$ on those triangular plaquettes.

Thus, $\langle T \rangle$ is to be evaluated as $Z_{\mathcal{T}}/Z_{\mathrm{T}^3}$; since $Z_{\mathrm{T}^3}$ is positive \cite{Xu:2024hyo}, the phase of $\langle T \rangle$ is the phase of $Z_{\mathcal{T}}$. More generally, to extract the full spectrum of $T$, we can evaluate the lattice version of $\langle T^m \rangle$ as $Z_{\mathcal{T}_m}/Z_{\mathrm{T}^3}$, where $Z_{\mathcal{T}_m}=\Tr\left( (T e^{-\beta H/m})^m \right)$, with $\mathcal{T}_m$ the closed manifold obtained by having $m$ copies of \cref{fig:T_path_int}, such that the top square grid of each $T$ is now glued to the bottom square grid of its next $e^{-\beta H/m}$---obviously, the ``next of the $m$th copy'' is the first. Unlike the $m=1$ case, in $\mathcal{T}_m$, a loop across the $x$ direction has nontrivial $\mathbb{Z}_m$ homological torsion, i.e. such a loop is non-contractible, but running around it $m$ times becomes contractible, see Supplemental Material \cite{SupplementalMaterial}. (Note that if $T$ exactly commutes with $H$, then $\Tr\left( (T e^{-\beta H/m})^m \right)=\Tr\left(T^m e^{-\beta H}\right)$, but now we do not expect this to exactly hold in the lattice realization, so each $e^{-\beta H/m}$ in between two $T$'s, with large enough $\beta$, is to relax the system back to the ground subspace after each $T$ operation.)

\

\noindent\emph{\bf Setup of calculation--- \hspace{.0cm}} The lattice model \cref{eqn:model} (and its fermionic version with $z_\chi$) is defined in a manifestly local manner. But to perform actual calculations, this is not the most convenient form. Rather, it is convenient to exploit the one-form $\mathbb{Z}$ gauge \cref{eqn:1-form_Z} to make $A_l$ $\mathbb{R}$-valued on most links while $s_p=0$ on most plaquettes. After doing so, the path integral becomes essentially a Gaussian integral, hence solvable, along with some extra treatments that depend manifestly on the spacetime topology---hence the price paid is that this alternative formulation is not manifestly local, but manageable. According to Sec. V and Appendix E of \cite{Xu:2024hyo}, after this procedure, on a spacetime that has no homological torsion---which is applicable to both $\mathrm{T}^3$ and $\mathcal{T}$ (see Supplemental Material \cite{SupplementalMaterial})---the path integral $Z$ defined by \cref{eqn:model} (or its fermionic version) is equal to $Z'$, where
\begin{align}
 Z'=&\left[\prod_{l} \int_{-\infty}^{\infty} \frac{dA_l'}{2\pi}\right]_{\mathrm{nlocFP}} \nonumber \\ 
 &\exp \left\{\frac{ik}{4\pi}\sum_c (A'\cup dA')_c - \frac{1}{2e^2}\sum_p (dA')_p^2\right\} \label{eqn:practical}
\end{align}
is a Gaussian integral of an $\mathbb{R}$-valued gauge field $A'_l$, but with an unusual \emph{nonlocal} Faddeev-Popov treatment ``$\mathrm{nlocFP}$'', which does the following: when we perform the Gaussian integral of the $\mathbb{R}$ gauge field, we ignore \emph{any} zero mode of $A'$, i.e., a closed one-form that satisfies $dA'=0$, regardless of whether the zero mode is an $\mathbb{R}$ gauge transformation (exact one-form, which is indeed removed in the usual Fadeev-Popov procedure, which is local) or a global holonomy (nonexact closed one-form, which is unusual to drop because it is nonlocal to do so). This $\mathrm{nlocFP}$ thus removes the diverging size of the 1-form $\mathbb{R}$ global symmetry that an actual $\mathbb{R}$ gauge theory would have had \cite{Chen:2019mjw}, rendering the partition function finite, as it should, because the theory really started out as \cref{eqn:model}, a manifestly local $U(1)$ rather than $\mathbb{R}$ gauge theory.

After having this alternative expression \cref{eqn:practical}, we can reseparate $\mathcal{T}$ into two parts as in \cref{fig:T_path_int}, and interpret the two equivalent expressions \cref{eqn:model} (or its fermionic version) and \cref{eqn:practical} of the path integral as
\begin{equation}
    \Tr\left( T e^{-\beta H} \right) = Z_{\mathcal{T}}= Z'_{\mathcal{T}}= \Tr\left( T' e^{-\beta H'} \right).
    \label{eqn:Z_Z'}
\end{equation}
In this alternative formulation, the lower part of \cref{fig:T_path_int} is constructing  $e^{-\beta H'}$ for the $\mathbb{R}$-valued $A'$. This is a Gaussian integral, and the result is a Gaussian function of those $A'_l$ on the links at the bottom ($\tau=0$) and top ($\tau=\beta$) boundaries of the path integral. We will present the details in Supplemental Material \cite{SupplementalMaterial}. Essentially, after Fourier transforming the $x$ and $y$ directions, the Gaussian coefficients in $e^{-\beta H'}$ have an analytical expression that is lengthy; alternatively, their numerical values can be evaluated algorithmically, with time cost scaling as $L^2\times \log_2 \beta$, where there are $L^2$ Fourier modes, and the $\log_2 \beta$ is due to iteratively doubling the $\tau$ direction size. For the ground subspace, we will take $\beta\rightarrow \infty$.

On the other hand, the upper part of \cref{fig:T_path_int} is now constructing the modular $T$ operator for the $\mathbb{R}$-valued $A'$, denoted as $T'$. Because of the complicated lattice in $T'$, although it is still a Gaussian integral, we do not have a closed form expression for the result, so we will algorithmically evaluate the numerical value of the desired $\Tr\left( T' e^{-\beta H'} \right)$; see Supplemental Material \cite{SupplementalMaterial}. More exactly, the $x$ direction in the $T'$ part of \cref{fig:T_path_int} can still be Fourier transformed, so only the $y$ direction needs to be handled numerically. The time cost scales as $L\times L^3$, where the $L$ is the number of Fourier modes in the $x$ direction, and, after that, the $L^3$ arises from the manipulations of matrices in the $y$ coordinates.

More generally, for $\mathcal{T}_m$, according to Appendix E of \cite{Xu:2024hyo}, we have
\begin{align}
    Z_{\mathcal{T}_m} = Z'_{\mathcal{T}_m} \sum_{j=0}^{m-1} e^{i\pi k j^2/m}(-1)^{kj}.
    \label{eqn:Z_Z'_general}
\end{align}
Here, the appearance of an additional factor [compared to \cref{eqn:Z_Z'}] is due to the extra $\mathbb{Z}_m$ torsion of a loop running across the $x$ direction. (For odd $k$ fermionic theories, recall $T$ is only defined if the fermions are periodic across the $x$ direction, and \cref{eqn:Z_Z'_general} is valid for this case. If the fermions are anti-periodic across the $x$ direction, we can nonetheless still define $T^m$ for even $m$, see Supplemental Material \cite{SupplementalMaterial}, and in this case we have \cref{eqn:Z_Z'_general} but with the $(-1)^{kj}$ factor removed.) So, still, our remaining task is to evaluate \cref{eqn:practical}, now decomposed as $Z'_{\mathcal{T}_m}=\Tr\left( (T' e^{-\beta H'/m})^m \right)$.

\

\noindent\emph{\bf Results--- \hspace{.0cm}} Numerical value of the complex phase of the path integral $Z_{\mathcal{T}}=\Tr\left( T e^{-\beta H} \right)=\Tr\left( T' e^{-\beta H'} \right)$ is evaluated using the setup above for various values of $k$, $e^2$, and $L$; since \cref{eqn:model} obviously becomes its complex conjugation upon flipping $k$, it suffices to consider $k>0$; and we take $\beta=2^{12}$, which is sufficiently large to reach the ground subspace. For each value of $k$ and $e^2$, we fit the phase with the ansatz
\begin{align}
    2\pi C_0 + 2\pi C_2 L^2
\end{align}
using the $L$'s in a range $L_1\leq L \leq L_2$ for some large enough $L_1, L_2$. Here $C_0$ is expected to be $-1/8+1/24=-1/12$ from the universal piece of \cref{eqn:expected_result}, and $C_2$ is from the imaginary part of the nonuniversal $\alpha$ in \cref{eqn:expected_result}. 

\begin{figure}[t]
    \centering
    \includegraphics[width=0.9\linewidth]{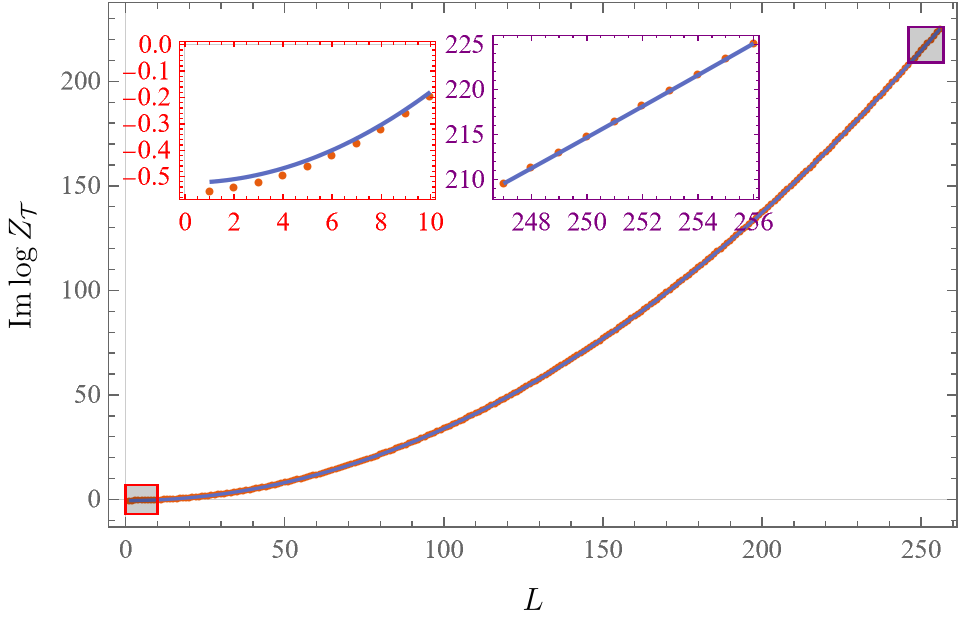}
    \caption{The orange dots are phases of $Z_{\mathcal{T}}$ for $k=1$, $e^2=1$ at different $L$'s. The blue curve is the quadratic fit using the $L$'s from $L_1=129$ to $L_2=256$.}
    \label{fig:Im_sample}
\end{figure}

\begin{figure}[t]
    \centering
    \includegraphics[width=.9\linewidth]{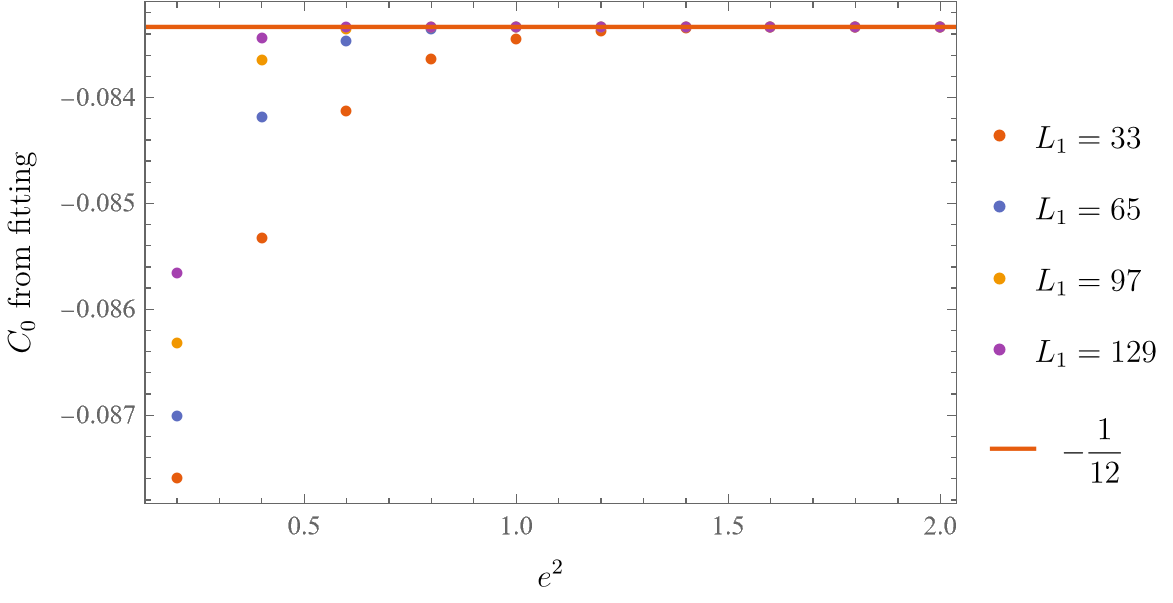}
    \caption{$C_0$ fitted for $k=1$ using $L_2=256$ and increasing values of $L_1$ and compared to the expected value $-1/12$.}
    \label{fig:Im_e2}
\end{figure}

\begin{figure}[t]
    \centering
    \includegraphics[width=.8\linewidth]{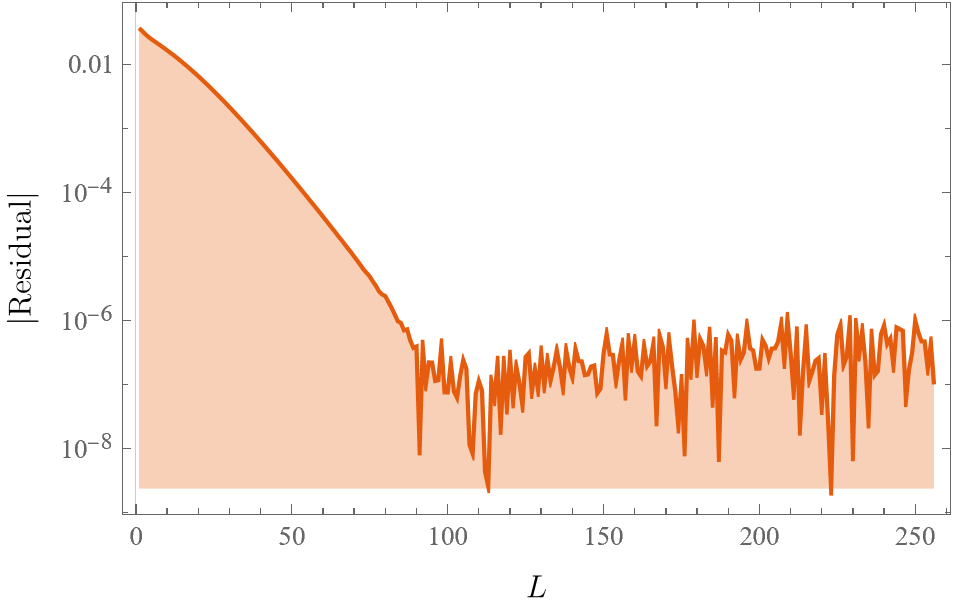}
    \caption{Deviations of the plotted orange points in \cref{fig:Im_sample} from the blue quadratic fitting curve.}
    \label{fig:Im_residue_sample}
\end{figure}

An example of such fit is shown in \cref{fig:Im_sample} and we can see the quadratic ansatz fits well overall. The $C_0$ fitted for $k=1$ and different values of $e^2$, fixing $L_2=256$ and using increasing values of $L_1$, is shown in \cref{fig:Im_e2}. As we can see, for larger values of $e^2$, the fitted $C_0$ agrees with the desired $-1/12$ very well. Meanwhile, for smaller $e^2$, there is some deviation when $L_1$ is small, but the deviation fades away as we increase $L_1$, the smallest system size used in the fitting. This suggests the deviation is a finite size effect. To confirm this, in \cref{fig:Im_residue_sample} we plot the fitting residual of \cref{fig:Im_sample}; alternatively, we can plot the deviation of $C_0$ from $-1/12$ in \cref{fig:Im_e2} at some fixed $e^2$ as a function of $L_1$, and we will get a similar plot. This confirms that the deviation is indeed a finite size effect that vanishes exponentially with $L$.

\begin{figure}[t]
    \centering
    \includegraphics[width=.9\linewidth]{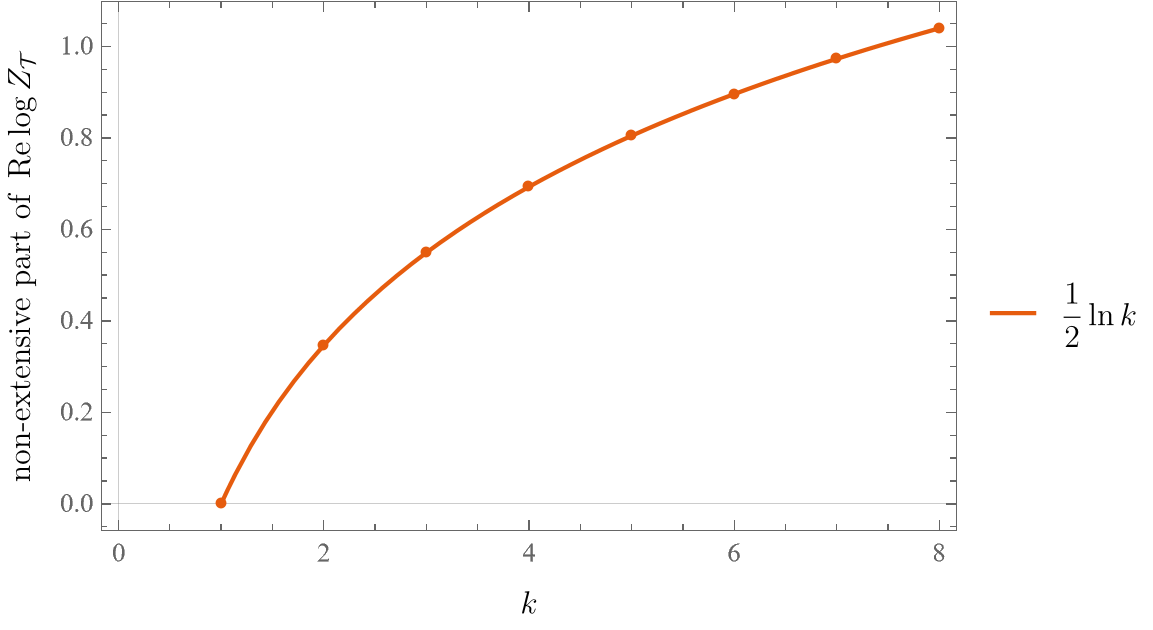}
    \caption{Numerical result for nonextensive part of $\operatorname{Re}\ln Z_\mathcal{T}$. We set $e^2=2$, $L_1=129$, $L_2=256$, $1\leq k\leq8$.}
    \label{fig:Re}
\end{figure}

Thus, we have confirmed that the evaluated complex phase agrees with the anticipated result \cref{eqn:expected_result}, with finite size effects vanishing exponentially. We have further confirmed that the magnitude of $Z_{\mathcal{T}}/Z_{\mathrm{T}^3}$, especially its universal part, also agrees with \cref{eqn:expected_result}, see \cref{fig:Re}.

\begin{figure}[t]
    \centering
    \includegraphics[width=.93\linewidth]{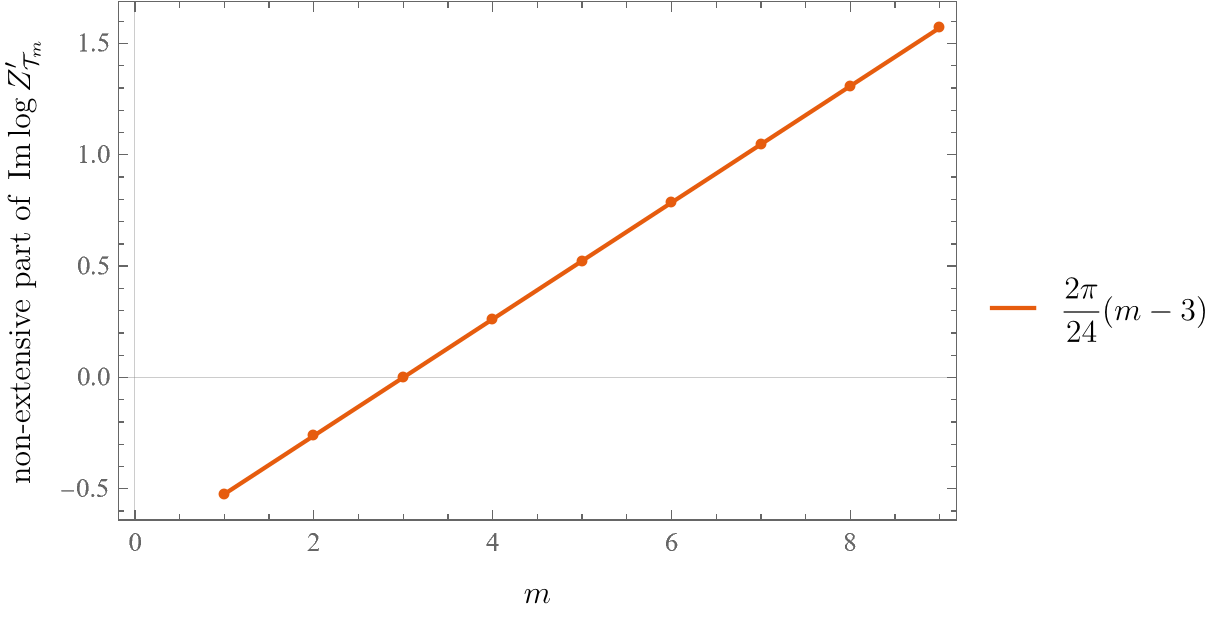}
    \includegraphics[width=.9\linewidth]{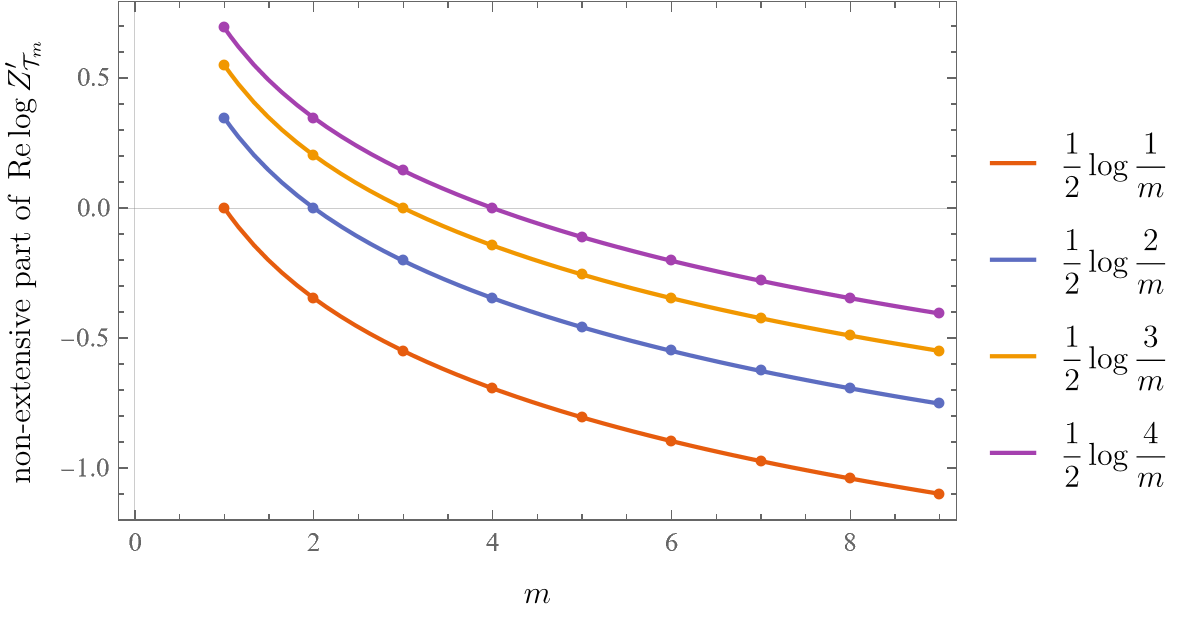}
    \caption{Numerical results for the nonextensive parts of $\operatorname{Im}\ln {Z'_{\mathcal{T}_m}}$ and $\operatorname{Re}\ln {Z'_{\mathcal{T}_m}}$. We set $e^2=2$, $L_1=129$, $L_2=256$, $1\leq m \leq 9$, and $k=1$ in the upper panel (the result is actually independent of $|k|$) and $1\leq k \leq 4$ in the lower panel.}
    \label{fig:Im_n_Re_n}
\end{figure}

To extract the full spectrum of $T$, we also evaluate $\langle T^m \rangle$ for $m>1$ as $Z_{\mathcal{T}_m}/Z_{\mathrm{T}^3}$ following \cref{eqn:Z_Z'_general}. Let us first explain what we shall expect for the $Z'_{\mathcal{T}_m}$ to be evaluated. We use the following facts: for even $k$,
\begin{equation}
    \frac{1}{\sqrt{|k|}}\sum_{n=0}^{|k|-1}e^{-i\pi \frac{m}{k}n^2} = e^{-i2\pi\operatorname{sgn}{k}/8}\frac{1}{\sqrt{m}}\sum_{j=0}^{m-1}e^{i\pi \frac{k}{m}j^2}
    \label{eqn:gauss_sum_even}
\end{equation}
and for odd $k$,
\begin{equation}
    \frac{1}{\sqrt{|k|}}\sum_{n=0}^{|k|
    -1}e^{-i\pi \frac{m}{k}(n+1/2)^2} = e^{-i2\pi\operatorname{sgn}{k}/8}\frac{1}{\sqrt{m}}\sum_{j=0}^{m-1}e^{i\pi \frac{k+m}{m}j^2}.
\end{equation}
Here, the left-hand sides are the anticipated $h_n$ contributions in \cref{eqn:modular_T}. Comparing the right-hand sides to \cref{eqn:Z_Z'_general}, what we shall anticipate would be
\begin{equation}
    Z'_{\mathcal{T}_m}=\sqrt{\frac{|k|}{m}}\: e^{-i2\pi \operatorname{sgn}k (1/8-m/24)}e^{-\epsilon_0 \beta L^2 - m\alpha L^2 + \cdots}.
\end{equation}
(For odd $k$ fermionic theories, recall the fermions must be periodic across the $x$ direction for $T$ to be defined. However, if we consider $T^m$ with even $m$ only, we can also allow the anti-periodic situation, in which case the $+1/2$ modification will be removed (see Supplemental Material \cite{SupplementalMaterial}), but the $(-1)^{kj}$ in \cref{eqn:Z_Z'_general} is also removed, resulting in the same anticipation for $Z'_{\mathcal{T}_m}$.) The numerical procedure for $m=1$ can now be applied to generic $m$. We indeed obtain the anticipated form of $Z'_{\mathcal{T}_m}$, see \cref{fig:Im_n_Re_n}, thus confirming the anticipated full spectrum of $T$ in the lattice Chern-Simons theory.

The code for these computations is available as a separate file.

\

\noindent\emph{\bf Conclusion--- \hspace{.0cm}} In this Letter we pinned down the last and perhaps the most crucial property desired for a ``successful'' realization of $U(1)$ Chern-Simons theory on the lattice \cite{Peng:2025nfa,Xu:2024hyo}---the framing anomaly (on top of the other key properties studied in \cite{Xu:2024hyo}). This not only paints a full picture for the decades-long story of ``how to realize $U(1)$ Chern-Simons on the lattice,'' but also serves as a concrete example of a UV-complete manifestation of a framing anomaly in a way that is different from the ones in which the chirality comes from fermion bands \cite{Kitaev:2005hzj}.

One immediate and important follow-up question is whether one can also pin down the chiral central charge of the proposed lattice realization of non-Abelian Chern-Simons \cite{Chen:2024ddr,Zhang:2024sgm}, which is necessarily interacting, at least in certain limits.

And it is desirable to cast what we have done in the Hamiltonian formalism. (Note that in the Hamiltonian formalism, there are other interesting ways to realize the modular $T$ operator \cite{You:2015cga}.) The main reason is that it has been fruitful to study chiral central charge through the lens of entanglement \cite{Kitaev:2005hzj}---for example, \cite{Kim:2021chiral,Kim:2021modular} showed chiral central charge is a bulk property using the modular commutator instead of framing anomaly, and see, e.g., Refs. \cite{Li:2024iwn,Li:2025sbv} for more recent developments. We hope our particular construction can be integrated as a concrete example into this general perspective.

\

\noindent\emph{Acknowledgments--- \hspace{.0cm}} We appreciate Peng Zhang for the suggestion to evaluate $\langle T^m \rangle$ to extract the full spectrum of $T$, and Meng Cheng for {discussions about the modular $T$ operator} in fermionic theories. This work is supported by NSFC under Grants No.~12447104, No.~12174213, and No.~12342501.

\

\noindent\emph{Data availability---} The data supporting the results in this paper were obtained through the Julia language script available at \cite{Xu:2026code}.

\bibliographystyle{apsrev4-2}
\bibliography{biblio}

\end{document}

\title{Supplemental Material}
\author{Ze-An Xu}
\affiliation{Institute for Advanced Study, Tsinghua University, Beijing, 100084, China}
\author{Jing-Yuan Chen}
\affiliation{Institute for Advanced Study, Tsinghua University, Beijing, 100084, China}
\maketitle

\section{Bosonic vs Fermionic Modular $T$ Operator}
\label{sec:mod_T}

In this section we present an intuitive understanding of the notion of ``a ground state's topological spin'' that appears in the modular $T$ operator, especially in the fermionic theories, from the perspective of continuum field theory.

\begin{figure}[htbp]
    \centering
   \includegraphics[width=0.3\linewidth]{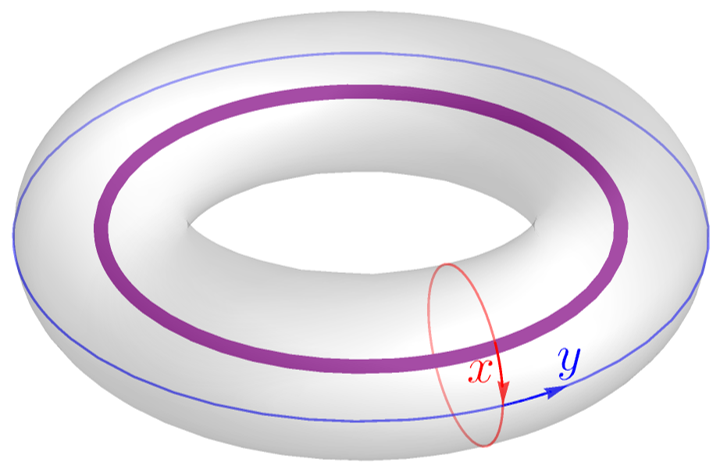}
   \caption{Path integral over a Euclidean solid torus spacetime with some loop operator (purple) insertion in the interior. This constructs a ground state wavefunction on its boundary torus.}
    \label{fig:solid_torus}
\end{figure}

In the continuum, it is well-known that a nice way to construct the ground states of a bosonic gapped field theory (whose IR limit is, generically, a bosonic topological field theory) on a spatial torus is to consider the path integral over a Euclidean solid torus with different loop operator insertions in the interior, see \cref{fig:solid_torus}, so that, for each given loop insertion, the resulting partition function as a function of the Dirichlet boundary values of the fields is interpreted as a ground state wavefunction \cite{Witten:1988hf}.

For bosonic $U(1)$ Chern-Simons in particular, the $|k|$ ground states, labeled as $|n\rangle$ $(n=0, \cdots, |k|-1)$, are obtained by such path integrals with Wilson loop (anyon worldloop) insertions of different powers, which we denote as $W^n$ $(n=0, \cdots, |k|-1)$, and $W^{|k|}$ is equivalent to the trivial loop $1$. From \cref{fig:solid_torus}, it is intuitive that if we perform a modular $T$ operation on the boundary torus, we will effectively accumulate a $2\pi$ twist of the inserted Wilson loop---or say a $2\pi$ rotation of the inserted anyon---as we go around the $y$-direction, and thus we will obtain a phase given by the spin of the inserted anyon, $e^{-i\pi n^2/k}$, which we shall regard as the topological spin of the state $|n\rangle$. Since $k$ is even, indeed $n+k$ is equivalent to $n$ in $e^{-i\pi n^2/k}$.

For fermionic theories, we may do the same solid torus path integrals, but then, as is well-known, on the boundary torus the fermions must be regarded as being \emph{anti-periodic} around the $x$-direction. To understand this, consider a fermion worldloop that ``travels along a straight line'' in the $x$-direction from the perspective of the boundary torus; but from the perspective of the solid torus, this is not a straight line---rather, the direction of travel has made a $2\pi$ rotation, leading to a $e^{i\pi}$ fermionic Berry phase. Hence in the boundary torus perspective we should interpret this $(-1)$ as the fermion being anti-periodic around the $x$-direction. (On the other hand, we are free to choose the fermions to be periodic or anti-periodic around the $y$-direction.) 

As we said in the main text, the modular $T$ operation is obviously not well-defined when the fermions on the torus are anti-periodic around the $x$-direction. If we pretend it is well-defined and do what we did for the bosonic case, then in the phase $e^{-i\pi n^2/k}$ that we get, $n$ is no longer equivalent to $n+k$ because $W^k$ is a fermion loop which has a $e^{i\pi}$ phase under the $2\pi$ twist. But there should really only be $|k|$ distinct ground states. Thus we run into contradiction. (On the other hand, if we only consider $T^m$ for even $m$, then there is no problem with the fermions being anti-periodic in the $x$-direction.)

So the problem becomes, how to construct the path integral so that the fermions on the boundary torus are \emph{periodic} in the $x$-direction. We should insert a special defect loop that is not a Wilson loop: the $(-1)\in\mathbb{Z}_2$ flux tube $w$ of (the background gauge field of) the fermion parity $\mathbb{Z}_2$ symmetry. Clearly, by definition, this $w$ insertion turns the fermions on the boundary torus from being anti-periodic in the $x$-direction to being periodic. The fermionic ground states that are periodic in the $x$-direction are therefore constructed by having $W^n w$ $(n=0, \cdots, |k|-1)$ loop insertions.

It turns out that, the fermion parity $\mathbb{Z}_2$ flux tube $w$ behaves as ``half of a Wilson loop'', $w\sim W^{1/2}$. To understand this, first recall that $W^k$ is a fermionic Wilson loop, whose braiding phase with any $W^n$ is trivially $e^{-i2\pi nk/k}=1$; on the other hand, by definition, the fermionic loop should braid with the fermion parity $\mathbb{Z}_2$ flux tube $w$ with phase $e^{i\pi}$, therefore $w$ appears as if having ``$n=1/2$''. More explicitly, and more familiarly in the quantum Hall context \cite{Lapa:2019uim,Cheng:2025ube}, in terms of the field theoretic Lagrangian, we add the term 
\begin{align}
   \frac{i}{2\pi}\int_{\text{3d}} \mathbf{A}\wedge dA = \frac{i}{2\pi}\int_{\text{3d}} A\wedge d\mathbf{A}
   \label{eqn:backgnd_U1}
\end{align}
where $\mathbf{A}$ is a background $U(1)$ gauge field that has a narrow $\pi$ flux tube threading around the solid torus, and flat elsewhere. Since a fermion is a $2\pi$ flux of the dynamical $U(1)$ gauge field $A$ in a fermionic Chern-Simons, such a coupling means the fermion has charge $1$ under $\mathbf{A}$, therefore a $\pi$ flux of $\mathbf{A}$ is indeed indistinguishable from $w$, traveling around which a fermion will pick up a $(-1)$ phase---this is the celebrated spin-charge relation in the quantum Hall context \cite{Wen:1995qn,Lapa:2019uim,Cheng:2025ube}. On the other hand, since a Wilson loop is to add $i\oint_{1d} A$, the coupling \cref{eqn:backgnd_U1} means a narrow $\pi$ flux tube of $\mathbf{A}$ is indeed like half of a Wilson loop, i.e. $w\sim W^{1/2}$; the other way around, it is also true that $W^n$ can be realized as a narrow $2\pi n$ flux tube of $\mathbf{A}$---this is how anyons can be adiabatically created in quantum Hall systems. Thus, for a ground state constructed by the insertion $W^n w\sim W^{n+1/2}$ $(n=0, \cdots, |k|-1)$ in the solid torus, when we perform the modular $T$ operation which effectively twists the insertion loop by $2\pi$, we will obtain a phase $e^{-i\pi (n+1/2)^2/k}$ \cite{Belov:2005ze,Lapa:2019uim,Cheng:2025ube}---in which $n$ and $n+k$ indeed become equivalent---that shall be regarded as the topological spin of the state $|n+1/2\rangle$. We can also decompose the phase as $e^{-i\pi n^2/k}e^{-i\pi n/k}e^{-i\pi^2/4\pi k}$, and interpret $e^{-i\pi n^2/k}$ as the spin of the $W^n$ anyon, $e^{-i\pi n/k}$ as the Aharonov-Bohm phase of the $W^n$ anyon of charge $n/k$ under $\mathbf{A}$ with a $\pi$ flux of $\mathbf{A}$, and $e^{-i\pi^2/4\pi k}$ as the $1/k$ Hall conductivity phase of a $\pi$ flux of $\mathbf{A}$.

So far we have focused on the insertion dependence, or say state dependence, of the modular $T$ operator, but $T$ also has a ubiquitous phase regardless of the insertion---the framing anomaly phase $e^{i2\pi\sgn(k)/24}$. Its appearance in the continuum solid torus setting is essentially explained by the computation that we demonstrated at the beginning of Section 6 of our previous work \cite{Xu:2024hyo}, setting $\delta x=L_x$.

Understanding the modular $T$ action on the boundary torus of a solid torus, as we did above, is very important, for this can generate a global change of framing on a generic 3d spacetime manifold \cite{Witten:1988hf}: Carve a solid torus region out of the spacetime manifold, perform a modular $T$ on the boundary of the solid torus that we carved out, and then glue it back. The topology of the spacetime manifold ends up unchanged, but the global framing has changed.

The discussions above are for continuum field theories. How about in our lattice construction? In our main text we did not consider the action of modular $T$ on the boundary of a lattice solid torus, but on one boundary of a lattice $S^1\times S^1 \times I$ (that constructs $e^{-\beta H}$) for technical conveniences. In principle we can also consider building a lattice solid torus, for example as a $I\times I \times S^1$ cubic lattice of size $L\times L\times 4L$, which has a $4L\times 4L$ square lattice boundary torus. We can include Wilson loop (or more optimally, lattice 1-form $\mathbb{Z}_{|k|}$ generator \cite{Jacobson:2023cmr}) insertion $W^n$ threading around the lattice solid torus. For fermionic theories, the lattice realization of $w$ will be explained at the end of \cref{sec:fermionic_path_integral}, see \cref{eqn:fermion_parity_defect} and \cref{fig:cross_section}.

Although for technical reasons we have not attempted the following evaluation, given our main results we expect the following to be true. Given two lattice solid tori realized as the above, gluing along their $4L\times 4L$ boundaries together without or with a $90$-degrees rotation (i.e. without or with a lattice modular $S$ transformation), we obtain a lattice realization of $S^2\times S^1$ or $S^3$ spacetime. Now, we can realize a lattice version of change of framing \cite{Witten:1988hf} by inserting our lattice modular $T$ operator in between the two $4L\times 4L$ boundaries that we are about to glue, and doing so does not change the $S^2\times S^1$ or $S^3$ topology of the resulting spacetime (in contrast to the change from $\mathrm{T}^3$ to $\mathcal{T}$ in the method we adopted in our main text). We expect the partition function to change with a universal phase given by the modular $T$ eigenvalue associated with the state we constructed that depends on the loop insertion, i.e. given by the framing anomaly phase $e^{i2\pi\sgn(k)/24}$ along with the insertion dependent topological spin phase $e^{-i\pi n^2/k}$ (bosonic) or $e^{-i\pi (n+1/2)^2/k}$ (fermionic) that we discussed above.

\section{Topology of $\mathcal{T}$ and $\mathcal{T}_m$}
\label{sec:top}

As we have seen in the main text, the computation of $\Tr((T e^{-\beta H/m})^m)$ has been translated into calculating the partition function $Z_{\mathcal{T}_m}$ on a ``twisted'' torus $\mathcal{T}_m$. We first consider the simplest case where $m=1$. It is not hard to see that, topologically, the twisted spacetime $\mathcal{T}=\mathcal{T}_1$ admits a discretization that involves only one cube as the above, where the eight vertices are identified as one, links of the same colors are identified, and faces are identified according to the arrangements of the links around. 

\begin{figure}[htbp]
    \centering
    \includegraphics[width=0.26\linewidth]{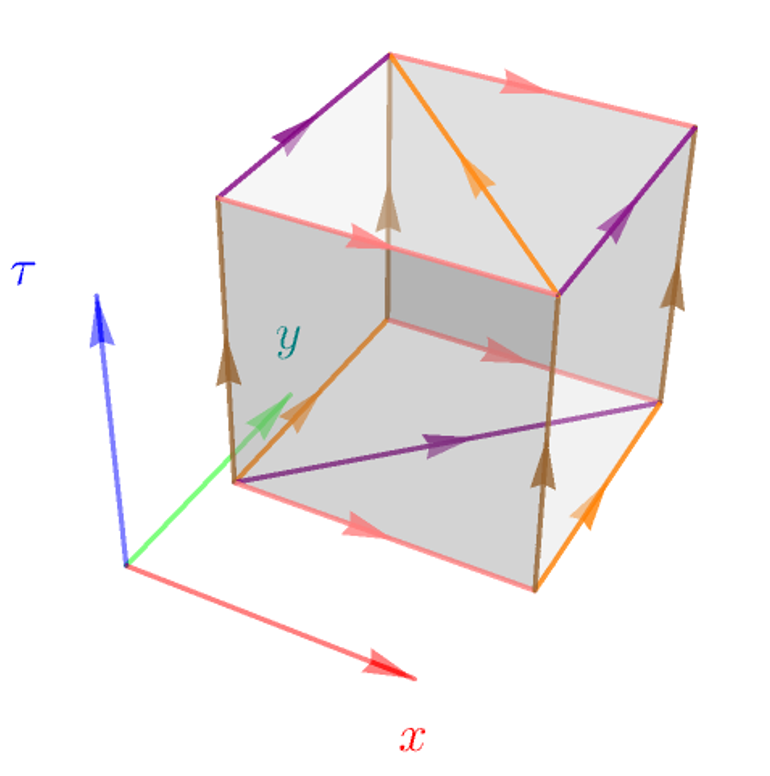}
   \caption{A discretization of the twisted closed manifold $\mathcal{T}$.}
    \label{fig:twisted_torus}
\end{figure}

Its homology can be easily calculated. $H_0(\mathcal{T})=\mathbb{Z}=H_3(\mathcal{T})$ as usual, while
\begin{equation}
H_1(\mathcal{T})=\mathbb{Z}\oplus\mathbb{Z}
\end{equation}
with one $\mathbb{Z}$ generated by the vertical brown link (a non-contractible loop), and the other $\mathbb{Z}$ generated by the purple or the orange link (a non-contractible loop)---the two are homologous to each other by sliding across the square on the right side; note the pink link is homologously trivial, by sliding across the surface formed by the triangle at the top right and the square on the right side. And
\begin{equation}
H_2(\mathcal{T})=\mathbb{Z}\oplus\mathbb{Z}
\end{equation}
with one $\mathbb{Z}$ generated by the square (a non-contractible closed surface) at the front, and the other $\mathbb{Z}$ generated by the square (a non-contractible closed surface) at the top. This shows that $\mathcal{T}$ is indeed topologically distinct from $\mathrm{T}^3$, since the latter has $H_1(\mathrm{T}^3)=H_2(\mathrm{T}^3)=\mathbb{Z}\oplus\mathbb{Z}\oplus\mathbb{Z}$. Notably, the homology groups of $\mathcal{T}$, like those of $\mathrm{T}^3$, have no torsion, i.e. no $\mathbb{Z}_p$ subgroups, so if a loop is non-contractible, going around it any number of times is still a non-contractible loop.

In physics, it is more often that we look at the cohomology. We usually take coefficient $\mathcal{A}=\mathbb{Z}$ or $\mathbb{Z}_n$ or $\mathbb{R}$, and we have
\begin{equation}
    H^1(\mathcal{T}; \mathcal{A}) = \mathcal{A}\oplus\mathcal{A}
\end{equation}
because a closed, non-exact $\mathcal{A}$-valued gauge field configuration on the links can take an arbitrary value on the brown link, and another arbitrary value simultaneously on the purple and the orange link (since the two are homologous), and $0$ on the pink link (since it is homologously trivial). And \begin{equation}
    H^2(\mathcal{T}; \mathcal{A}) = \mathcal{A}\oplus\mathcal{A}
\end{equation} 
because a closed, non-exact $\mathcal{A}$-valued 2-form gauge field configuration on the faces can take one arbitrary value on the front square, and another arbitrary value on the top square (how the value distributes between the two triangles is an exact difference). By contrast, $H^1(\mathrm{T}^3; \mathcal{A})=H^2(\mathrm{T}^3; \mathcal{A}) = \mathcal{A}\oplus\mathcal{A}\oplus\mathcal{A}$.

Let us have a simple check of the validity of our setup, using a simpler non-chiral theory. Consider a $\mathbb{Z}_n$ lattice gauge theory with flat gauge field (which is the Lagrangian version for the ground subspace of the $\mathbb{Z}_n$ toric code). Based on the above, the configuration space on our one-cube-discretization of $\mathcal{T}$ has a size of $|H^1(\mathcal{T}; \mathbb{Z}_n)|=n^2$, and the gauge redundancy is $n$, so the partition function on $\mathcal{T}$ is $Z_{\mathcal{T}}=n^2/n=n$, in contrast to $Z_{\mathrm{T}^3}=n^3/n=n^2$.
(More generally, say there are $N_0$ vertices, then there are a total of $n^{N_0}$ gauge redundancies, but only $n^{N_0}/|H_0|$ will lead to a non-trivial gauge transformation on the gauge field, while the remaining $|H_0|$ are global transformations that do not actually change the gauge field. So the partition function for flat $\mathbb{Z}_n$ gauge field on a spacetime without torsion is $Z=|H_1| n^{N_0}/|H_0|$, if we do not remove gauge redundancy. But local product contribution to $Z$ is unimportant as they can be viewed as local counter terms in the Hamiltonian density, so if we want, we can ignore the local product factor $n^{N_0}$, which is the size of gauge redundancy, leading to $Z=|H_1|/|H_0|$.) In comparison, we can evaluate $\langle T \rangle$ using the ground states on a spatial torus, which are labeled by $(e,m)\in \mathbb{Z}_n^2$:
\begin{equation}
    \langle T \rangle = \frac{\sum_{e, m =0}^{n-1} e^{i2\pi em/n}}{\sum_{e, m =0}^{n-1}1} = \frac{n}{n^2} \ .
\end{equation}
This provides the desired simple consistency check, that $\langle T \rangle=Z_{\mathcal{T}}/Z_{\mathrm{T}^3}$.

More generally, consider the case where $m>1$. Previously, when we discretized $\mathcal{T}$, we identified the top and bottom faces of a $T^2 \times [0,1]$ in a ``twisted'' manner according to the colors of the links in \cref{fig:twisted_torus}. Now, $\mathcal{T}_m$ can be viewed as gluing $m$ copies of $T^2 \times [0,1]$ sequentially, where the top face of each copy is identified with the bottom face of the next in the same ``twisted'' manner.

Let us denote the $x$- and $y$-direction loops at the bottom face of the $j$th copy of $T^2 \times [0,1]$ as $x_j$ and $y_j$, shown in pink and orange respectively at the bottom of \cref{fig:twisted_torus}. Sliding $x_j$ along the $z$-direction, we see all $x_j$'s are homologous to each other, with the class denoted as $[x]$. On the other hand, sliding $y_l$ along the $z$-direction, it becomes the purple loop, and thus we observe $[y_j]=[y_{j+1}]+[x]$. Repeating this process $m$ times, we eventually return to the initial copy of $T^2 \times [0,1]$, yielding $[y_j]=[y_j]+m[x]$. This shows $\mathcal{T}_m$ has a $\mathbb{Z}_m$ torsion, with the pink loop $[x]$ the corresponding generator. That is, compared to $\mathcal{T}=\mathcal{T}_1$, now we in general have \begin{align}
    H_1(\mathcal{T}_m)=\mathbb{Z}_m\oplus\mathbb{Z}\oplus\mathbb{Z},
\end{align}
while $H_0, H_2, H_3$ are the same as before. Correspondingly, in the cohomology of $\mathcal{T}^m$, compared to that of $\mathcal{T}$, by the universal coefficient theorem $H^1(\mathcal{T}_m; \mathcal{A})$ has an additional torsion part $\mathrm{Hom}(\mathbb{Z}_m, \mathcal{A})$ and $H^2(\mathcal{T}_m; \mathcal{A})$ has an additional torsion part $\mathcal{A}/m\mathcal{A}$.

\section{Cup and Higher Cup Products on General Lattices}
\label{sec:cup}

Practically, this section is a preparation for the next section on the fermionic Chern-Simons theories---our lattice modular $T$ operator involves lattice cells of unusual shapes, so we need an approach to define the cup product structures on general lattices other than the usual simplicial or cubic ones. On the other hand, formally, the approach we develop in this section, inspired by \cite{Thorngren:2018ziu}, is interesting in its own right, and does not seem to have been explicitly introduced in the literature.

The definition of continuous Chern-Simons action involves the wedge product $\wedge$. A corresponding ``lattice wedge product'' is naturally needed for the lattice version. The cup product, roughly speaking, is the lattice version of the wedge product. However, unlike the wedge product, $a\cup b\neq \pm b\cup a$ in general, because the cup product also encodes point-splitting regularization---which in the continuum is a separate subtle prescription. The full information of point-splitting is captured by the higher cup product system which satisfies
\begin{equation}
    a\cup_i b-(-1)^{pq+i}b\cup_ia=(-1)^{q+i}a\cup_{i+1} db-(-1)^{p+q+i+1}(da\cup_{i+1} b-d(a\cup_{i+1} b)).
    \label{eqn:cup_relation}
\end{equation}
where $a,b$ are lattice $p, q$-forms respectively, $\cup_0=\cup$ is the cup product and $\cup_i, i>0$ are higher cup products. When we take $i+1=0$, we define ``$\cup_{-1}$'' to be trivial, so this equation just becomes the Leibniz rule of the cup product $\cup=\cup_0$. When we take $i=0$, this equation means that $a \cup b$ and $b \cup a$ are ``equal only up to homotopy'', with the specific difference characterized by $\cup_1$; the non-commutativity of a higher cup product is captured by even higher-order cup products. 

For practical purpose, we will first present the results we need: the $\cup_{0}$ and $\cup_{1}$ products constructed on the cubic lattice as well as the special layer of lattice in the modular $T$ operator, which are directly used in our lattice model, including both the bosonic theory in main text and the fermionic theory elaborated in the next section. Then, we will explain the rationale of how these $\cup_i$ products are systematically constructed, in hope to make the highly technical process somewhat geometrically intuitive---especially its relation to the subtle continuum prescription of point-splitting.

\subsection{On Regular Cubic Lattice}
\label{sec:cup_on_cube}
For lattice $p$-form $a^p$ and $q$-form $b^q$, the value of $a^p\cup_i b^q$ on a $(p+q-i)$-cell is (term by term) pictured as the product the $a^p$ value on the purple $p$-cell and the $b^q$-value on the orange $q$-cell(s). We will only use $i=0, 1$.

\begin{table}[H]
    \begin{subtable}[t]{0.24\textwidth}
    \centering
    \begin{tabular}{c|c}
    \hline
         ${\color{purple}a^0}\cup{\color{orange}b^0}$ & $\vcenter{\hbox{\includegraphics[width=0.5\linewidth]{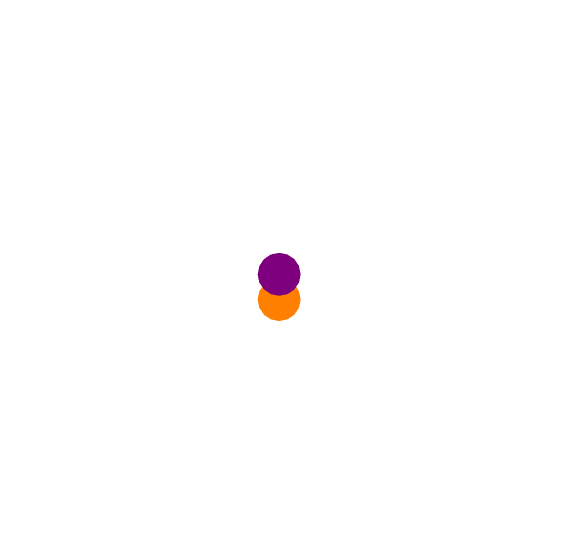}}}$\\
    \hline
    \end{tabular}
    \caption*{On a vertex}
    \end{subtable}
    \hfill
    \begin{subtable}[t]{0.24\textwidth}
    \centering
    \begin{tabular}{c|c}
         \hline
         ${\color{purple}a^1}\cup{\color{orange}b^0}$ & $\vcenter{\hbox{\includegraphics[width=0.5\linewidth]{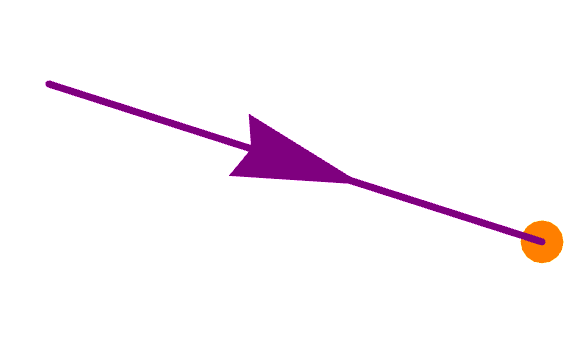}}}$\\
         \hline
         ${\color{purple}a^0}\cup{\color{orange}b^1}$ & $\vcenter{\hbox{\includegraphics[width=0.5\linewidth]{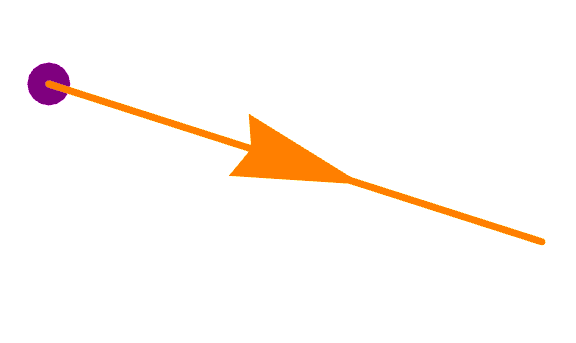}}}$\\
         \hline
         ${\color{purple}a^1}\cup_1{\color{orange}b^1}$ & $\vcenter{\hbox{\includegraphics[width=0.5\linewidth]{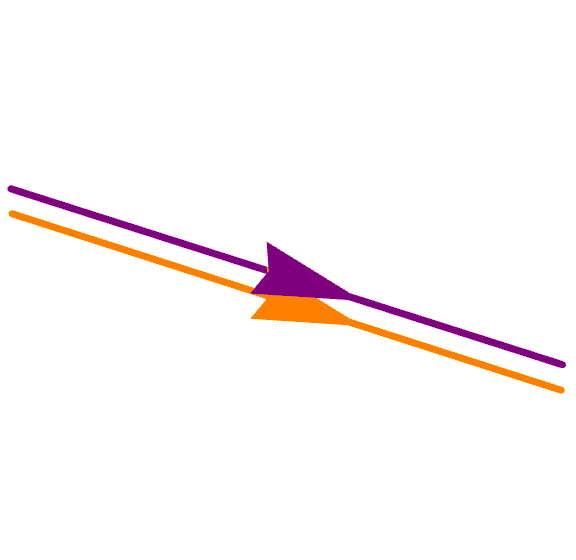}}}$\\
         \hline
    \end{tabular}
    \caption*{On a link}
    \end{subtable}
    \hfill
    \begin{subtable}[t]{0.48\textwidth}
    \centering
    \begin{tabular}{c|c}
         \hline
         ${\color{purple}a^2}\cup{\color{orange}b^0}$ & $\vcenter{\hbox{\includegraphics[width=0.25\linewidth]{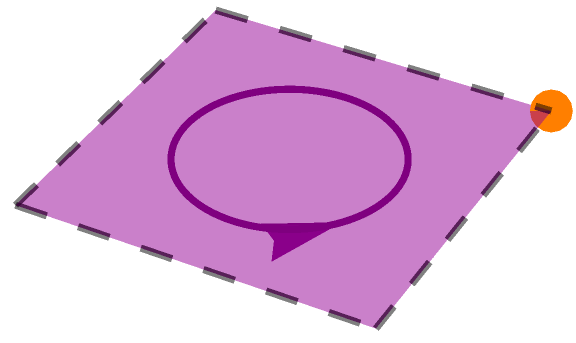}}}$\\
         \hline
         ${\color{purple}a^1}\cup{\color{orange}b^1}$& $\vcenter{\hbox{\includegraphics[width=0.25\linewidth]{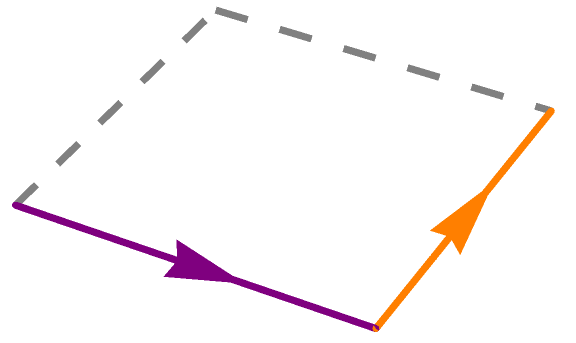}}}
        +\vcenter{\hbox{\includegraphics[width=0.25\linewidth]{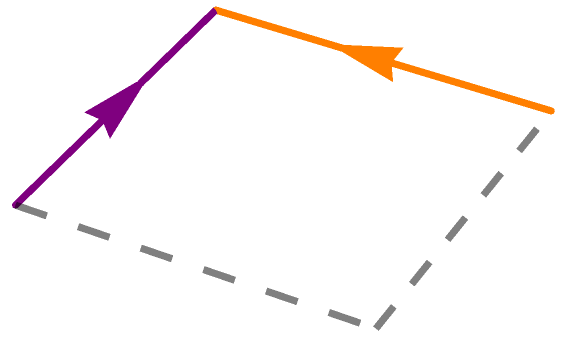}}}$\\
         \hline
         ${\color{purple}a^0}\cup{\color{orange}b^2}$ & $\vcenter{\hbox{\includegraphics[width=0.25\linewidth]{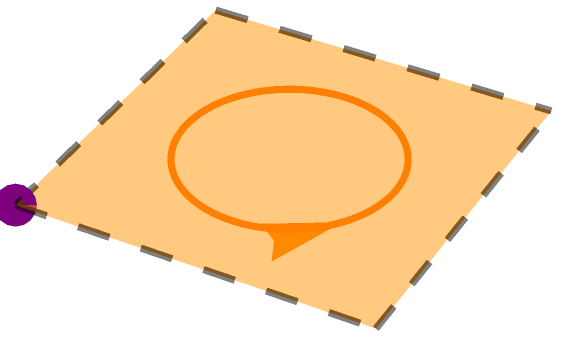}}}$\\
         \hline
         ${\color{purple}a^1}\cup_1{\color{orange}b^2}$ & $\vcenter{\hbox{\includegraphics[width=0.25\linewidth]{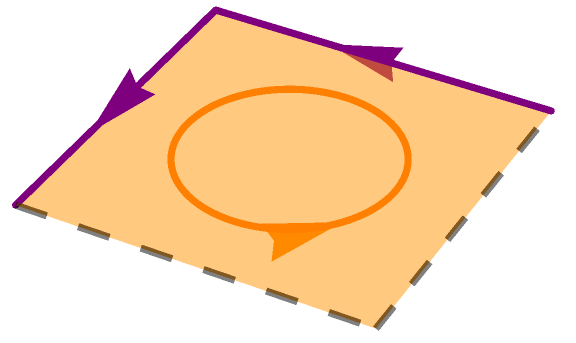}}}$\\
         \hline
         ${\color{purple}a^2}\cup_1{\color{orange}b^1}$ & $\vcenter{\hbox{\includegraphics[width=0.25\linewidth]{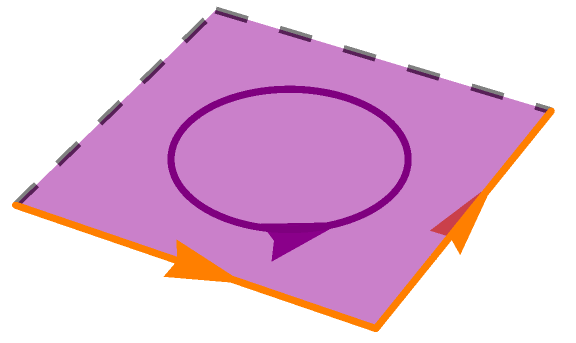}}}$\\
         \hline
    \end{tabular}
    \caption*{On a plaquette}
    \end{subtable}
    \label{tab:cup_on_v_l_p}
\end{table}

\begin{table}[H]
    \centering
    \begin{tabular}{c|c|c|c}
         \hline
         ${\color{purple}a^3}\cup{\color{orange}b^0}$ & $\vcenter{\hbox{\includegraphics[width=0.1\linewidth]{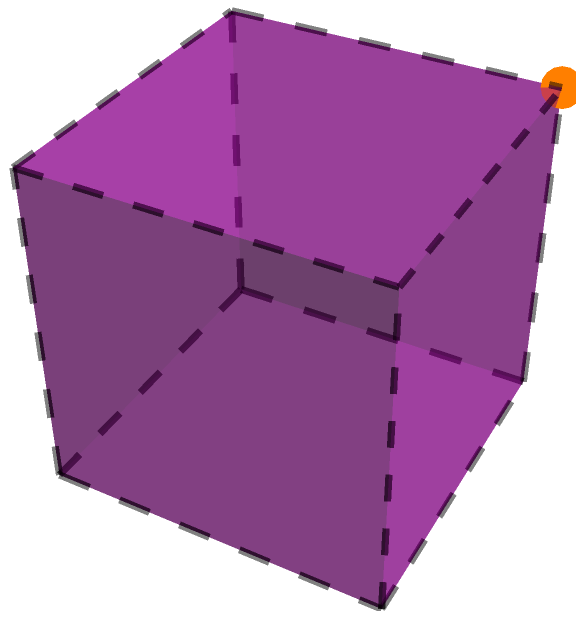}}}$ & ${\color{purple}a^3}\cup_1{\color{orange}b^1}$ & $\vcenter{\hbox{\includegraphics[width=0.1\linewidth]{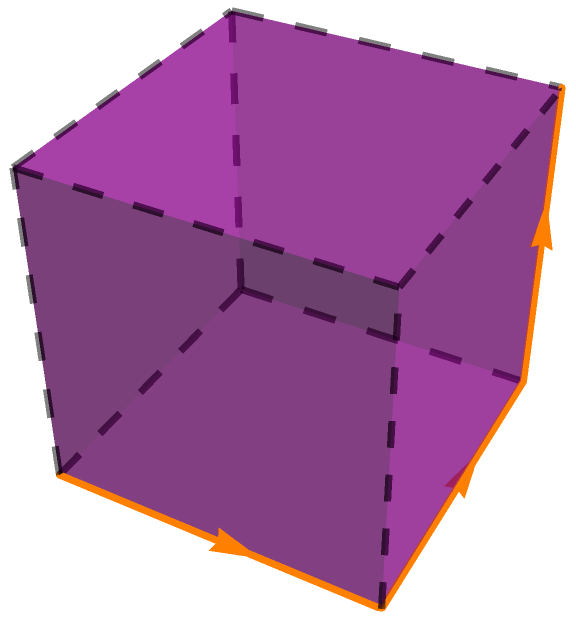}}}$\\
         \hline
         ${\color{purple}a^2}\cup{\color{orange}b^1}$& $\vcenter{\hbox{\includegraphics[width=0.1\linewidth]{plot/cup_cube_0_2_1_x.png}}}
        +\vcenter{\hbox{\includegraphics[width=0.1\linewidth]{plot/cup_cube_0_2_1_y.png}}}+\vcenter{\hbox{\includegraphics[width=0.1\linewidth]{plot/cup_cube_0_2_1_z.png}}}$ & ${\color{purple}a^2}\cup_1{\color{orange}b^2}$ & $\vcenter{\hbox{\includegraphics[width=0.1\linewidth]{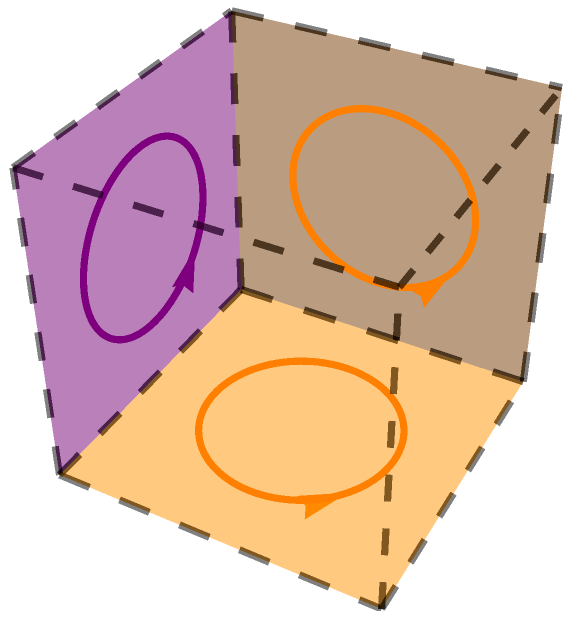}}}
         +\vcenter{\hbox{\includegraphics[width=0.1\linewidth]{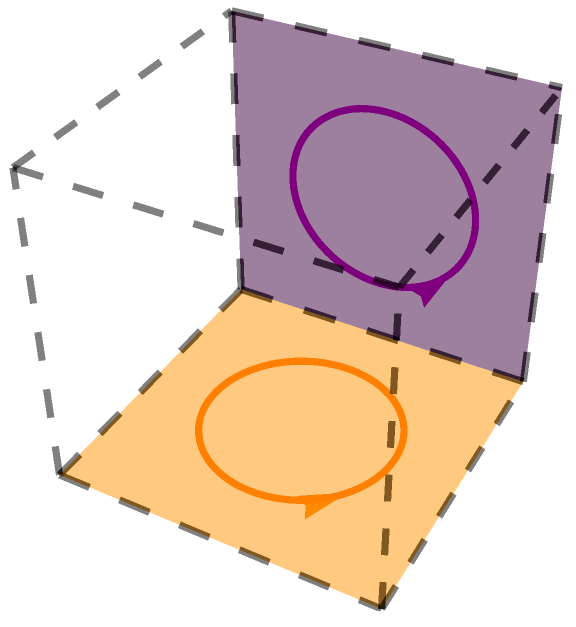}}}
         +\vcenter{\hbox{\includegraphics[width=0.1\linewidth]{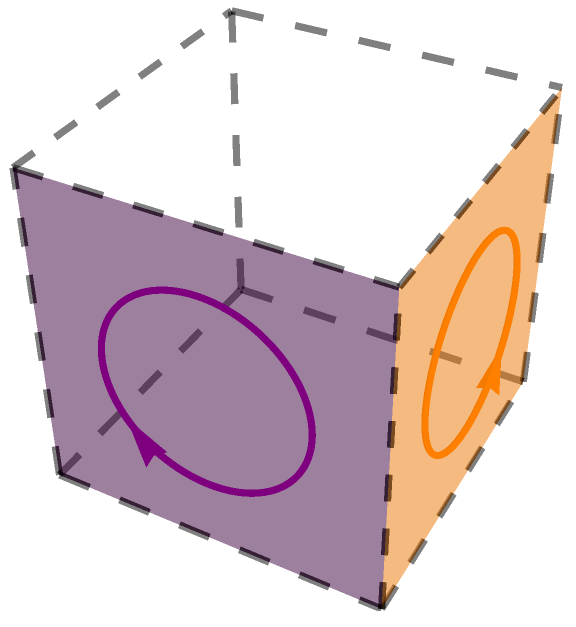}}}
         +\vcenter{\hbox{\includegraphics[width=0.1\linewidth]{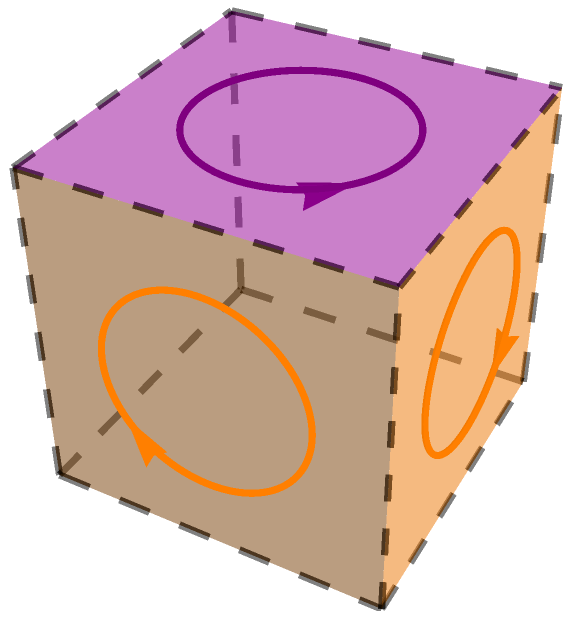}}}$\\
         \hline
         ${\color{purple}a^1}\cup{\color{orange}b^2}$ & $\vcenter{\hbox{\includegraphics[width=0.1\linewidth]{plot/cup_cube_0_1_2_x.png}}}
        +\vcenter{\hbox{\includegraphics[width=0.1\linewidth]{plot/cup_cube_0_1_2_y.png}}}+\vcenter{\hbox{\includegraphics[width=0.1\linewidth]{plot/cup_cube_0_1_2_z.png}}}$ & ${\color{purple}a^1}\cup_1{\color{orange}b^3}$ & $\vcenter{\hbox{\includegraphics[width=0.1\linewidth]{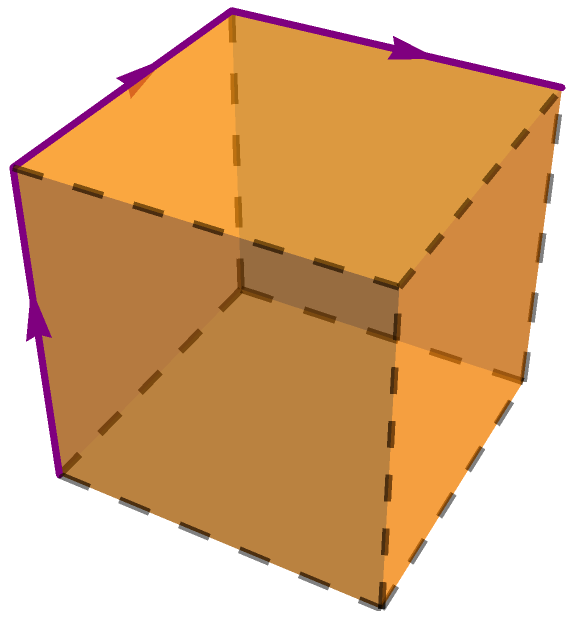}}}$\\
         \hline
         ${\color{purple}a^0}\cup{\color{orange}b^3}$ & $\vcenter{\hbox{\includegraphics[width=0.1\linewidth]{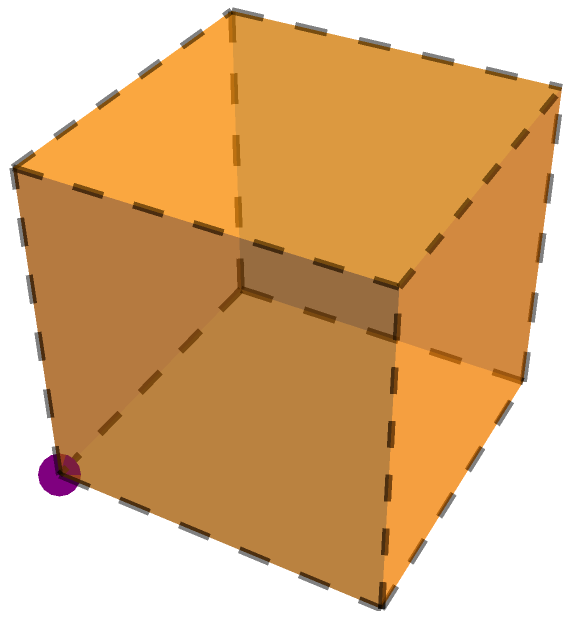}}}$ & &\\
         \hline
    \end{tabular}
    \caption*{On a cube}
    \label{tab:cup_on_cube}
\end{table}

\subsection{On the Special Layers of the Modular $T$ Operator}

\begin{table}[H]
    \centering
    \begin{tabular}{c|c|c|c|c|c}
         \hline
         ${\color{purple}a^2}\cup{\color{orange}b^0}$ & $\vcenter{\hbox{\includegraphics[width=0.12\linewidth]{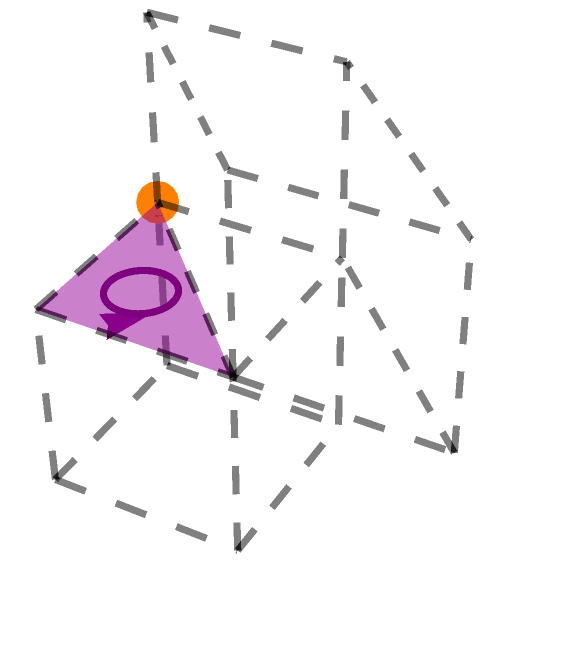}}}$ & $\vcenter{\hbox{\includegraphics[width=0.12\linewidth]{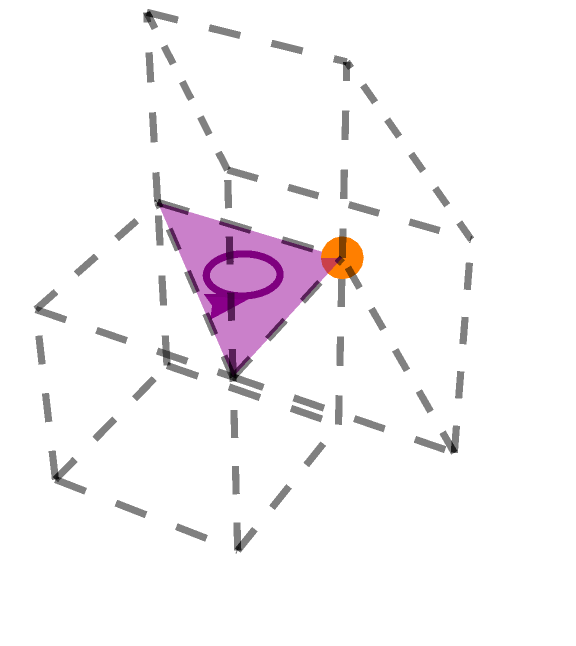}}}$ & ${\color{purple}a^2}\cup_1{\color{orange}b^1}$ & $\vcenter{\hbox{\includegraphics[width=0.12\linewidth]{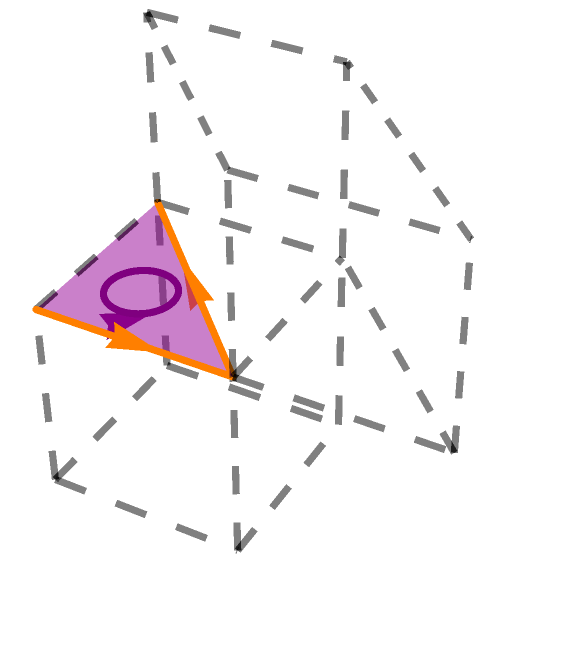}}}$ & $\vcenter{\hbox{\includegraphics[width=0.12\linewidth]{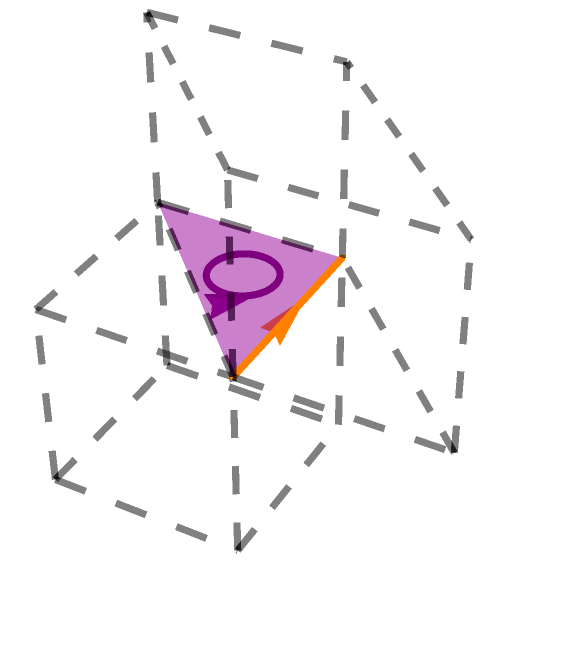}}}$\\
         \hline
         ${\color{purple}a^1}\cup{\color{orange}b^1}$& $\vcenter{\hbox{\includegraphics[width=0.12\linewidth]{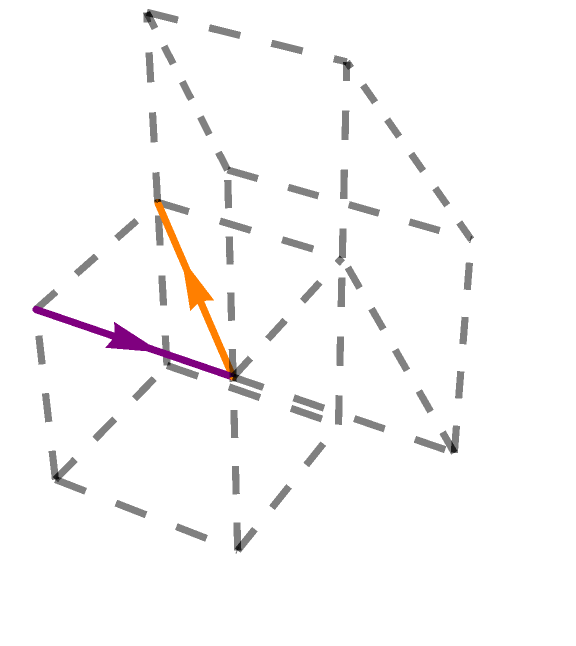}}}$ & $\vcenter{\hbox{\includegraphics[width=0.12\linewidth]{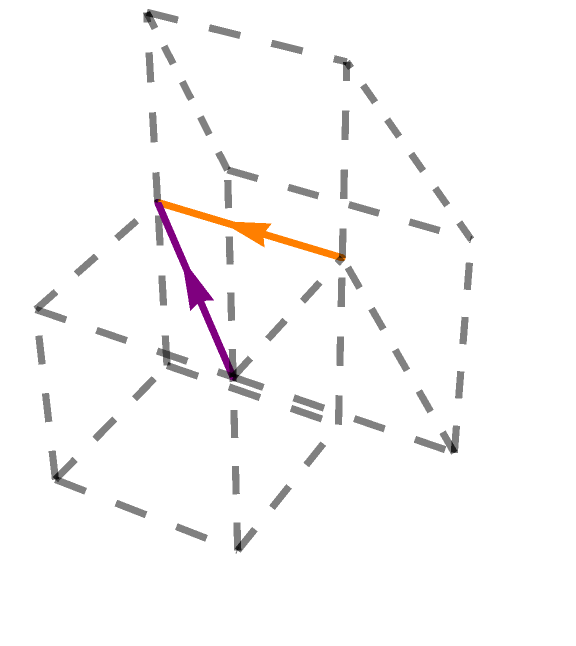}}}$& ${\color{purple}a^1}\cup_1{\color{orange}b^2}$ & $\vcenter{\hbox{\includegraphics[width=0.12\linewidth]{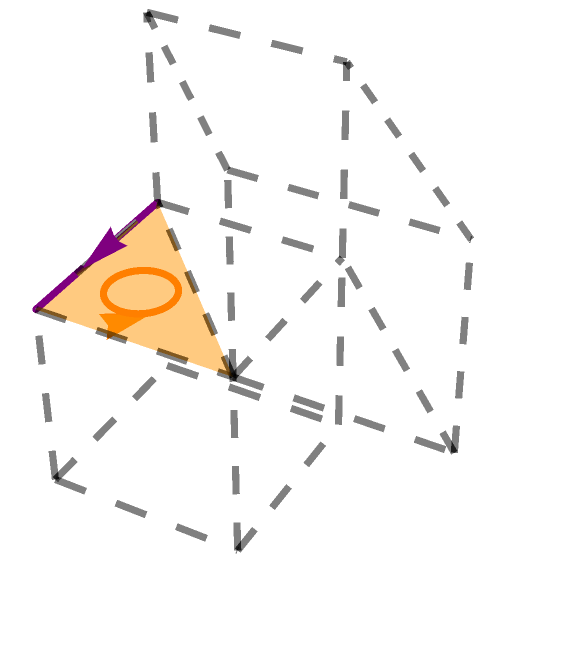}}}$ & $\vcenter{\hbox{\includegraphics[width=0.12\linewidth]{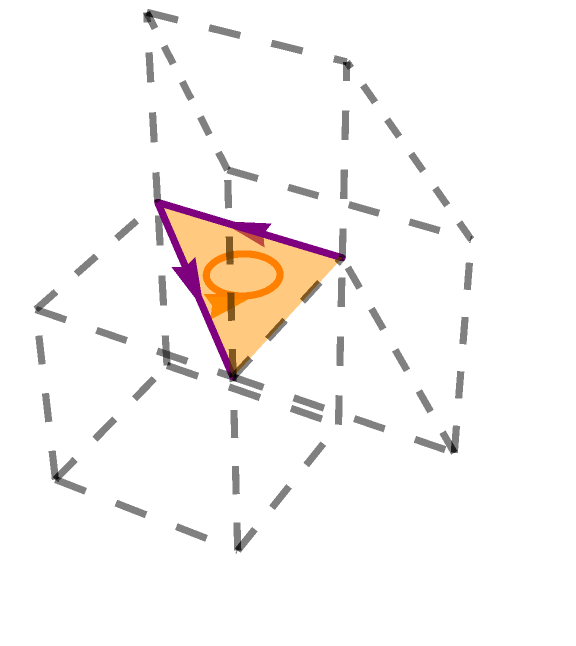}}}$\\
         \hline
         ${\color{purple}a^0}\cup{\color{orange}b^2}$& $\vcenter{\hbox{\includegraphics[width=0.12\linewidth]{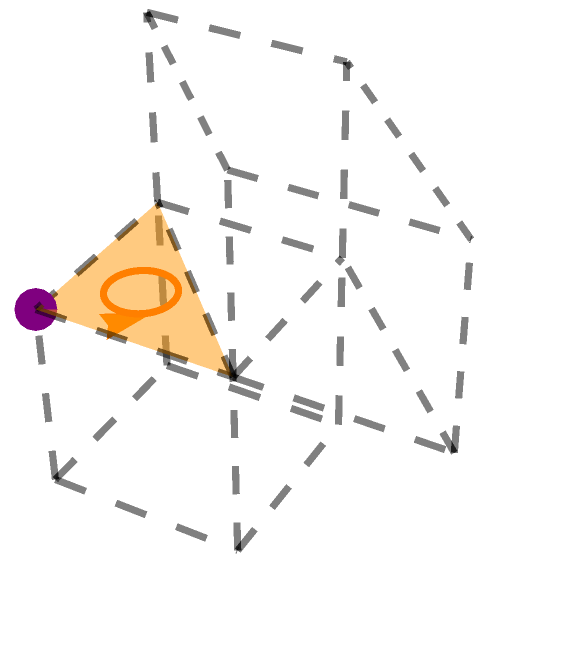}}}$ & $\vcenter{\hbox{\includegraphics[width=0.12\linewidth]{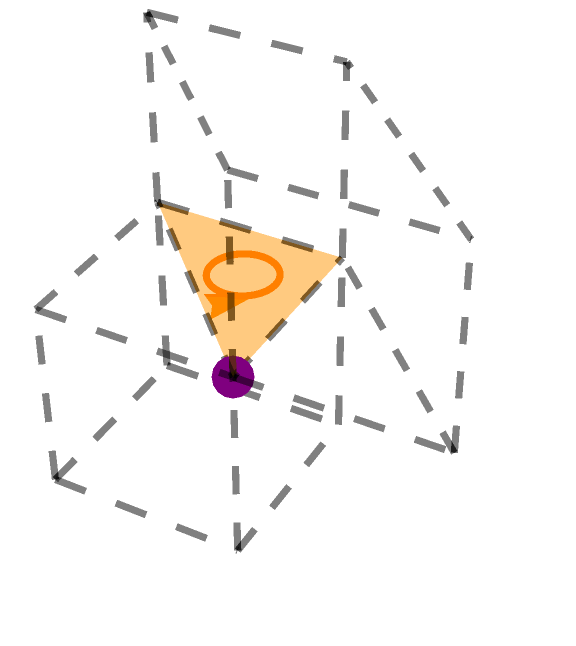}}}$ & & \\
         \hline
    \end{tabular}
    \caption*{On the triangular plaquettes}
    \label{tab:cup_on_triangular_plaquettes}
\end{table}

\begin{table}[H]
    \centering
    \begin{tabular}{c|c|c}
         \hline
         & On a lower layer cube & On an upper layer cube \\
         \hline
         ${\color{purple}a^2}\cup{\color{orange}b^1}$ & $\vcenter{\hbox{\includegraphics[width=0.12\linewidth]{plot/cup_0_2_1_x.png}}}
         +\vcenter{\hbox{\includegraphics[width=0.12\linewidth]{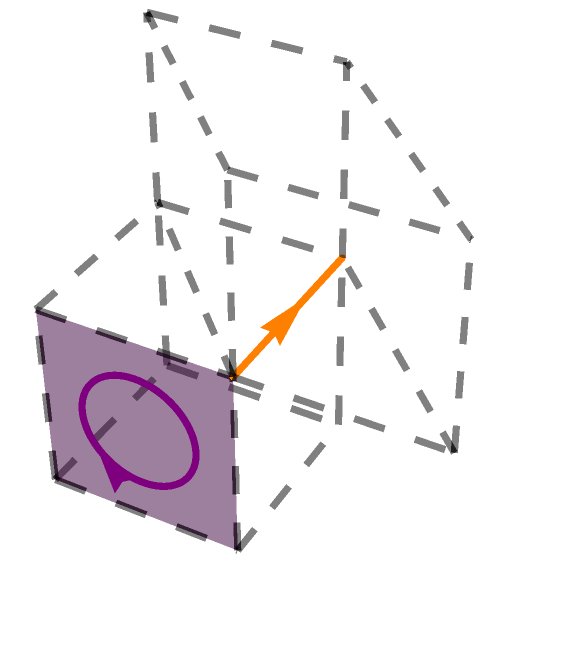}}}
         +\vcenter{\hbox{\includegraphics[width=0.12\linewidth]{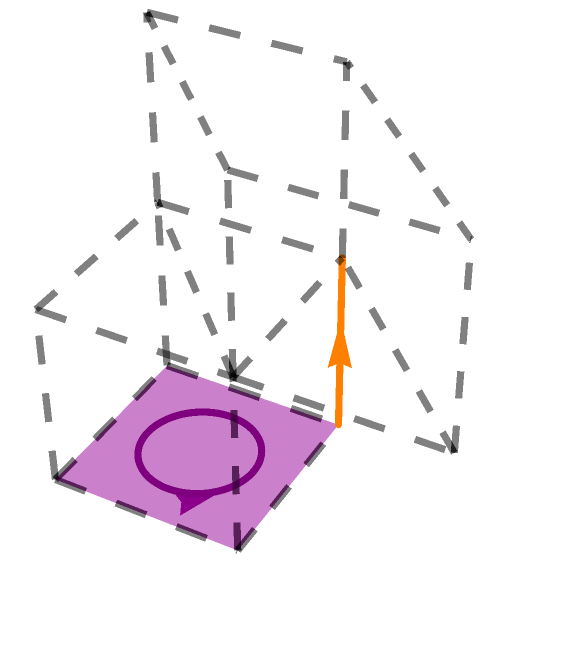}}}$ & same as on regular cubic lattice \\
         \hline
         ${\color{purple}a^1}\cup{\color{orange}b^2}$& 
         \makecell{$\vcenter{\hbox{\includegraphics[width=0.12\linewidth]{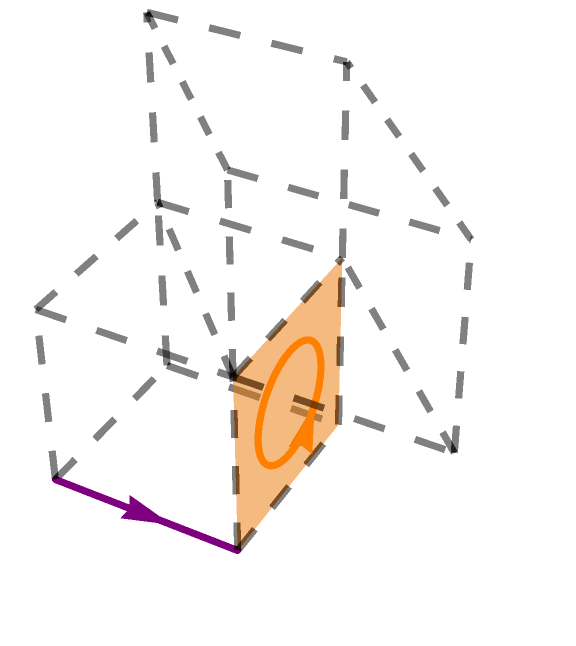}}}
         +\vcenter{\hbox{\includegraphics[width=0.12\linewidth]{plot/cup_0_1_2_x_2.png}}}$\\
         $+\vcenter{\hbox{\includegraphics[width=0.12\linewidth]{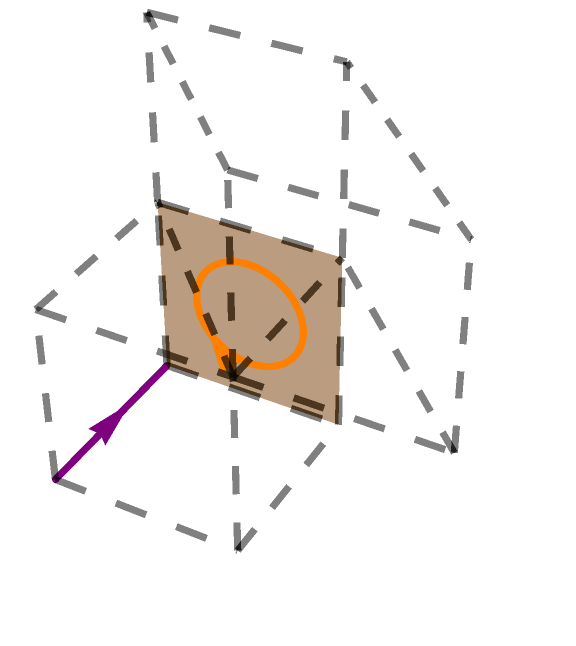}}}
         +\vcenter{\hbox{\includegraphics[width=0.12\linewidth]{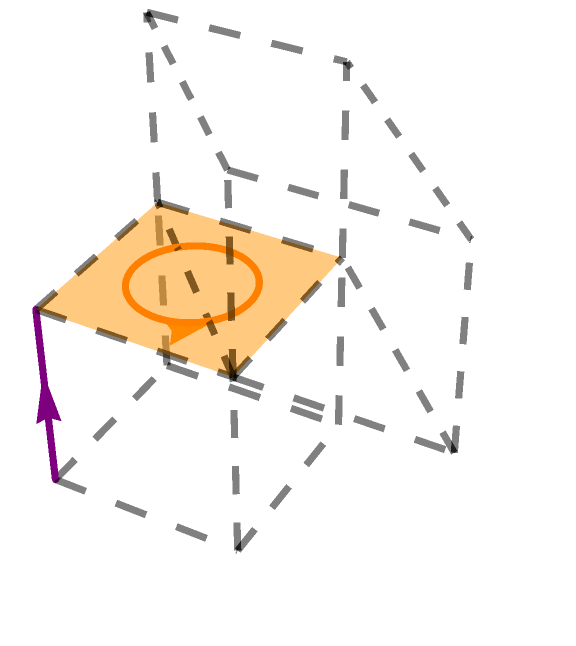}}}$} & same as on regular cubic lattice\\
         \hline
         ${\color{purple}a^2}\cup_1{\color{orange}b^2}$ & 
         \makecell{$\vcenter{\hbox{\includegraphics[width=0.12\linewidth]{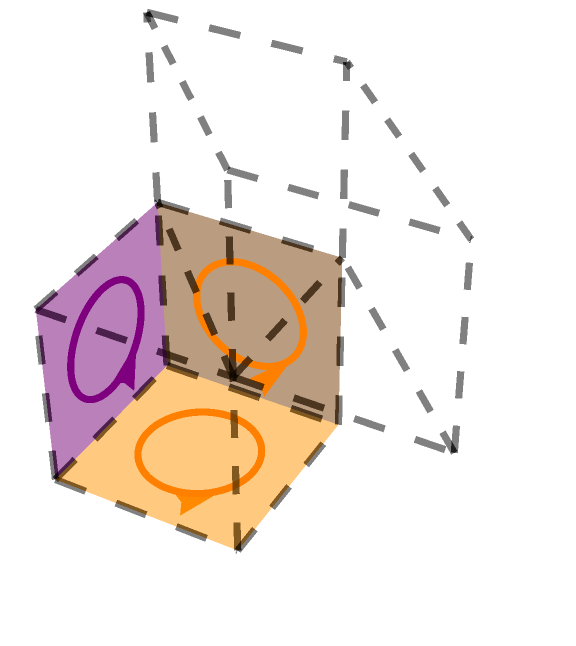}}}
         +\vcenter{\hbox{\includegraphics[width=0.12\linewidth]{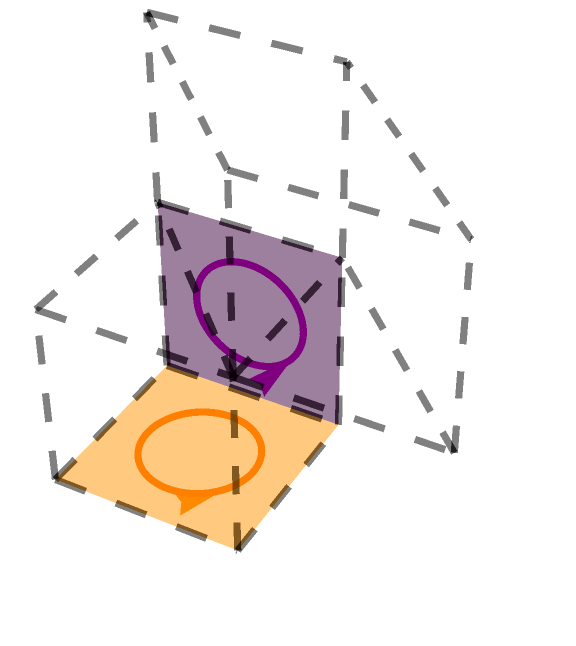}}}
         +\vcenter{\hbox{\includegraphics[width=0.12\linewidth]{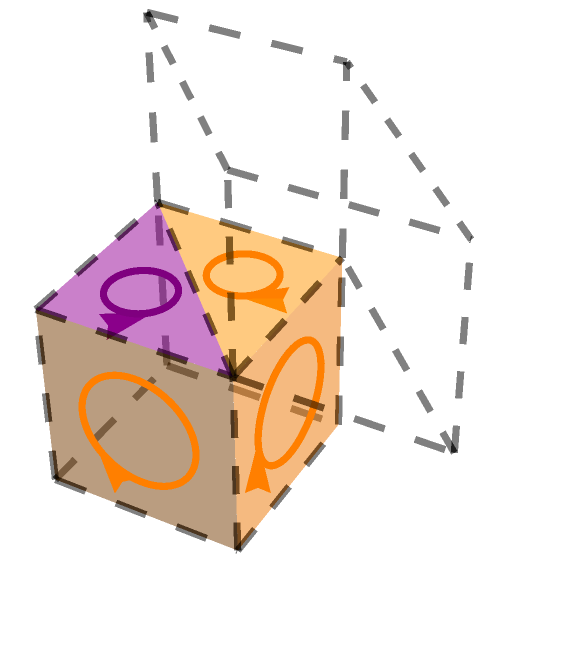}}}$\\
         $+\vcenter{\hbox{\includegraphics[width=0.12\linewidth]{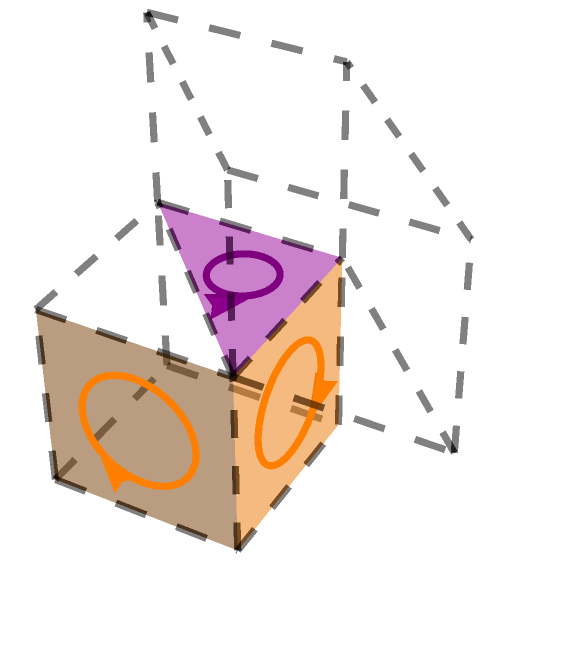}}}
         +\vcenter{\hbox{\includegraphics[width=0.12\linewidth]{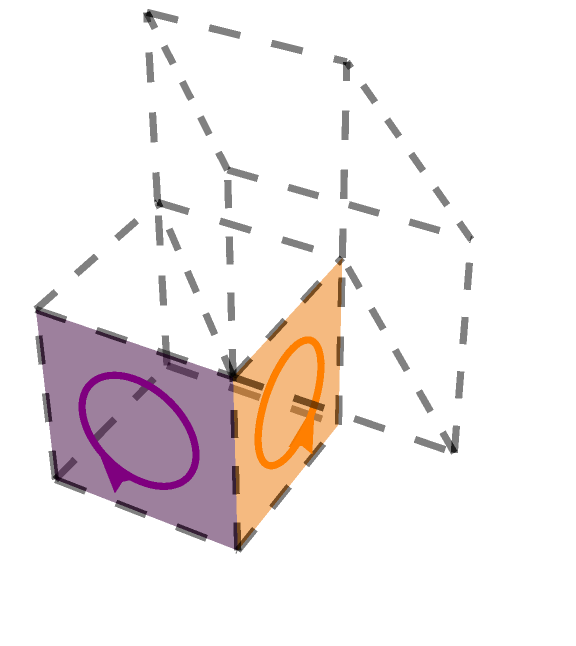}}}$} & 
         \makecell{$\vcenter{\hbox{\includegraphics[width=0.12\linewidth]{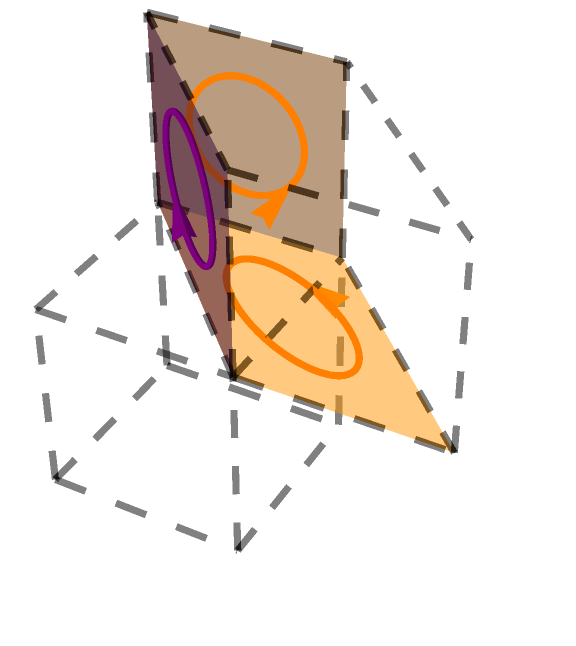}}}
         +\vcenter{\hbox{\includegraphics[width=0.12\linewidth]{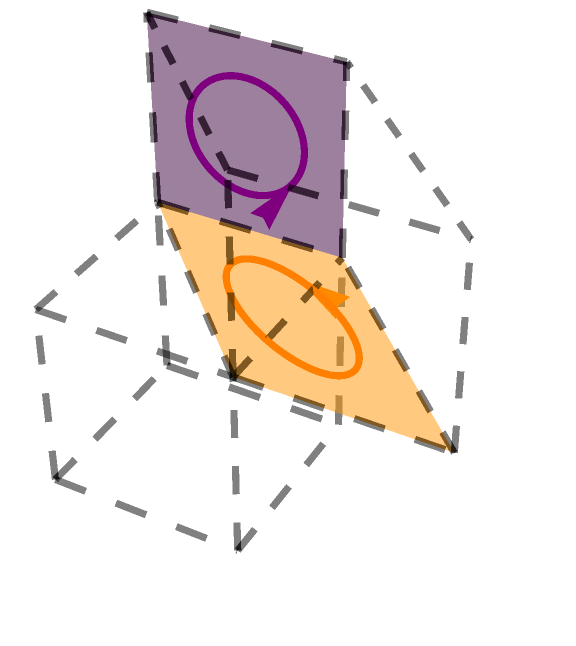}}}
         +\vcenter{\hbox{\includegraphics[width=0.12\linewidth]{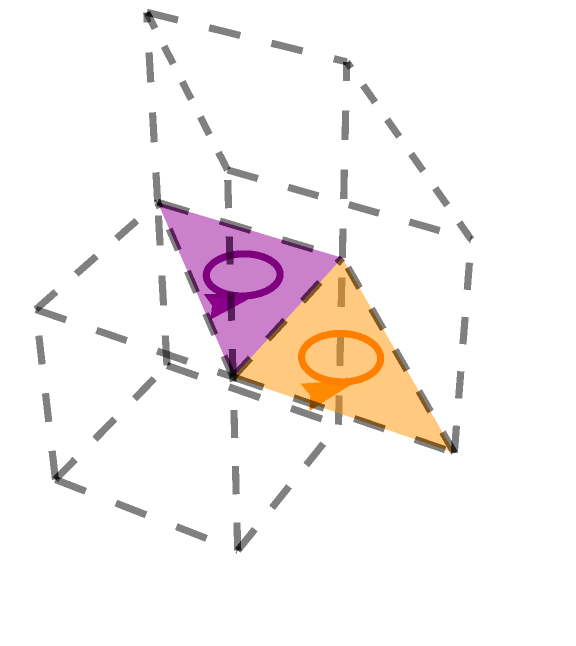}}}$\\
         $+\vcenter{\hbox{\includegraphics[width=0.12\linewidth]{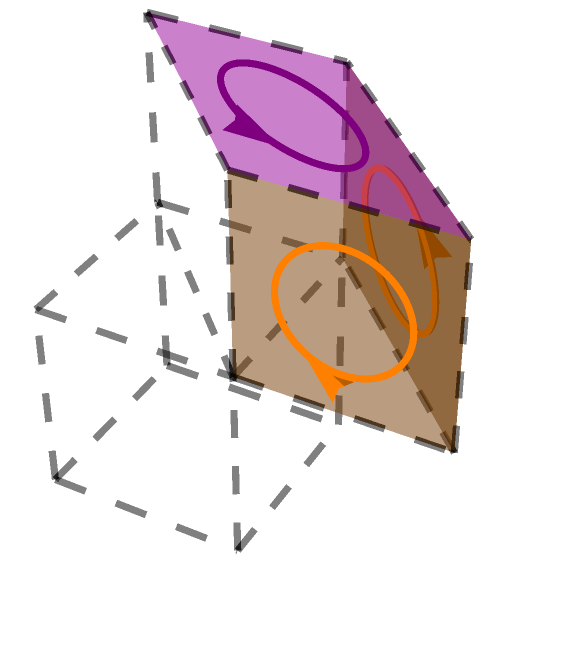}}}
         +\vcenter{\hbox{\includegraphics[width=0.12\linewidth]{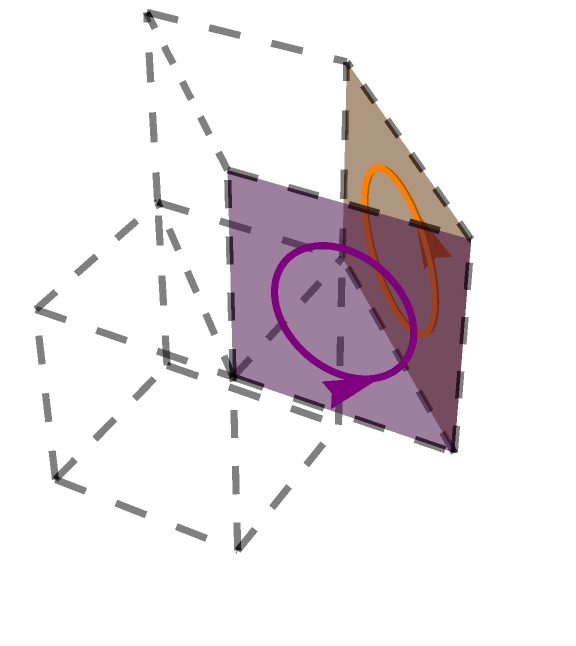}}}$} \\
         \hline
    \end{tabular}
    \caption*{On the cubes}
    \label{tab:cup_on_cube_special_layer}
\end{table}

\subsection{General Construction}
\label{sec:general_construction_of_cup}
Inspired by \cite{Thorngren:2018ziu}, we propose a general method for constructing cup products on arbitrary an lattice $\mathcal{M}$. The language of chain and cochain complex will be used.

We shall first try to define the cup product on a basis and then extend it linearly to arbitrary forms. That is, we first consider a lattice $p$-form (formally called a $p$-\emph{cochain}) $a$ that is non-zero only on a single $p$-cell $\sigma_a$, and a $(d-p)$-form $b$ that is non-zero only on a single $(d-p)$-cell $\sigma_b$. Furthermore we set this non-zero value to be simply $1$ (resembling delta functions). To find a lattice version of ``$a\wedge b$'' for such $a$ and $b$, we need a way to relate the $p$-cell $\sigma_a$ and the $(d-p)$-cell $\sigma_b$. Note that a $p$-cell $\sigma_a$ naturally corresponds to a $(d-p)$-cell $\sigma_a^\vee$ on the dual lattice $\mathcal{M}^\vee$. Hence, if we have a linear map $f_{\text{naive}}$ from the \emph{chain complex} $C_{\bullet}(\mathcal{M}^\vee)$ to $C_{\bullet}(\mathcal{M})$, that maps $q$-\emph{chains} (formal sum of $q$-cells) on the dual lattice to $q$-chains on the original lattice (mapping to zero is also allowed), then $f_{\text{naive}}(\sigma_a^\vee)$ would be some $(d-p)$-chain of the original lattice, and thus we can use the coefficient of $\sigma_b$ in $f_{\text{naive}}(\sigma_a^\vee)$ as the desired output (see below for details).

We can also motivate the introduction of $f_{\text{naive}}$ from another point of view. The cup product we want to establish is a product between twp cochains on the lattice. Meanwhile, we know that there is a natural pairing between a cochain and a chain, and there is a natural correspondence between a chain on the original lattice and a cochain on the dual lattice. So we have a natural multiplication between a cochain on the original lattice and a cochain on the dual lattice. Thus, if we have an $f_{\text{naive}}$ that maps from the dual lattice to the original lattice, it will induce a product between two cochains on the original lattice.

Importantly, it is not hard to see the natural pairing between a cochain on the original lattice and a cochain on the dual lattice has the property that
\begin{equation}
    (-1)^p\sum_{\text{$p$-cell }\sigma}a_\sigma b^\vee_{\partial\sigma^\vee}+\sum_{\text{$(p+1)$-cell }\tilde\sigma}a_{\partial\tilde\sigma} b^\vee_{\tilde\sigma^\vee}=0.
    \label{eqn:lattice_integration_by_part}
\end{equation}
for any $p$-cochain $a$ on the original lattice and $(d-p)$-cochain $b^\vee$ on the dual lattice. \cref{eqn:lattice_integration_by_part} is the lattice version of integration by part, but recall it is about the pairing between a cochain on the original lattice and a cochain on the dual lattice, meanwhile what we are interested in is the Leibniz rule of the cup product between two cochains both on the original lattice, i.e. \cref{eqn:cup_relation} when $i+1=0$. We can relate the latter to \eqref{eqn:lattice_integration_by_part} via $f_\text{naive}$. More exactly, we will soon see that \eqref{eqn:lattice_integration_by_part} will lead to an ``integrated version'' of the Leibniz rule of the cup product induced by $f_\text{naive}$, if and only if $f_\text{naive}$ meets the conditions of what is known as a chain map. Moreover, in the subsequent more complete construction, this property \eqref{eqn:lattice_integration_by_part} is also crucial for the full construction that satisfies \cref{eqn:cup_relation}.

Now we proceed with more details. We first consider the simplest case where $f_{\text{naive}}(\sigma_a)$ is either $0$ or contains only one cell with coefficient $1$ (as opposed to a linear combination of multiple cells). In such case, we can define
\begin{equation}
    \int a\cup b:=\begin{cases}
    1,\quad f_{\text{naive}}(\sigma_a^\vee)=\sigma_b\\
    0,\quad f_{\text{naive}}(\sigma_a^\vee)\neq\sigma_b
    \end{cases}.
\end{equation}
For the general case where $f_{\text{naive}}(\sigma_a^\vee)$ is a sum of multiple cells, we define $\int a\cup b$ as the coefficient of $\sigma_b$ in $f(\sigma_a^\vee)$. Extending the definition by linearity, we have
\begin{equation}
    \int a\cup b:=\sum_{p\mbox{\scriptsize{-cells}}\: \sigma} a_\sigma b_{f_{\text{naive}}(\sigma^\vee)}
\end{equation}
for arbitrary $p$-form $a$ and $(d-p)$-form $b$. If $f_{\text{naive}}(\sigma^\vee)$ is a sum of multiple cells, say $f_{\text{naive}}(\sigma^\vee)=\lambda\sigma'+\mu\sigma''+\cdots$, we define $b_{f_{\text{naive}}(\sigma^\vee)}=\lambda b_{\sigma'}+\mu b_{\sigma''}+\cdots$.

Let us see what conditions the Leibniz rule, i.e. \cref{eqn:cup_relation} when $i+1=0$, impose of $f_{\text{naive}}$. Observe that by integrating the Leibniz rule, we have
\begin{equation}
\begin{aligned}
    0&=\int (-1)^pa\cup db+ \int da\cup b\\
    &=(-1)^p\sum_{\text{$p$-cell }\sigma} a_\sigma(db)_{f_{\text{naive}}(\sigma^\vee)}+\sum_{\text{$(p+1)$-cell }\tilde\sigma}(da)_{\tilde\sigma}b_{f_{\text{naive}}(\tilde\sigma^\vee)}\\
    &=(-1)^p\sum_\sigma a_\sigma b_{\partial f_{\text{naive}}(\sigma^\vee)}+\sum_{\tilde\sigma}a_{\partial \tilde\sigma}b_{f_{\text{naive}}(\tilde\sigma^\vee)}.
\end{aligned}
\end{equation}
If we define a dual $(d-p-1)$-form $b^\vee$ as $b^\vee_{\tilde\sigma^\vee}=b_{f_{\text{naive}}(\tilde\sigma^\vee)}$, the above equation becomes
\begin{equation}
    0=(-1)^p\sum_\sigma a_\sigma b_{\partial f_{\text{naive}}(\sigma^\vee)}+\sum_{\tilde\sigma}a_{\partial \tilde\sigma}b^\vee_{\tilde\sigma^\vee}.
\end{equation}
By \cref{eqn:lattice_integration_by_part}, we have
\begin{equation}
\begin{aligned}
    0&=(-1)^p\sum_\sigma a_\sigma b_{\partial f_{\text{naive}}(\sigma^\vee)}-(-1)^p\sum_{\sigma}a_{\sigma}b^\vee_{\partial \sigma^\vee}.\\
    &=(-1)^p\sum_\sigma a_\sigma b_{\partial f_{\text{naive}}(\sigma^\vee)}-(-1)^p\sum_\sigma a_\sigma b_{f_{\text{naive}}(\partial \sigma^\vee)}.\\
\end{aligned}
\end{equation}
Since $a,b$ are arbitrary forms, we need
\begin{equation}
    \partial f_{\text{naive}} - f_{\text{naive}} \partial=0
    \label{eqn:chain_map_naive},
\end{equation}
which is nothing but the \emph{chain map} condition. Therefore, requiring $f_\text{naive}$ to be a chain map is a \emph{necessary} condition for satisfying the Leibniz rule.

However, this naive approach encounters several issues:
\begin{enumerate}
    \item In this formulation, $\int a \cup b$ is only defined for a $p$-form $a$ and a $(d-p)$-form $b$. Can we also handle forms $b$ of degree $d-p-i$?
    \item Knowing only the integral $\int a \cup b$ is insufficient---the explicit expression for $a \cup b$ itself remains unknown. Although one may say that $a\cup b$ must be nonzero on some $(d-i)$-cell(s) adjacent to $\sigma_a$ (in the case when $a_{\sigma_a}=1$ and $0$ elsewhere), it is impossible to determine directly from the action of $f_{\text{naive}}$ exactly which $(d-i)$-cell(s) adjacent to $\sigma_a$ it/they would be.
    \item How can we generalize this to higher cup products and how to find a sufficient condition that ensures the Leibniz rule ($i+1=0$) and ``generalized Leibniz rule'' ($i\geq0$) given by \cref{eqn:cup_relation} to be satisfied?
    \item How to ensure that the cup product is ``local''? This means, if we glue two lattice systems along a common boundary into one, as long as their respective definitions of cup products on the common boundary are consistent, we should get a cup product that is well-defined over the resulting lattice system.
\end{enumerate}

\begin{figure}[!htbp]
    \begin{subfigure}[b]{0.4\linewidth}
    \centering
    \includegraphics[width=0.6\linewidth]{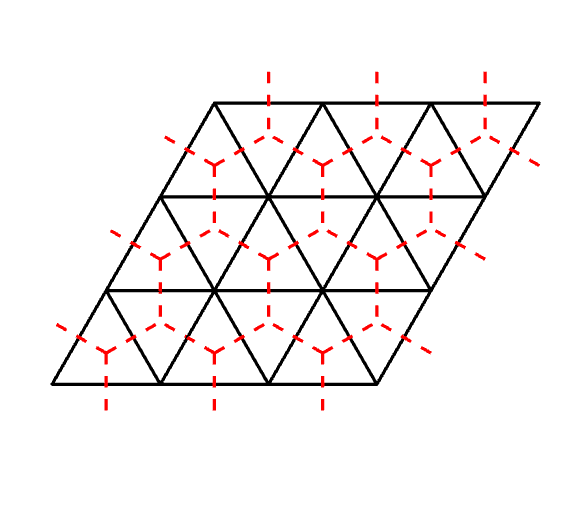}
    \caption{dual lattice for triangular lattice}
    \label{fig:dual_lattice_triangle}
    \end{subfigure}
    \begin{subfigure}[b]{0.4\linewidth}
    \centering
    \includegraphics[width=0.6\linewidth]{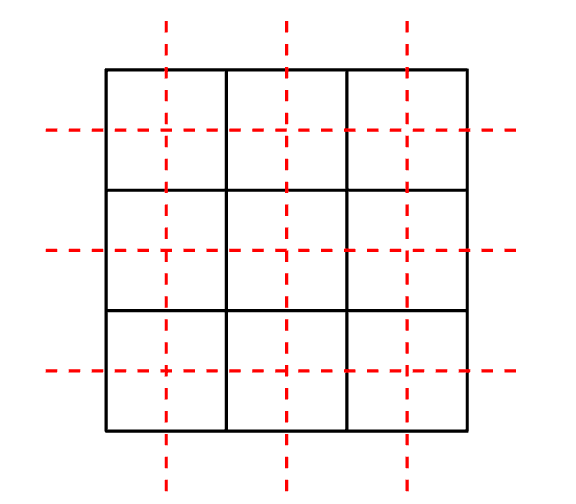}
    \caption{dual lattice for square lattice}
    \label{fig:dual_lattice_square}
    \end{subfigure}

    \begin{subfigure}[b]{0.4\linewidth}
    \centering
    \includegraphics[width=0.6\linewidth]{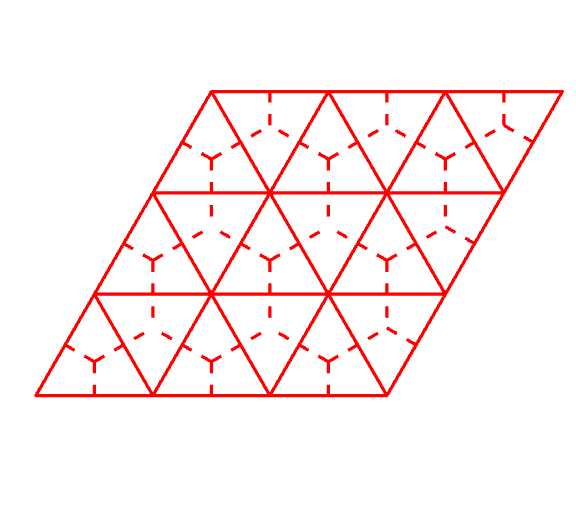}
    \caption{refined dual lattice for triangular lattice}
    \label{fig:refined_dual_lattice_triangle}
    \end{subfigure}
    \begin{subfigure}[b]{0.4\linewidth}
    \centering
    \includegraphics[width=0.6\linewidth]{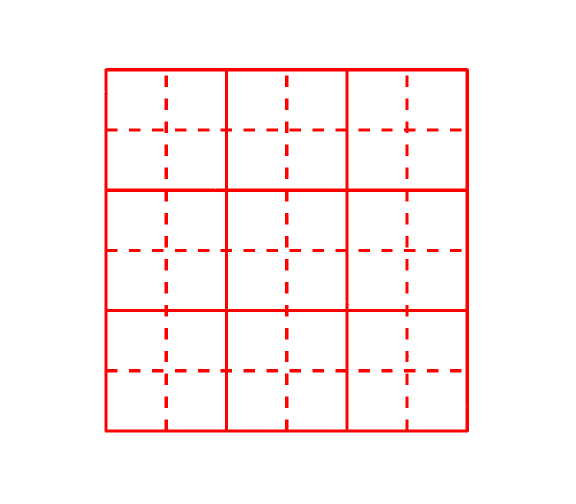}
    \caption{refined dual lattice for square lattice}
    \label{fig:refined_dual_lattice_square}
    \end{subfigure}
    \caption{In \cref{fig:dual_lattice_triangle,fig:dual_lattice_square}, the original triangular and square lattices are drawn in black solid lines, and their corresponding dual lattices in red dashed lines. In \cref{fig:refined_dual_lattice_triangle,fig:refined_dual_lattice_square} the refined dual lattices consist of both the red solid and dashed lines, as if we overlay the original lattice onto the dual lattice. And notice that the refined dual lattice for an original lattice with smooth boundary has no dangling links, i.e. not rough on the boundary, unlike the usual dual lattice.
    }
    \label{fig:refined_dual_lattice_single_cell}
\end{figure}

\begin{figure}[!htbp]
    \centering
    \includegraphics[width=0.5\linewidth]{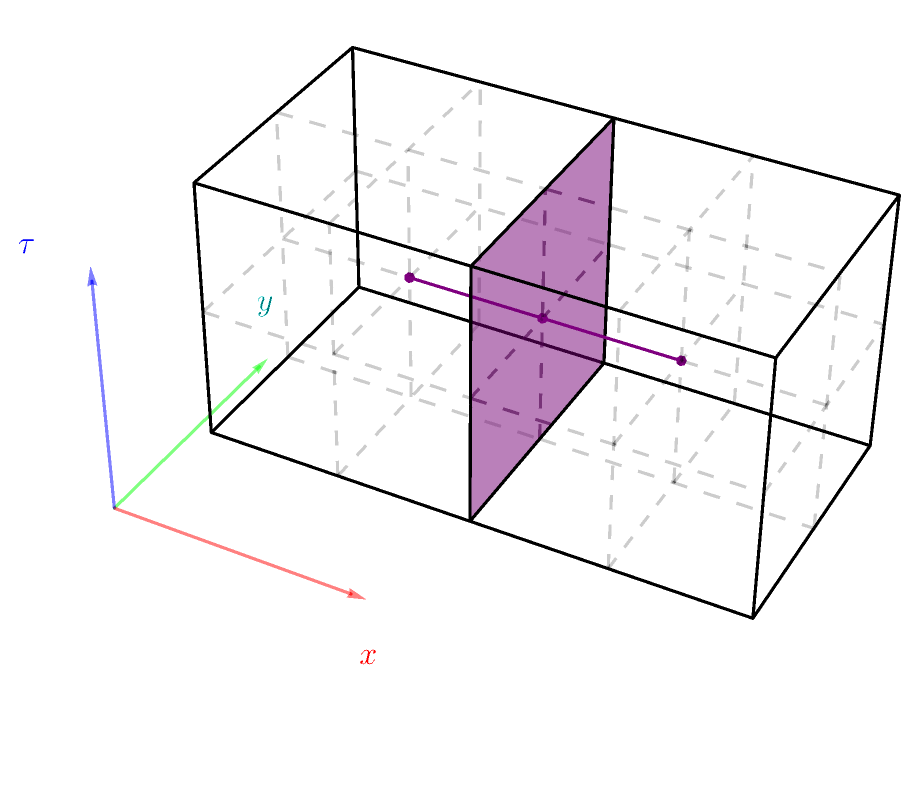}
    \caption{Consider two cubes in the original lattice, with the right one denoted as $c_+$ and the left one $c_-$. The purple plaquette (a $2$-cell) is denoted as $\sigma$. The purple link is the dual $1$-cell $\sigma^\vee$. After the refinement, $\sigma^\vee$ now consists of two segments (two refined dual $1$-cells), and a point in the middle (refined dual $0$-cell). Since each segment is the intersection of $\sigma^\vee$ with $c_+$ or $c_-$, we denote them as $\sigma^\vee_{c_\pm}$ respectively; and the middle point is the intersection of $\sigma^\vee$ with $\sigma$ itself, so we denote it as $\sigma^\vee_{\sigma}$.}
    \label{fig:refined_dual_lattice}
\end{figure}

To address these issues, we must consider a \emph{refined} version of the dual lattice, which will be denoted as $\mathcal{M}^\vee_{\text{refined}}$ (on the other hand, the original lattice cells are not refined), and accordingly a refined version of the map $f_{\text{naive}}$, which we shall call $f$. We shall see a subdivision similar to the barycentric subdivision provides a desired refinement. This subdivision like overlaying the original lattice on top of the usual dual lattice, see \cref{fig:refined_dual_lattice_single_cell,fig:refined_dual_lattice} (this is actually slightly ``coarser'' than barycentric subdivision). We denote by $C_{\bullet}(\mathcal{M}^\vee_{\text{refined}})$ the chain complex of the refined dual lattice, and the refined $f$ is a map from the chain complex $C_{\bullet}(\mathcal{M}^\vee_{\text{refined}})$ of the refined dual lattice to the chain complex $C_{\bullet}(\mathcal{M})$ of the original lattice which we do not refine. Now let us explain why we refined the dual lattice in such a way, and what conditions should be imposed on $f$.

\begin{enumerate}
    \item  Consider a $p$-cell $\sigma$ on the original lattice. $\sigma^\vee$, which used to be just a $(d-p)$-cell on the dual lattice, is now, geometrically, a composite of several refined $(d-p)$-cells; moreover, at where these refined dual $(d-p)$-cells join, there will appear refined dual $(d-p-i)$-cells. For example, in \cref{fig:refined_dual_lattice}, $\sigma$ is the purple plaquette (original $2$-cell), and the purple dual link $\sigma^\vee$ is now a composite of two segments (refined dual $(3-2=1)$-cells), joining at the mid-point (a refined dual $(3-2-1=0)$-cell). The appearance of refined dual $(d-p-i)$-cells inside $\sigma^\vee$ will allow us to handle issue 1 above.

    \item The particular refinement of the dual lattice is motivated by issue 2 above, which we now explain.
    
    For a $p$-cell $\sigma$ on the original lattice, consider all the $(d-i)$-cells ($p\leq d-i\leq d$) $\sigma'$ on the original lattice that contains $\sigma$: For example, in \cref{fig:refined_dual_lattice}, each of the two cubes (i.e. $(3-0=3)$-cells) on the two sides of the purple plaquette $\sigma$ contains $\sigma$ as part of its boundary, and $\sigma$ itself (a $(3-1=2)$-cell) obviously contains itself. Now note that, with the way we refined the dual lattice, each of those refined dual $(d-p-i)$-cells in the unrefined $\sigma^\vee$ (see previous paragraph) is nothing but the geometric intersection of $\sigma^\vee$ (unrefined dual $(d-p)$-cell) with some $\sigma'$ (original $(d-i)$-cell)---thus, we can denote such a refined dual $(d-p-i)$-cell as $\sigma^\vee_{\sigma'}$, see \cref{fig:refined_dual_lattice} for example. This is the key idea behind the refinement.
    
    With this notation, if we have some specified $f$ from $C_{\bullet}(\mathcal{M}^\vee_{\text{refined}})$ to $C_{\bullet}(\mathcal{M})$, then we can define
    \begin{equation}
        (a\cup b)_{\sigma'}:=\sum_{\text{$p$-cell }\sigma} a_{\sigma}b_{f(\sigma_{\text{$\sigma'$}}^\vee)}
        \label{eqn:def_cup}
    \end{equation}
    for a generic $p$-form $a$, $(d-p-i)$-form $b$ and $(d-i)$-cell $\sigma'$. If $\sigma$ is not contained in $\sigma'$, we set $\sigma_{\text{$\sigma'$}}^\vee=0$. This resolves issue 2 above. (Let us remark on the sign convention. For any $(d-i)$-cell $\sigma'$ containing $\sigma$, we first think of a lattice consisting of a single cell $\sigma'$, then the refined dual $(d-p-i)$-cell $\sigma_{\sigma'}^\vee$ in $\sigma^\vee$ should correspond to the dual cell of $\sigma$ in this single-cell lattice with the usual sign convention. For example, in \cref{fig:refined_dual_lattice}, if the two cubes are oriented by the right-hand-rule, and the purple plaquette is pointing towards $+\hat x$ under right-hand-rule, then the two purple line segments---refined dual cells---will both orient towards $+\hat x$.)

    \item Similar to what we have derived for $f_\text{naive}$ before, here, per issue 3, again $f$ must be a chain map satisfying
    \begin{equation}
        \partial f - f \partial=0.
        \label{eqn:chain_map}
    \end{equation}
    However, in the derivation for $f_\text{naive}$, we only used the integrated version of the Leibniz rule, so the term $d(a\cup b)$ does not appear. Now we want to consider the Leibniz rule without the integration, and we must consider the refinement of the dual lattice. A crucial property of the refined dual lattice is that for any $p$-cell $\sigma$ and one of its adjacent cell $\sigma'$ in the original lattice, we have
    \begin{equation}
        \partial(\sigma^\vee_{\sigma'})=(\partial\sigma^\vee)_{\sigma'}+(-1)^{p}\sigma^\vee_{\partial\sigma'},
        \label{eqn:dual_lattice_property}
    \end{equation}
    where we have extended all definitions by linearity. For example, in \cref{fig:refined_dual_lattice}, consider the purple plaquette as $\sigma$ (pointing $+\hat x$) and the $c_+$ cube as $\sigma'$. $\partial(\sigma^\vee_{\sigma'})$ is given by the central point of the $c_+$ cube minus the midpoint of $\sigma^\vee$. On the other hand, $(\partial\sigma^\vee)_{\sigma'}$ is the central point of the $c_+$ cube and $\sigma^\vee_{\partial\sigma'}$ is negative of the midpoint of $\sigma^\vee$. The negative sign is because $\partial\sigma'$ contains $-\sigma$, so $\sigma^\vee_{-\sigma}$ gives negative of the midpoint of $\sigma^\vee$. And as $p=2$, this result agrees with \cref{eqn:dual_lattice_property}.
    
    Now we are ready to proof the Leibniz rule, which says
    \begin{equation}
        [d(a\cup b)]_{\sigma'}=(a\cup b)_{\partial\sigma'}=(-1)^p(a\cup db)_{\sigma'}+(da\cup b)_{\sigma'}.
        \label{eqn:Leibniz_rule}
    \end{equation}
    Using our definition of cup product \cref{eqn:def_cup}, we need to prove
    \begin{equation}
        \sum_{\text{$p$-cell }\sigma} a_\sigma b_{f(\sigma^\vee_{\partial\sigma'})}=(-1)^p\sum_{\text{$p$-cell }\sigma} a_{\sigma} b_{\partial f(\sigma^\vee_{\sigma'})}+\sum_{\text{$(p+1)$-cell }\tilde\sigma} a_{\partial\tilde\sigma} b_{f(\tilde\sigma^\vee_{\sigma'})}
        \label{eqn:proof_of_Leibniz}
    \end{equation}
    If we define a dual $(d-p-1)$-form $b^\vee$ as $b^\vee_{\tilde\sigma^\vee}=b_{f(\tilde\sigma^\vee_{\sigma'})}$, the right-hand-side of \cref{eqn:proof_of_Leibniz} becomes
    \begin{equation}
    \begin{aligned}
        \text{RHS}&=(-1)^p\sum_\sigma a_{\sigma} b_{\partial f(\sigma^\vee_{\sigma'})}+\sum_{\tilde\sigma} a_{\partial\tilde\sigma} b^\vee_{\tilde\sigma^\vee}\\
        &=(-1)^p\sum_\sigma a_{\sigma} b_{\partial f(\sigma^\vee_{\sigma'})}-(-1)^p\sum_\sigma a_{\sigma} b^\vee_{\partial\sigma^\vee}\\
        &=(-1)^p\sum_\sigma a_{\sigma} b_{\partial f(\sigma^\vee_{\sigma'})}-(-1)^p\sum_\sigma a_{\sigma} b_{f((\partial\sigma^\vee)_{\sigma'})},
    \end{aligned}
    \end{equation}
    where the second ``$=$'' is due to the lattice integration by parts \cref{eqn:lattice_integration_by_part}. Thus the difference between the LHS and RHS of \cref{eqn:proof_of_Leibniz} is
    \begin{equation}
        \begin{aligned}
        \text{LHS}-\text{RHS}&=\sum_{\text{$p$-cell }\sigma} a_\sigma b_{f(\sigma^\vee_{\partial\sigma'})}
        -(-1)^p\sum_\sigma a_{\sigma} b_{\partial f(\sigma^\vee_{\sigma'})}+(-1)^p\sum_\sigma a_{\sigma} b_{f((\partial\sigma^\vee)_{\sigma'})}\\
        &=\sum_\sigma a_\sigma b_{f((-1)^p(\partial\sigma^\vee)_{\sigma'}+\sigma^\vee_{\partial\sigma'}-(-1)^p\partial (\sigma^\vee_{\sigma'}))}-(-1)^p\sum_\sigma a_\sigma b_{\partial f(\sigma^\vee_{\sigma'})- f(\partial(\sigma^\vee_{\sigma'}))}\\
        &=-(-1)^p\sum_\sigma a_\sigma b_{\partial f(\sigma^\vee_{\sigma'})- f(\partial(\sigma^\vee_{\sigma'}))}=0.
        \end{aligned}.
    \end{equation}
    Thus, at the last step, we see that the chain map property of $f$ leads to the Leibniz's rule \cref{eqn:Leibniz_rule}, i.e. \cref{eqn:cup_relation} with $i+1=0$.

    While the cup product satisfies the Leibniz rule like the continuum wedge product does, unlike the latter it is not antisymmetric. This leads to the $i = 0$ equation in \cref{eqn:cup_relation}, which involves the notion of $\cup_1$. So next we to consider the cases of $i \geq 0$, namely, the higher cup products and the ``generalized Leibniz rule'' among them \cref{eqn:cup_relation}. 
    
    For $i = 0$, the RHS of \cref{eqn:cup_relation} closely resembles that of the case $i+1=0$, except that $\cup_1$ is used in place of $\cup_0$. This form suggests that $\cup_1$ can be defined in a way similarly to \cref{eqn:def_cup} for $\cup_0$, using a map $h$ from $C_{\bullet}(\mathcal{M}^\vee_{\text{refined}})$ to $C_{\bullet+1}(\mathcal{M})$:
    \begin{equation}
        (a\cup_1 b)_{\sigma'}:=(-1)^{q+1}\sum_\sigma a_{\sigma}b_{h(\sigma_{\sigma'}^\vee)}
    \end{equation}
    for any $p$-form $a$ and $q$-from $b$, on any cell $\sigma'$ (we will explain the reason for the appearance of the sign $(-1)^{q+1}$ shortly). Note that since the ``$\cup_1$'' product of $p$-form $a$ with $q$-form $b$ will result in a $(p+q+1)$-from, we require $h$ to map $q$-chains to $(q+1)$-chains. By this definition, the right hand side of \cref{eqn:cup_relation} for $i=0$ on any cell $\sigma'$ reads
    \begin{equation}
    \begin{aligned}
        \text{RHS}_{\sigma'}
        &=\sum_\sigma a_\sigma b_{\partial h(\sigma^\vee_{\sigma'})}-
        (-1)^p\sum_{\tilde\sigma}a_{\partial\tilde\sigma}b_{h(\tilde\sigma^\vee_{\sigma'}))}+(-1)^p\sum_\sigma a_\sigma b_{h(\sigma^\vee_{\partial\sigma'})}\\
        &=\sum_\sigma a_\sigma b_{\partial h(\sigma^\vee_{\sigma'})}+\sum_{\sigma}a_{\sigma}b_{h((\partial\sigma^\vee)_{\sigma'}))}+(-1)^p\sum_\sigma a_\sigma b_{h(\sigma^\vee_{\partial\sigma'})}\\
        &=\sum_\sigma a_\sigma b_{\partial h(\sigma^\vee_{\sigma'})+h(\partial(\sigma^\vee_{\sigma'}))}
    \end{aligned},
    \end{equation}
    where we use the same trick as in proving the conventional Leibniz rule.
    
    The appearance of $\partial h + h \partial$ suggests that $h$ is likely a homotopy between some chain maps. To see what these chain maps are, let us turn to the LHS. We have already defined $a\cup b$ by a chain map $f$. Now we may define a $\tilde{f}$, which satisfies a relation analogously to $f$ but with the order of arguments reversed, i.e.
    \begin{equation}
        (b\cup a)_{\sigma'}=(-1)^{pq}\sum_\sigma a_\sigma b_{\tilde f(\sigma_{\sigma'}^\vee)},
        \label{eqn:tilde_f}
    \end{equation}
    Note that once we have constructed the cup product using $f$, $\tilde f$ could be read out. Heuristically speaking, $f$ ``defines'' a certain ``lattice vector field'' along which any dual lattice cell $\sigma^\vee$ is mapped to $f(\sigma^\vee)$, while $\tilde{f}$ maps $\sigma^\vee$ in the opposite direction. After substituting \cref{eqn:tilde_f}, the LHS of \cref{eqn:cup_relation} for $i=0$ becomes 
    \begin{equation}
        \text{LHS}_{\sigma'}=\sum_\sigma a_\sigma b_{f(\sigma^\vee_{\sigma'})}-\sum_\sigma a_\sigma b_{\tilde f(\sigma^\vee_{\sigma'})}.
    \end{equation}
    Thus \cref{eqn:cup_relation} for $i=0$ is satisfied if and only if
    \begin{equation}
        f-\tilde f=\partial h+h\partial,
    \end{equation}
    which means that $h$ is a homotopy between $f$ and $\tilde f$. Now we can also see that the additional sign $(-1)^{q+1}$ introduced when defining $\cup_1$ using $h$ earlier is merely to align with the condition for the homotopy between chain maps. 

    Inductively, higher cup products are realized as higher homotopies (homotopies between homotopies). To match the sign in Leibniz rule, we define
    \begin{equation}
       (a\cup_i b)_{\sigma'}=(-1)^{i(q+i)}\sum_\sigma a_\sigma b_{h^{(i)}(\sigma^\vee_{{\sigma'}})},
       \label{eqn:def_cup_i_by_h}
    \end{equation}
    and
    \begin{equation}
        (b\cup_i a)_{\sigma'}=(-1)^{pq+i}(-1)^{i(q+i)}\sum_\sigma a_\sigma b_{\tilde h^{(i)}(\sigma^\vee_{\sigma'})},
    \end{equation}
    where we have denoted $i$-th homotopy by $h^{(i)}, \tilde h^{(i)}$, which are maps from $C_{\bullet}(\mathcal{M}^\vee_{\text{refined}})$ to $C_{\bullet+i}(\mathcal{M})$, satisfying
    \begin{equation}
        h^{(i-1)}-\tilde h^{(i-1)}=\partial h^{(i)}-(-1)^ih^{(i)}\partial
        \label{eqn:homotopy}
    \end{equation}
    and in particular, $h^{(-1)}=0,h^{(0)}=f, h^{(1)}=h$. With such a set of (higher) homotopies $h^{(i)}$, we can construct a collection of $\cup_i$ that satisfy \cref{eqn:cup_relation}. Furthermore, it is evident that, if we have a set of $\cup_i$ satisfying \cref{eqn:cup_relation}, we can also construct a set of (higher) homotopies $h^{(i)}$ conversely.
    \item To ensure locality of the cup product, we require that $h^{(i)}(\sigma^\vee_{\sigma'})$ must be geometrically contained in $\sigma'$ for any $\sigma,\sigma'$ of the original lattice. In particular, this means for each vertices ($0$-cell) of the original lattice, $f=h^{(0)}$ must act as identity on them. It is easy to see why: if we want to glue two lattices together at a point, we need a definition of the cup product at that point, and this definition must not involve any other points. For general $p$-cells, the reasoning is similar.
\end{enumerate}

In summary, we define $\cup_i$ product using \cref{eqn:def_cup_i_by_h}. The condition \cref{eqn:cup_relation} is equivalent to \cref{eqn:homotopy}. The locality requirement becomes that $h^{(i)}(\sigma^\vee_{\sigma'})$ must be geometrically contained in $\sigma'$ for any $\sigma,\sigma'$ of the original lattice.

As an example, we use the construction of the cup products on regular cubic lattice to illustrate our method. We will construct the action of $f$ on $0$-, $1$-, $2$-, and $3$-cells sequentially from lower to higher dimensions. This is because once the action of $f$ on $p$-cells is obtained, the chain map condition that $f$ must satisfy imposes constraints on the action of $f$ on $(p+1)$-cells: namely, the boundary of the image of a $(p+1)$-cell under $f$ must equal the image under $f$ of the boundary of that $(p+1)$-cell. And since the cup product is local, we can construct $f$ first on a single cube, and then extend the definition by gluing. 

We first determine the action of $f$ on vertices of the original cube, now seen as vertices on the refined dual lattice. As we have argued before, $f$ must act as identity on them. This gives the cup product between $0$-form $a$ and $0$-form $b$. 
Next, we let 
\begin{equation}
    f(\text{center of the $x,y,z$-direction edge of the original cube})=\text{$+\hat x,+\hat y,+\hat z$ end point of such edge}.
\end{equation}
One way to visualize it is to think of us moving these points by $+\hat{x}/2$, $+\hat{y}/2$, and $+\hat{z}/2$ respectively. Now we have the cup product between $1$-form $a$ and $0$-form $b$. 
For face centers, we set
\begin{equation}
    f(\text{center of the $xy,yz,zx$-direction face of the original cube})=\text{$+\hat x+\hat y,+\hat y+\hat z,+\hat z+\hat x$ vertex of such face}.
\end{equation}
Also one may visualize this as moving these points by $+\hat{x}/2+\hat y/2$, $+\hat{y}/2+\hat{z}/2$, and $+\hat{z}/2+\hat{x}/2$ respectively. And thus we have the cup product between $2$-form $a$ and $0$-form $b$.
The final type of $0$-cell is the center of the original cube. We ``move'' it by $+\hat{x}/2+\hat y/2+\hat z/2$, i.e.,
\begin{equation}
    f(\text{center of the original cube})=\text{$+\hat x+\hat y+\hat z$ vertex of the original cube}.
\end{equation}
This gives the definition for the cup product between $3$-form $a$ and $0$-form $b$.

For $1$-cells, we use the simplest choice that satisfy the Leibniz rule. That is if the boundary points of the $1$-cell have been mapped to the same vertex, we set $f$ on such $1$-cell to be $0$, or if not, we set $f$ on such $1$-cell to be the shortest path between the image of $f$ on its boundary points. For example, we set
\begin{equation}
    f\left(\vcenter{\hbox{\includegraphics[width=0.15\linewidth]{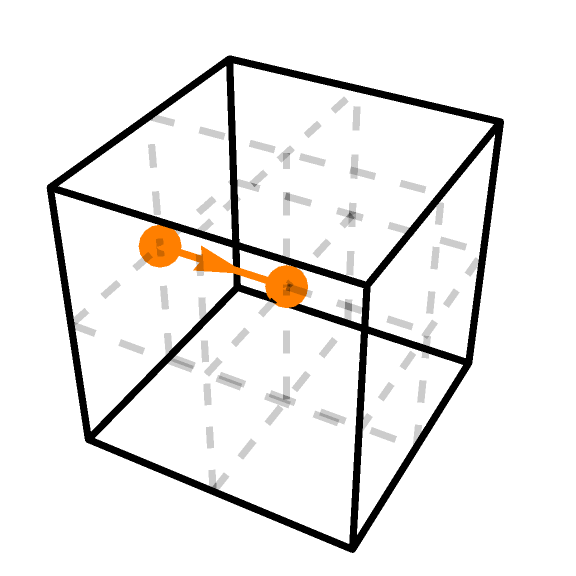}}}\right)=\vcenter{\hbox{\includegraphics[width=0.15\linewidth]{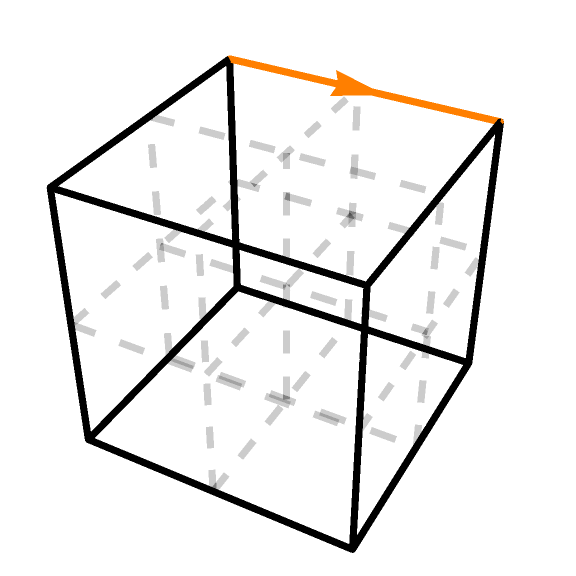}}},
\end{equation}
This gives the first term in the cup product of $2$-form $a$ and $1$-from $b$. For $2$-cells and $3$-cells, we define $f$ using the same principle. The resulting cup product is precisely the one we defined in \cref{sec:cup_on_cube}.

To define the cup-$1$ product, we first read out $\tilde f$ using the definition \cref{eqn:tilde_f}. For example, the central point of a cube is mapped to the vertex in the $-\hat x-\hat y - \hat z$ direction. Then, following the same order in which we defined $f$, we sequentially define the action of $h$ on the $0$-, $1$-, and $2$-cells. Note that, since our lattice is three-dimensional, $h$ can only yield $0$ for any $3$-cell. 

For any point, $h\partial$ is always $0$, so the Leibniz rule is just $\partial h=f-\tilde f$. For vertices of the original cube, $f=\tilde f$, thus we simply set $h=0$ on them. For midpoints of edges of the original cube, action of $f$ and $\tilde f$ gives the two endpoints of such edges. The simplest choice is to set $h$ on these central points to be such edges. The action of $h$ on face centers is a little bit more non-trivial. For example, we shall set
\begin{equation}
    h\left(\vcenter{\hbox{\includegraphics[width=0.15\linewidth]{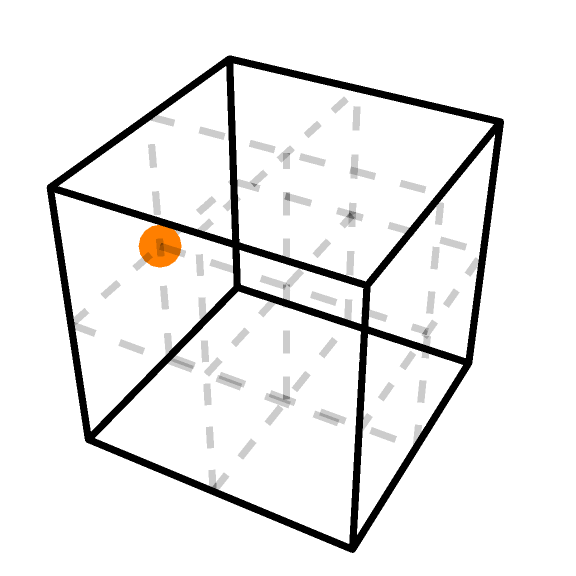}}}\right)=\vcenter{\hbox{\includegraphics[width=0.15\linewidth]{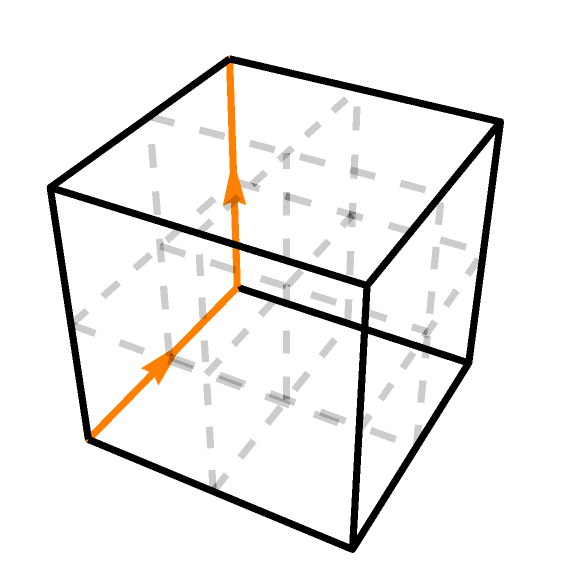}}}.
\end{equation}
For centers of $xy,yz,zx$ faces, we set $h$ on them to be the $+\hat x-\hat y,+\hat y-\hat z,+\hat x-\hat z$ half of the boundary of such faces. Next, for the center of the original cube, we define
\begin{equation}
    h\left(\vcenter{\hbox{\includegraphics[width=0.15\linewidth]{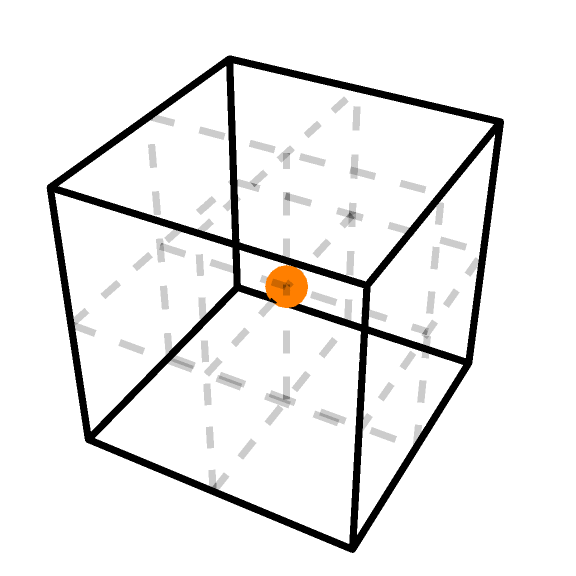}}}\right)=\vcenter{\hbox{\includegraphics[width=0.15\linewidth]{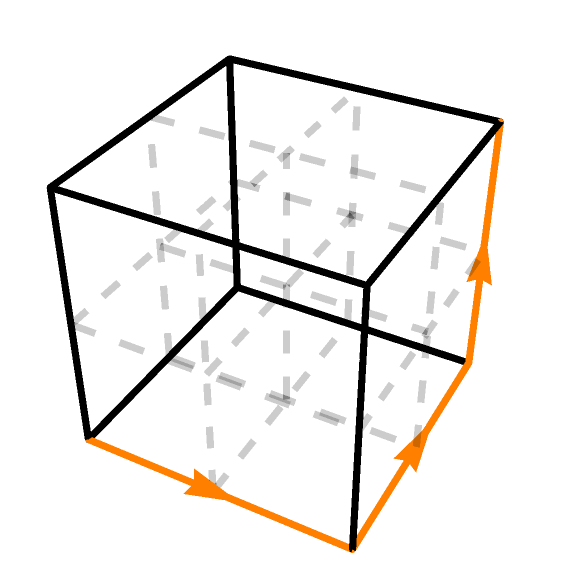}}}.
\end{equation}

For $1$-cells, we again use the simplest choice that satisfy the Leibniz rule. For example, we set
\begin{equation}
    h\left(\vcenter{\hbox{\includegraphics[width=0.15\linewidth]{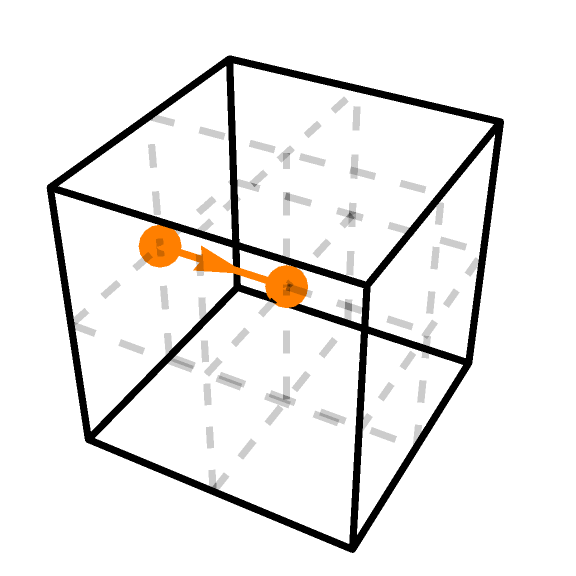}}}\right)=\vcenter{\hbox{\includegraphics[width=0.15\linewidth]{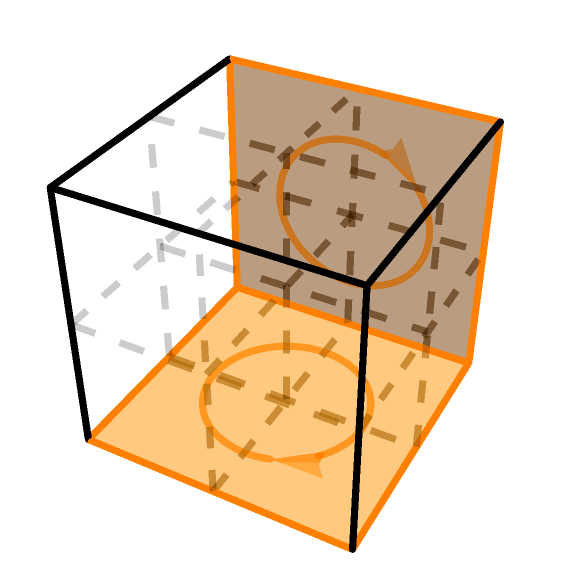}}}.
\end{equation}
That is because
\begin{equation}
    f\left(\vcenter{\hbox{\includegraphics[width=0.15\linewidth]{plot/cup_cube_1_2_2_h_1_l.png}}}\right)
    -\tilde f\left(\vcenter{\hbox{\includegraphics[width=0.15\linewidth]{plot/cup_cube_1_2_2_h_1_l.png}}}\right)=
    \vcenter{\hbox{\includegraphics[width=0.15\linewidth]{plot/cup_cube_0_2_1_f_1_r.png}}}
\end{equation}
and 
\begin{equation}
    h\partial\left(\vcenter{\hbox{\includegraphics[width=0.15\linewidth]{plot/cup_cube_1_2_2_h_1_l.png}}}\right)=
    \vcenter{\hbox{\includegraphics[width=0.15\linewidth]{plot/cup_cube_1_3_1_h_0_r.png}}}-\vcenter{\hbox{\includegraphics[width=0.15\linewidth]{plot/cup_cube_1_2_2_h_0_r.png}}}.
\end{equation}
Following this pattern, we can define all ``$\cup_{0,1}$'' products on a regular cubic lattice as shown in \cref{tab:cup_on_cube}. 

On the special layers of the lattice modular $T$ operator, similar construction can be made. First, we define $f$ on the vertices of the original lattice, then on the midpoints of edges, the centers of faces, and finally on the center of the cube. Among these, only the centers of the two triangular faces are not included in the regular cubic lattice. We define the result of $f$ acting on these two triangular face centers as the orange points in the two diagrams corresponding to $a^2\cup b^0$ in \cref{tab:cup_on_cube_special_layer}. Then, for the action of $f$ on $1,2$-cells and $3$-cells, we still follow the previous principles of choosing the shortest path, minimal area, and minimal volume. As for $h$, the non-trivial choices still appear in its action on the two triangular face centers: we select the orange links in the two diagrams corresponding to $a^1\cup_1 b^1$ in \cref{tab:cup_on_cube_special_layer}. Then we can continue to use the principles of minimal area and minimal volume to define the action of $h$ on $1,2$-cells. 

Furthermore, ``$\cup_2$'' can also be obtained in a similar way. First, the action of $h^{(2)}$ on the face centers is obvious (similar to how the action of $h$ on edge midpoints is evident), so the non-trivial definition arises from the action of $h^{(2)}$ on cube centers. After specifying the result of this action, the action of $h^{(2)}$ on $1$-cells can be determined by applying the minimal volume principle. Since the Chern-Simons theory discussed in this paper does not involve ``$\cup_2$'', we will not provide the detailed construction here.

\section{Fermionic Theories}
\label{sec:fermionic_path_integral}

In this section we provide the details of the lattice construction of fermionic Chern-Simons-Maxwell theories. While the construction on cubic lattice has already been introduced in \cite{Xu:2024hyo} citing some technical details from \cite{Chen:2019mjw}, now in the lattice modular $T$ operator we have lattice cells of more general shapes, so we need to introduce the more general construction. Also we would like to introduce the lattice realization of the fermion parity $\mathbb{Z}_2$ flux tube $w$ that is crucial in \cref{sec:mod_T} (though not directly used in the main text).

When $k$ is an odd integer, under the $1$-form $\mathbb{Z}$ transformation
\begin{equation}
\left\{
    \begin{aligned}
        A&\mapsto A+2\pi m\\
        s&\mapsto s+dm
    \end{aligned}\right.\ ,
\end{equation}
lattice Chern-Simons part changes by a sign
\begin{equation}
    \exp\left(\int\frac{ik}{4\pi}A\cup (dA-2\pi s)-\frac{ik}{4\pi}2\pi s \cup A\right) \ \mapsto \ (-1)^{\int m\cup dm+s\cup dm + dm\cup s} \times \left(\phantom{\frac{}{}}\mbox{iteself}\phantom{\frac{}{}}\right) \ .
    \label{eqn:bosonic_part}
\end{equation}
This sign ambiguity is to be compensated by a fermionic factor \cite{Gu:2012ib,Gaiotto:2015zta}
\begin{align}
    z_\chi[s]=\sigma_\chi[s]\: (-1)^{\sum_p s_p\eta_p}=\pm 1
\end{align}
which is realized as a path integral $\sigma_\chi[s]$ of fermion worldlines along $s \mod 2$, together with a phase $(-1)^{\int s\eta}$, where $\eta$ is a $\mathbb{Z}_2$-valued $1$-from on the dual lattice that will encode the spin structure. We will first construct $\sigma_\chi[s]$ so that under the gauge transformation $s\mapsto s+dm$ it will have the desired factor $(-1)^{\int m\cup dm+s\cup dm + dm\cup s}$ compensating that in \cref{eqn:bosonic_part}. But aside from this desired factor, from $\sigma_\chi[s]$ there arises yet another gauge ambiguity factor $(-1)^{\int m {w_2}_{\text{rep}}}$ where ${w_2}_{\text{rep}}$, a $2$-form on the dual lattice, is a particular representative element of the second Stiefel-Whitney class \cite{Gaiotto:2015zta}. To remove this remaining gauge ambiguity, the extra phase $(-1)^{\int s\eta}$ with $\tilde d\eta={w_2}_{\text{rep}} \mod 2$ (where $\tilde{d}$ is like $d$ but on the dual lattice) is introduced. Now there is no gauge ambiguity, however this factor will in turn introduce a spin structure dependence.

Now we begin the concrete construction of $\sigma_\chi[s]$. Following \cite{Gu:2012ib,Gaiotto:2015zta}, $\sigma_\chi[s]$ is defined as a fermionic partition function. We denote the mod $2$ reduction $\mathtt s_p\in \{0, 1\}, \ (-1)^{\mathtt s_p}=(-1)^{s_p}$ on the plaquettes, and the plaquettes on which $\mathtt s_p=1$ form closed loops on the dual lattice because $ds=0$; then we put Majorana fermion worldloops on top of these $\mathtt s_p=1$ loops, and these worldloops will contribute some $\pm 1$ phase. More exactly, each plaquette $p$ is associated with a pair Grassmann fields $\chi_p$ and $\bar{\chi}_p$, one residing on each side of the plaquette (which one resides on which side will be specified below). In each cube we have a weight $h_c[s,\chi,\bar\chi]$, which depends on those $\chi,\bar\chi$ living inside the cube and the $\mathtt s$ on the plaquettes of the boundary of the cube. The partition function is then given by:
\begin{equation}
    \sigma_\chi[s] = \left(\prod_p \int  d\chi_p \, d\bar\chi_p \right)
    \left(\prod_c h_c[s,\chi,\bar\chi]\right)
    \left(\prod_p e^{\bar\chi_p \chi_p}\right),
\end{equation}
where:
\begin{itemize}
    \item the first product represents the Grassmann integral measure over all the plaquette fields;
    \item the second product is to absorb the Grassmann integrals on those plaquettes where the $\mathtt s_p=1$ worldloops (on dual lattice) go through, generating some $(\pm 1)$ factor that depends on the detailed design of $h$ and the detailed shape of the worldloops;
    \item the third product is to absorb the remaining Grassmann integrals on those plaquettes with no worldloop going through, i.e. where $\mathtt s_p=0$.
\end{itemize}
Thus, after integrating out the Grassmann fields on those plaquettes with $\mathtt{s}_p=0$, we are left with
\begin{equation}
    \sigma_\chi[s] = \left(\prod_{\text{$p$ s.t. $\mathtt{s}(p)=1$}} \int  d\chi_p \, d\bar\chi_p \right)
    \left(\prod_c h_c[s,\chi,\bar\chi]\right).
    \label{eqn:z_chi_no_mass}
\end{equation}
We now explicitly construct the $h_c[s,\chi,\bar\chi]$ for each cube $c$ and specify which of the Grassmann fields $\chi_p, \bar\chi_p$ lives on which side of each plaquette $p$.
\begin{enumerate}
    \item 

For a regular lattice cube $c$, we define $h_c[s,\chi,\bar\chi]$ as \cite{Chen:2019mjw}
\begin{equation}
    h_c[s,\chi,\bar\chi]=\chi_{-z}^{\mathtt{s}_{-z}}\chi_{+y}^{\mathtt{s}_{+y}}\chi_{-x}^{\mathtt{s}_{-x}}\bar\chi_{+z}^{\mathtt{s}_{+z}}\bar\chi_{-y}^{\mathtt{s}_{-y}}\bar\chi_{+x}^{\mathtt{s}_{+x}},
\end{equation}
where we use $\pm \mu$ ($\mu=x,y,z$) to denote the plaquette $p$ whose center is $\pm \hat{\mu}/2$ from the center of the cube $c$. And as the $\chi,\bar\chi$ in the $h_c$ indicate, $\chi_{-z}$, $\chi_{+y}$, $\chi_{-x}$, $\bar\chi_{+z}$, $\bar\chi_{-y}$, $\bar\chi_{+x}$ reside on the interior-facing sides of their respective plaquettes, while their conjugate partners $\bar\chi_{-z}$, $\bar\chi_{+y}$, $\bar\chi_{-x}$, $\chi_{+z}$, $\chi_{-y}$, $\chi_{+x}$ are on the exterior-facing sides and thus do not appear in $h_c$.

The design of $h_c$ looks complicated. An intuitive Berry phase interpretation is given in \cite{Chen:2019mjw}. But more concretely, the design is guided by the following mathematical feature: Look at the definition of $a^2 \cup_1 b^2$ in \cref{sec:cup_on_cube}. In each of the first two terms there, the Grassmann field $\chi$ associated with a purple plaquette must appear in $h_c$ to the right of the $\chi$ associated with an orange plaquette. Similarly, in each of the last two terms there, the Grassmann field $\bar\chi$ associated with a purple plaquette must appear in $h_c$ to the left of the $\chi$ associated with an orange plaquette. This design is to ensure $\sigma_\chi[s]$ has the crucial property
\begin{equation}
    \sigma_\chi[s+s']=\sigma_\chi[s]\sigma_\chi[s'](-1)^{\int s\cup_1s'},
    \label{eqn:quadratic_refinement_property}
\end{equation}
which will be very useful later. To verify this property, we first evaluate the product of two weight functions on the same cube (which will appear in $\sigma_\chi[s]\sigma_\chi[s']$):
\begin{equation}
    \begin{aligned}
    h_c&[{s},\chi,\bar\chi]h_c[{s}',\chi',\bar{\chi'}]\\
    &=\chi_{-z}^{\mathtt{s}_{-z}}\chi_{+y}^{\mathtt{s}_{+y}}\chi_{-x}^{\mathtt{s}_{-x}}\bar\chi_{+z}^{\mathtt{s}_{+z}}\bar\chi_{-y}^{\mathtt{s}_{-y}}\bar\chi_{+x}^{\mathtt{s}_{+x}}
    {\chi'}_{-z}^{{\mathtt{s}'}_{-z}}{\chi'}_{+y}^{{\mathtt{s}'}_{+y}}{\chi'}_{-x}^{{\mathtt{s}'}_{-x}}\bar{\chi'}_{+z}^{{\mathtt{s}'}_{+z}}\bar{\chi'}_{-y}^{{\mathtt{s}'}_{-y}}\bar{\chi'}_{+x}^{{\mathtt{s}'}_{+x}}\\
    &=(-1)^{\mathtt{s}_{+x} {\mathtt{s}'}_{+x}+\mathtt{s}_{-y} {\mathtt{s}'}_{-y}+\mathtt{s}_{+z}{\mathtt{s}'}_{+z}}(-1)^{\mathtt{s}_{-y}{\mathtt{s}'}_{+x}+\mathtt{s}_{+z}({\mathtt{s}'}_{-y}+{\mathtt{s}'}_{+x})}(-1)^{\mathtt{s}_{-x}({\mathtt{s}'}_{-z}+{\mathtt{s}'}_{+y})+\mathtt{s}_{+y}{\mathtt{s}'}_{-z}}\\
    &\quad\quad
    (\chi_{-z}^{\mathtt{s}_{-z}}{\chi'}_{-z}^{{\mathtt{s}'}_{-z}})
    (\chi_{+y}^{\mathtt{s}_{+y}}{\chi'}_{+y}^{{\mathtt{s}'}_{+y}})
    (\chi_{-x}^{\mathtt{s}_{-x}}{\chi'}_{-x}^{{\mathtt{s}'}_{-x}})
    (\bar\chi_{+z}^{\mathtt{s}_{+z}}\bar{\chi'}_{+z}^{{\mathtt{s}'}_{+z}})
    (\bar\chi_{-y}^{\mathtt{s}_{-y}}\bar{\chi'}_{-y}^{{\mathtt{s}'}_{-y}})
    (\bar\chi_{+x}^{\mathtt{s}_{+x}}\bar{\chi'}_{+x}^{{\mathtt{s}'}_{+x}})\\
    &=(-1)^{\mathtt{s}_{+x} {\mathtt{s}'}_{+x}+\mathtt{s}_{-y} {\mathtt{s}'}_{-y}+\mathtt{s}_{+z}{\mathtt{s}'}_{+z}}
    (-1)^{\mathtt{s}\cup_1 \mathtt{s}'}\\
    &\quad\quad
    (\chi_{-z}^{\mathtt{s}_{-z}}{\chi'}_{-z}^{{\mathtt{s}'}_{-z}})
    (\chi_{+y}^{\mathtt{s}_{+y}}{\chi'}_{+y}^{{\mathtt{s}'}_{+y}})
    (\chi_{-x}^{\mathtt{s}_{-x}}{\chi'}_{-x}^{{\mathtt{s}'}_{-x}})
    (\bar\chi_{+z}^{\mathtt{s}_{+z}}\bar{\chi'}_{+z}^{{\mathtt{s}'}_{+z}})
    (\bar\chi_{-y}^{\mathtt{s}_{-y}}\bar{\chi'}_{-y}^{{\mathtt{s}'}_{-y}})
    (\bar\chi_{+x}^{\mathtt{s}_{+x}}\bar{\chi'}_{+x}^{{\mathtt{s}'}_{+x}})
    \end{aligned},
    \label{eqn:h_c_h_c_cubic}
\end{equation}
where we have used the condition $ds_c=ds'_c=0$ to simplify the result. In \cref{eqn:h_c_h_c_cubic}, the previously mentioned feature causes the $\cup_1$ to naturally arise when the order of the Grassmann variables is interchanged. Notice that to relate $\sigma_\chi[s]\sigma_\chi[s']$ to $\sigma_\chi[s+s']$, using \cref{eqn:z_chi_no_mass}, we need to integrate out Grassmann variables $\chi_p,\bar{\chi}_p,\chi_p',\bar{\chi'}_p$ on plaquette $p$ where $\mathtt s(p)\mathtt s'(p)=1$. Since we need to exchange the Grassmann variables in the integral like $\int d\chi d\bar\chi d\chi'd\bar{\chi'}(\chi\chi')(\bar\chi\bar{\chi'})$, we have an extra sign $\prod_p(-1)^{\mathtt s_p\mathtt s'_p}$, which cancels with the factor $(-1)^{\mathtt{s}_{+x} {\mathtt{s}'}_{+x}+\mathtt{s}_{-y} {\mathtt{s}'}_{-y}+\mathtt{s}_{+z}{\mathtt{s}'}_{+z}}$ in \cref{eqn:h_c_h_c_cubic}. So we have proven \cref{eqn:quadratic_refinement_property}. This derivation is parallel to that presented in \cite{Gaiotto:2015zta} for simplicial complex.

\item
Now we define $h_c$ on the two special layers in the lattice modular $T$ operator, guided by the same feature in relation to $\cup_1$ as before. On the lower layer we define
\begin{equation}
    h_c[{s},\chi,\bar\chi]=\chi_{-z}^{\mathtt{s}_{-z}}\chi_{+y}^{\mathtt{s}_{+y}}\chi_{-x}^{\mathtt{s}_{-x}}\bar\chi_{+z_1}^{\mathtt{s}_{+z_1}}\bar\chi_{+z_2}^{\mathtt{s}_{+z_2}}\bar\chi_{-y}^{\mathtt{s}_{-y}}\bar\chi_{+x}^{\mathtt{s}_{+x}},
\end{equation}
where $+z_{1,2}$ represent the two triangular plaquettes in the $+\hat{z}$ direction from the cube center, $+z_1$ for the one located on the $-\hat{x}-\hat{y}$ side, $+z_2$ for the one located on the $\hat{x}+\hat{y}$ side.
Using this definition of $h_c[s,\chi,\bar\chi]$, we have
\begin{equation}
    \begin{aligned}
    h_c&[{s},\chi,\bar\chi]h_c[{s}',\chi,\bar{\chi'}]\\
    &=\chi_{-z}^{\mathtt{s}_{-z}}\chi_{+y}^{\mathtt{s}_{+y}}\chi_{-x}^{\mathtt{s}_{-x}}\bar\chi_{+z_1}^{\mathtt{s}_{+z_1}}\bar\chi_{+z_2}^{\mathtt{s}_{+z_2}}\bar\chi_{-y}^{\mathtt{s}_{-y}}\bar\chi_{+x}^{\mathtt{s}_{+x}}
    {\chi'}_{-z}^{{\mathtt{s}'}_{-z}}{\chi'}_{+y}^{{\mathtt{s}'}_{+y}}{\chi'}_{-x}^{{\mathtt{s}'}_{-x}}\bar{\chi'}_{+z_1}^{{\mathtt{s}'}_{+z_1}}\bar{\chi'}_{+z_2}^{{\mathtt{s}'}_{+z_2}}\bar{\chi'}_{-y}^{{\mathtt{s}'}_{-y}}\bar{\chi'}_{+x}^{{\mathtt{s}'}_{+x}}\\
    &=(-1)^{\mathtt{s}_{+x} {\mathtt{s}'}_{+x}+\mathtt{s}_{-y} {\mathtt{s}'}_{-y}+\mathtt{s}_{+z_2}{\mathtt{s}'}_{+z_2}+\mathtt{s}_{+z_1}{\mathtt{s}'}_{+z_1}}\\
    &\quad\quad(-1)^{\mathtt{s}_{-y}{\mathtt{s}'}_{+x}+\mathtt{s}_{+z_2}({\mathtt{s}'}_{-y}+{\mathtt{s}'}_{+x})+\mathtt{s}_{+z_1}({\mathtt{s}'}_{+z_2}+{\mathtt{s}'}_{-y}+{\mathtt{s}'}_{+x})}
    (-1)^{\mathtt{s}_{-x}({\mathtt{s}'}_{-z}+{\mathtt{s}'}_{+y})+\mathtt{s}_{+y}{\mathtt{s}'}_{-z}}\\
    &\quad\quad\quad
    (\chi_{-z}^{\mathtt{s}_{-z}}{\chi'}_{-z}^{{\mathtt{s}'}_{-z}})
    (\chi_{+y}^{\mathtt{s}_{+y}}{\chi'}_{+y}^{{\mathtt{s}'}_{+y}})
    (\chi_{-x}^{\mathtt{s}_{-x}}{\chi'}_{-x}^{{\mathtt{s}'}_{-x}})\\
    &\quad\quad\quad\quad
    (\bar\chi_{+z_1}^{\mathtt{s}_{+z_1}}\bar{\chi'}_{+z_1}^{{\mathtt{s}'}_{+z_1}})
    (\bar\chi_{+z_2}^{\mathtt{s}_{+z_2}}\bar{\chi'}_{+z_2}^{{\mathtt{s}'}_{+z_2}})
    (\bar\chi_{-y}^{\mathtt{s}_{-y}}\bar{\chi'}_{-y}^{{\mathtt{s}'}_{-y}})
    (\bar\chi_{+x}^{\mathtt{s}_{+x}}\bar{\chi'}_{+x}^{{\mathtt{s}'}_{+x}})\\
    &=(-1)^{\mathtt{s}_{+x} {\mathtt{s}'}_{+x}+\mathtt{s}_{-y} {\mathtt{s}'}_{-y}+\mathtt{s}_{+z_2}{\mathtt{s}'}_{+z_2}+\mathtt{s}_{+z_1}{\mathtt{s}'}_{+z_1}}
    (-1)^{\mathtt{s}\cup_1 \mathtt{s}'}\\
    &\quad\quad
    (\chi_{-z}^{\mathtt{s}_{-z}}{\chi'}_{-z}^{{\mathtt{s}'}_{-z}})
    (\chi_{+y}^{\mathtt{s}_{+y}}{\chi'}_{+y}^{{\mathtt{s}'}_{+y}})
    (\chi_{-x}^{\mathtt{s}_{-x}}{\chi'}_{-x}^{{\mathtt{s}'}_{-x}})\\
    &\quad\quad\quad(\bar\chi_{+z_1}^{\mathtt{s}_{+z_1}}\bar{\chi'}_{+z_1}^{{\mathtt{s}'}_{+z_1}})
    (\bar\chi_{+z_2}^{\mathtt{s}_{+z_2}}\bar{\chi'}_{+z_2}^{{\mathtt{s}'}_{+z_2}})
    (\bar\chi_{-y}^{\mathtt{s}_{-y}}\bar{\chi'}_{-y}^{{\mathtt{s}'}_{-y}})
    (\bar\chi_{+x}^{\mathtt{s}_{+x}}\bar{\chi'}_{+x}^{{\mathtt{s}'}_{+x}})\\
    \end{aligned}
\end{equation}

While on the upper layer we define
\begin{equation}
    h_c[{s},\chi,\bar\chi]=\chi_{-z_1}^{\mathtt{s}_{-z_1}}\chi_{-z_2}^{\mathtt{s}_{-z_2}}\chi_{+y}^{\mathtt{s}_{+y}}\chi_{-x}^{\mathtt{s}_{-x}}\bar\chi_{+z}^{\mathtt{s}_{+z}}\bar\chi_{-y}^{\mathtt{s}_{-y}}\bar\chi_{+x}^{\mathtt{s}_{+x}},
\end{equation}
where $-z_{1,2}$ represents the two triangular plaquettes in the $-\hat{z}$ direction from the cube center, $-z_1$ for the one located on the $\hat{x}-\hat{y}$ side, $-z_2$ for the one located on the $-\hat{x}+\hat{y}$ side (matching our previous definition). Now we have
\begin{equation}
    \begin{aligned}
    h_c&[{s},\chi,\bar\chi]h_c[{s}',\chi,\bar{\chi'}]\\
    &=\chi_{-z_1}^{\mathtt{s}_{-z_1}}\chi_{-z_2}^{\mathtt{s}_{-z_2}}\chi_{+y}^{\mathtt{s}_{+y}}\chi_{-x}^{\mathtt{s}_{-x}}\bar\chi_{+z}^{\mathtt{s}_{+z}}\bar\chi_{-y}^{\mathtt{s}_{-y}}\bar\chi_{+x}^{\mathtt{s}_{+x}}
    {\chi'}_{-z_1}^{\mathtt{s}_{-z_1}}{\chi'}_{-z_2}^{\mathtt{s}_{-z_2}}{\chi'}_{+y}^{{\mathtt{s}'}_{+y}}{\chi'}_{-x}^{{\mathtt{s}'}_{-x}}\bar{\chi'}_{+z}^{{\mathtt{s}'}_{+z}}\bar{\chi'}_{-y}^{{\mathtt{s}'}_{-y}}\bar{\chi'}_{+x}^{{\mathtt{s}'}_{+x}}\\
    &=(-1)^{\mathtt{s}_{+x} {\mathtt{s}'}_{+x}+\mathtt{s}_{-y} {\mathtt{s}'}_{-y}+\mathtt{s}_{+z}{\mathtt{s}'}_{+z}}\\
    &\quad\quad(-1)^{\mathtt{s}_{-y}{\mathtt{s}'}_{+x}+\mathtt{s}_{+z}({\mathtt{s}'}_{-y}+{\mathtt{s}'}_{+x})}
    (-1)^{\mathtt{s}_{-x}({\mathtt{s}'}_{-z_1}+{\mathtt{s}'}_{-z_2}+{\mathtt{s}'}_{+y})+\mathtt{s}_{+y}({\mathtt{s}'}_{-z_1}+{\mathtt{s}'}_{-z_2})+\mathtt{s}_{-z_2}{\mathtt{s}'}_{-z_1}}\\
    &\quad\quad\quad
    (\chi_{-z_1}^{\mathtt{s}_{-z_1}}{\chi'}_{-z_1}^{{\mathtt{s}'}_{-z_1}})
    (\chi_{-z_2}^{\mathtt{s}_{-z_2}}{\chi'}_{-z_2}^{{\mathtt{s}'}_{-z_2}})
    (\chi_{+y}^{\mathtt{s}_{+y}}{\chi'}_{+y}^{{\mathtt{s}'}_{+y}})
    (\chi_{-x}^{\mathtt{s}_{-x}}{\chi'}_{-x}^{{\mathtt{s}'}_{-x}})\\
    &\quad\quad\quad\quad
    (\bar\chi_{+z}^{\mathtt{s}_{+z}}\bar{\chi'}_{+z}^{{\mathtt{s}'}_{+z}})
    (\bar\chi_{-y}^{\mathtt{s}_{-y}}\bar{\chi'}_{-y}^{{\mathtt{s}'}_{-y}})
    (\bar\chi_{+x}^{\mathtt{s}_{+x}}\bar{\chi'}_{+x}^{{\mathtt{s}'}_{+x}})\\
    &=(-1)^{\mathtt{s}_{+x} {\mathtt{s}'}_{+x}+\mathtt{s}_{-y} {\mathtt{s}'}_{-y}+\mathtt{s}_{+z}{\mathtt{s}'}_{+z}}
    (-1)^{\mathtt{s}\cup_1 \mathtt{s}'}\\
    &\quad\quad
    (\chi_{-z_1}^{\mathtt{s}_{-z_1}}{\chi'}_{-z_1}^{{\mathtt{s}'}_{-z_1}})
    (\chi_{-z_2}^{\mathtt{s}_{-z_2}}{\chi'}_{-z_2}^{{\mathtt{s}'}_{-z_2}})
    (\chi_{+y}^{\mathtt{s}_{+y}}{\chi'}_{+y}^{{\mathtt{s}'}_{+y}})
    (\chi_{-x}^{\mathtt{s}_{-x}}{\chi'}_{-x}^{{\mathtt{s}'}_{-x}})\\
    &\quad\quad\quad
    (\bar\chi_{+z}^{\mathtt{s}_{+z}}\bar{\chi'}_{+z}^{{\mathtt{s}'}_{+z}})
    (\bar\chi_{-y}^{\mathtt{s}_{-y}}\bar{\chi'}_{-y}^{{\mathtt{s}'}_{-y}})
    (\bar\chi_{+x}^{\mathtt{s}_{+x}}\bar{\chi'}_{+x}^{{\mathtt{s}'}_{+x}})\\
    \end{aligned}.
\end{equation}
Use the same argument for regular cubic lattice, we can again prove the crucial property \cref{eqn:quadratic_refinement_property}.
\end{enumerate}

Now, returning to our original motivation of compensating for the anomalous factor in the gauge transformation of \cref{eqn:bosonic_part}, \cref{eqn:quadratic_refinement_property} implies that
\begin{equation}
    \sigma_\chi[s+dm]=\sigma_\chi[s]\sigma_\chi[dm](-1)^{\int s\cup_1dm}=\sigma_\chi[s]\sigma_\chi[dm](-1)^{\int s\cup dm+dm\cup s}.
\end{equation}
We might expect that $\sigma_\chi[dm]$ satisfies $\sigma_\chi[dm] \text{``}=\text{''}(-1)^{\int m \cup dm}$, which would completely cancel the additional $(-1)$ factor. However, using \cref{eqn:quadratic_refinement_property}, we can only prove that 
\begin{equation}
    \sigma_\chi[dm] = (-1)^{\int m \cup dm + (\text{linear term in } m)}.
\end{equation}
To determine the actual value of the linear term, one may compute $\sigma_\chi[dm]$ for $m$ supported on a single link. The result for the linear term depends on the lattice and the cup product, and it turns out to be a representative of the second Stiefel-Whitney class ${w_2}_{\text{rep}}$ of the dual lattice \cite{Gaiotto:2015zta}: 
\begin{equation}
    \sigma_\chi[dm]=(-1)^{\int m\cup dm+m{w_2}_{\text{rep}}}.
\end{equation}
Notice that here we use the nature product between a ($\mathbb{Z}_2$-valued) $2$-form on the dual lattice (${w_2}_{\text{rep}}$) and a $1$-form on the original lattice ($m$). 

\begin{figure}[!htbp]
    \begin{subfigure}[b]{0.45\linewidth}
    \centering
    \includegraphics[width=0.6\linewidth]{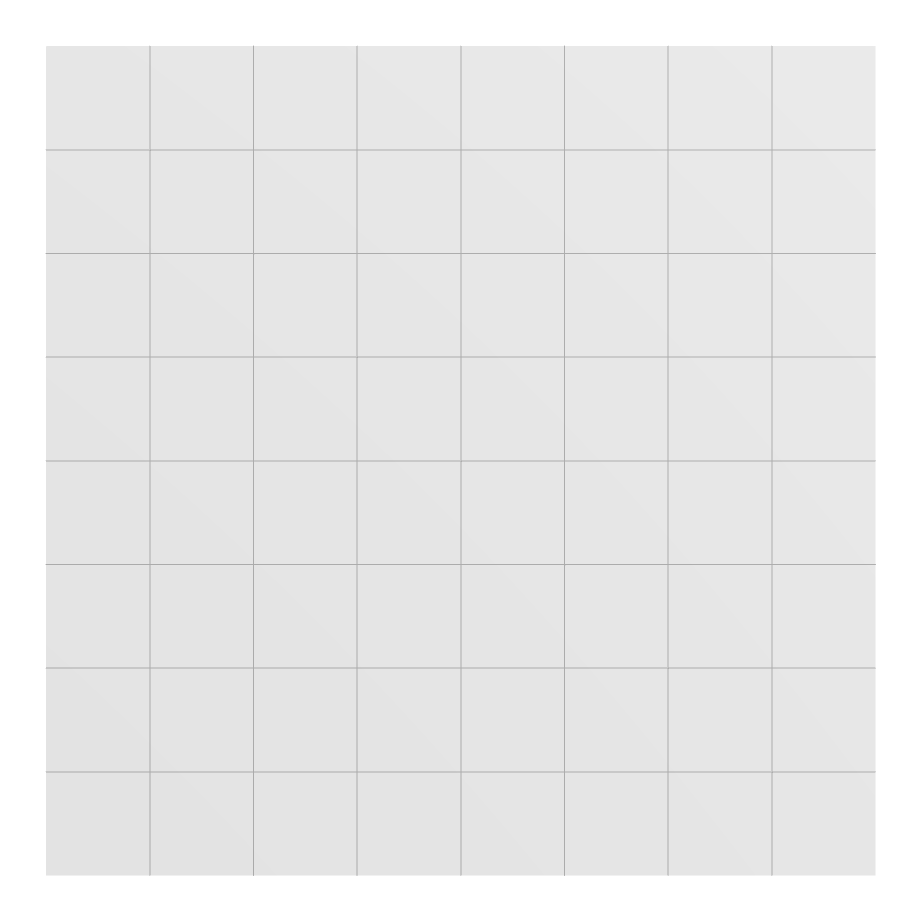}
    \caption{fermions anti-periodic around circumference}
    \label{fig:cartesian}
    \end{subfigure}
    \hspace{1cm}
    \begin{subfigure}[b]{0.45\linewidth}
    \centering
    \includegraphics[width=0.6\linewidth]{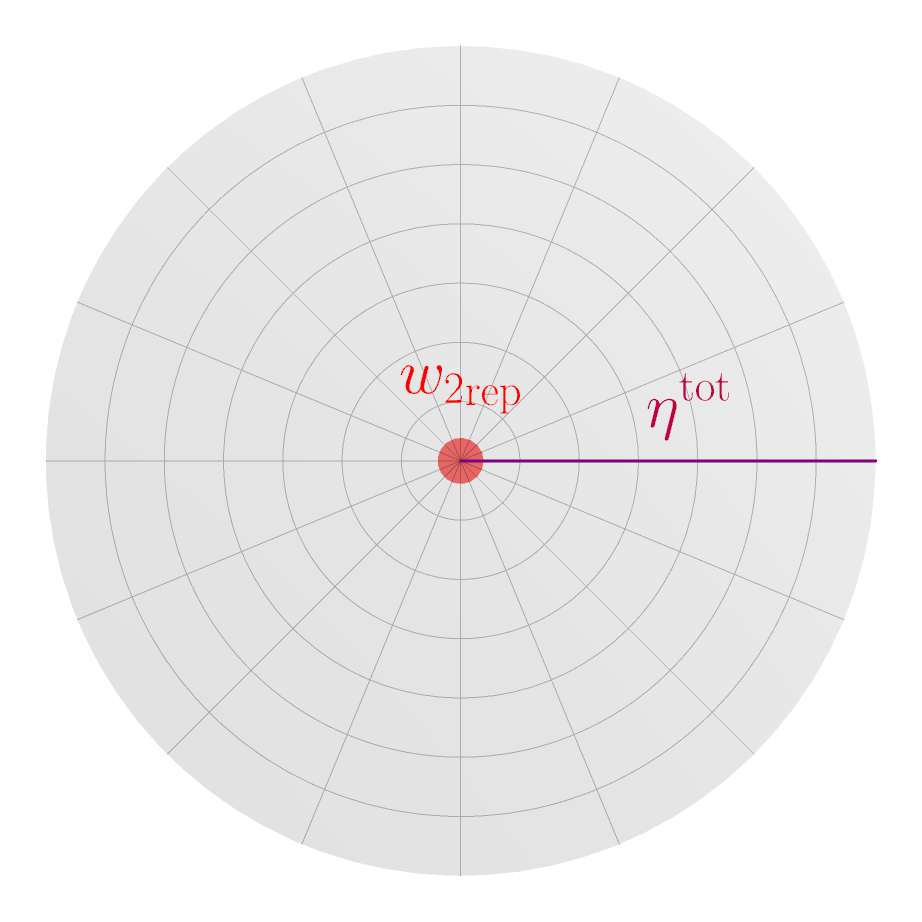}
    \caption{fermions anti-periodic around circumference}
    \label{fig:polar}
    \end{subfigure}

    \begin{subfigure}[b]{0.45\linewidth}
    \centering
    \includegraphics[width=0.6\linewidth]{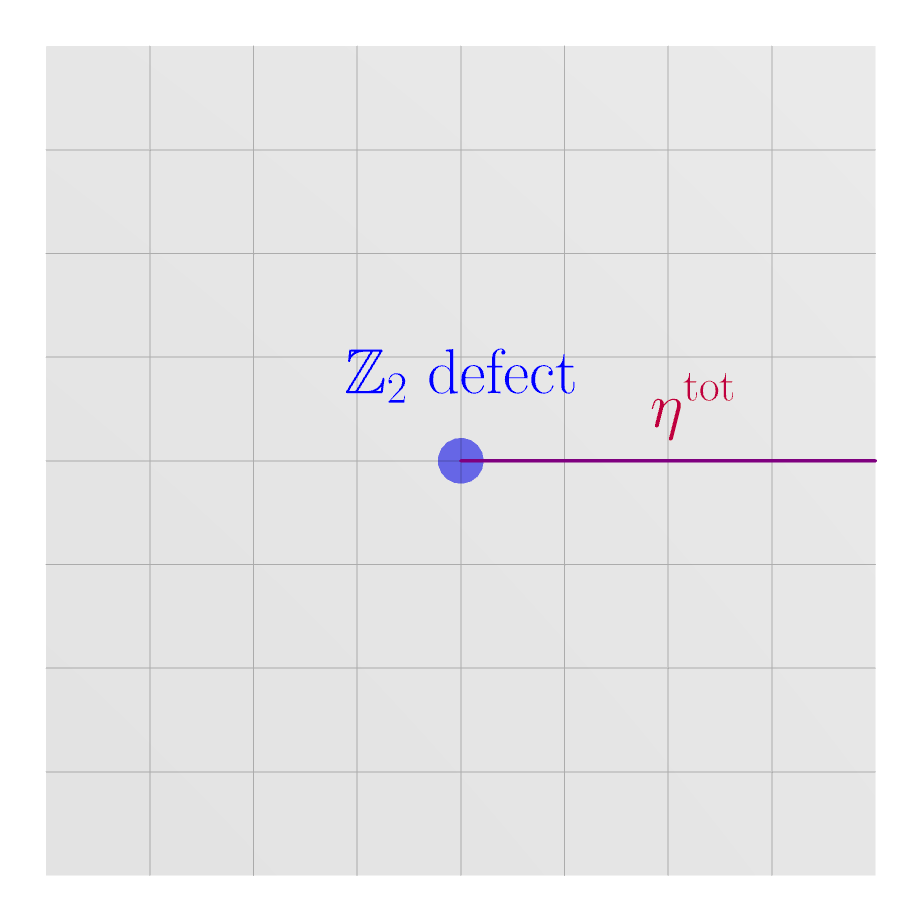}
    \caption{fermions periodic around circumference}
    \label{fig:cartesian_defect}
    \end{subfigure}
    \hspace{1cm}
    \begin{subfigure}[b]{0.45\linewidth}
    \centering
    \includegraphics[width=0.6\linewidth]{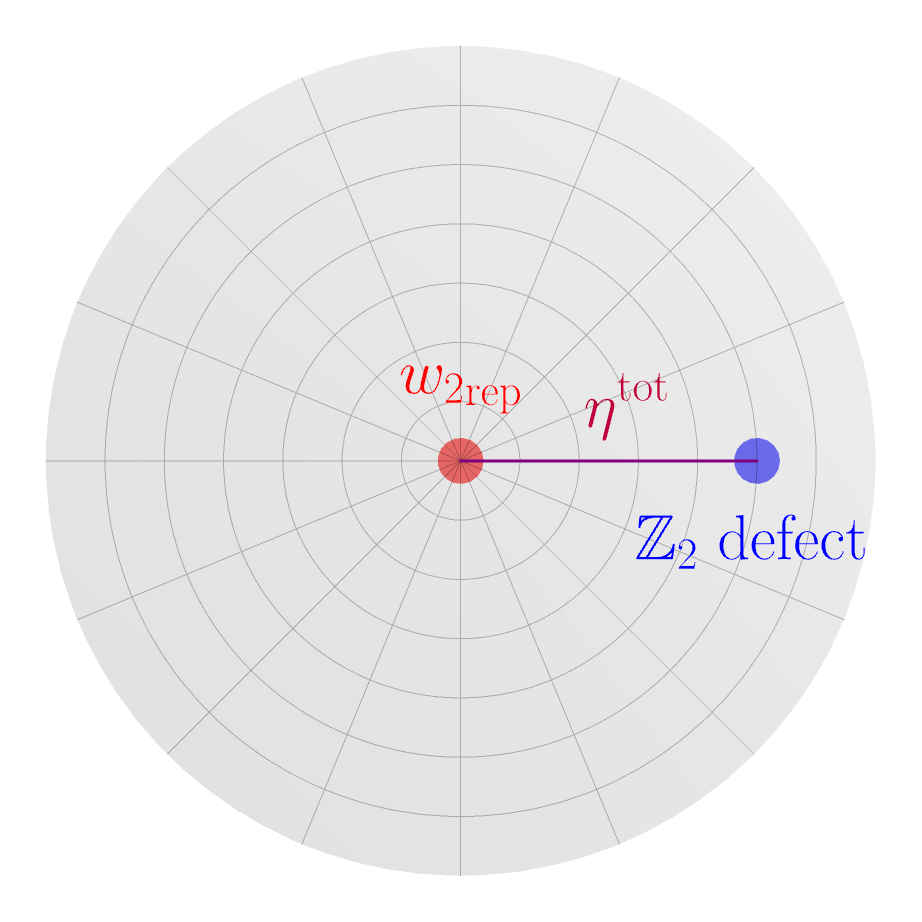}
    \caption{fermions periodic around circumference}
    \label{fig:polar_defect}
    \end{subfigure}
    \caption{Both \cref{fig:cartesian} and \cref{fig:polar} have \emph{no} defect in the cross section, and both lead to \emph{anti-periodic} fermions around the circumference, although the details work out differently: In \cref{fig:cartesian}, we use the familiar framing on the square lattice (see the cup product on plaquettes in \cref{sec:cup_on_cube}), so when we go around the circumference once, a $2\pi$ rotation relative to this framing is made, leading to a $(-1)$ phase for a fermion traveling around the circumference. In \cref{fig:polar}, by contrast, the framing looks like that of a square lattice but growing radially outwards, so when we go around the circumference, no rotation relative to the framing is made, but on the other hand this framing inevitably has a non-trivial $w_{2\mathrm{rep}}$ at the origin, so we have a non-trivial domain wall $\eta^{\mathrm{tot}}=\eta$ extending from the origin to the boundary, leading to a $(-1)$ phase for a fermion traveling around the circumference. In \cref{fig:cartesian_defect} and \cref{fig:polar_defect}, we inserted a fermion parity $\mathbb{Z}_2$ defect \cref{eqn:fermion_parity_defect} inside the solid torus. The introduction of $\eta'$ leads to change of $\eta^{\mathrm{tot}}$ at the boundary, turning the fermions around the circumference from being anti-periodic to being \emph{periodic}.}
    \label{fig:cross_section}
\end{figure}

The appearance of second Stiefel-Whitney class is actually not surprising. On the dual lattice, $dm$ represents some closed loops. The fermions in $\sigma_\chi[dm]$ will move along these loops. Fermions acquire a phase change of $(-1)$ under a $2\pi$ rotation, which is represented by $m\cup dm$, the self-linking number of $dm$ regarded as a loop on the dual lattice. However if the background framing (implicitly given by the cup product on the lattice) around such loop undergoes a $2\pi$ rotation, then we should obtain an extra phase of $(-1)$---much like the effect of a spin connection background in the continuum. And ${w_2}_{\text{rep}}$ is precisely the location around which the local framing undergoes a $2\pi$ rotation, see \cref{fig:polar} to be introduced below for an example. 

Now the $1$-form $\mathbb{Z}$ gauge invariance is violated by $(-1)^{m{w_2}_{\text{rep}}}$. Fortunately, the second Stiefel-Whitney class of any orientable $3$-manifold vanishes, so ${w_2}_{\text{rep}}$ must be exact, i.e. there must exist some $\mathbb{Z}_2$-valued $1$-form $\eta$ on the dual lattice such that $\tilde d\eta={w_2}_{\text{rep}} \mod 2$ (where $\tilde{d}$ is like $d$ but on the dual lattice), and we introduce an extra $(-1)^{\int s\eta}$ in the path integral, so that the gauge invariance is restored by $(-1)^{\int (dm)\eta}=(-1)^{\int m\tilde d\eta}=(-1)^{\int m{w_2}_{\text{rep}}}$ \cite{Gaiotto:2015zta}. This $1$-form field $\eta$ on the dual lattice can be viewed as a domain wall on the original lattice, and the domain wall ends on ${w_2}_{\text{rep}}$. Each time a fermion worldline represented by $\mathtt{s}$ crosses this domain wall, it obtains a $(-1)$ phase.

Note that given the lattice and the cup product, ${w_2}_{\text{rep}}$ is determined, but it does not uniquely determine $\eta$. First consider $\eta\mapsto \eta+\tilde d\kappa$ for any $\mathbb{Z}_2$-valued $0$-form $\kappa$ on the dual lattice. This is completely equivalent to $\eta$ because $(-1)^{\int s \eta}$ changes by $(-1)^{\int ds \kappa}=1$ due to the closeness of $s$; this can be pictured as a local deformation of the domain wall. On the other hand, consider $\eta$ versus $\eta+\xi$ where $\xi$ is a closed but non-exact $\mathbb{Z}_2$-valued $1$-form on the dual lattice, $\tilde d\xi=0 \mod 2$ but $\xi\neq \tilde d\kappa \mod 2$. Both of them satisfy $\tilde d\eta=\tilde d(\eta+\xi)={w_2}_{\text{rep}} \mod 2$, so both serve our purpose. But they are physically distinct when coupled to non-contractible loop $s$. Therefore, different $[\xi]\in H^1(\mathcal{M}^\vee; \mathbb{Z}_2)\cong H_2(\mathcal{M}; \mathbb{Z}_2)$ correspond to different \emph{spin structures}, i.e. different choices fermion boundary conditions that we can make.

In conclusion, if we define the fermionic Maxwell-Chern-Simons theory as
\begin{align}
    Z[\eta]=&\left[\prod_{\text{link} \: l} \int_{-\pi}^\pi \frac{dA_l}{2\pi}\right] \left[\prod_{\text{plaq} \: p} \sum_{s_p\in\mathbb{Z}} \right]
    \left[\prod_{\text{cube} \: c} \int_{-\pi}^\pi \frac{d\lambda_c}{2\pi} \: e^{i\lambda_c ds_c}\right] \nonumber\\
    &\exp \left\{\frac{ik}{4\pi}\sum_c \left[(A\cup dA)_c - (A\cup 2\pi s)_c - (2\pi s\cup A)_c\right] \right. \left. -\frac{1}{2e^2}\sum_p F_p^2 \right\} z_\chi[s],
\end{align}
where $z_\chi[s]=\sigma_\chi[s](-1)^{\sum_p s_p\eta_p}$, the theory becomes invariant under 1-form $\mathbb{Z}$ transformation. The theory now depends on the spin structure data $\eta$, where different choices can be made between $\eta$ vs $\eta+\xi$ for $[\xi]\in H^1(\mathcal{M}^\vee; \mathbb{Z}_2)\cong H_2(\mathcal{M}; \mathbb{Z}_2)$. For our definition of cup products on the twisted spacetime $\mathcal{T}_m$, one can check that ${w_2}_{\text{rep}}$ is simply $0$, so $\eta$ itself canonically takes value in $[\eta]\in H^1(\mathcal{T}_m; \mathbb{Z}_2)$, and $H^1$ for $\mathcal{T}_m$ has been computed in \cref{sec:top}: for odd $m$, there is only the choice of periodic fermion boundary condition around the $x$-direction, while for odd $m$ both periodic and anti-periodic choices can be made around the $x$-direction.

Finally, we would like to explain how the fermion parity $\mathbb{Z}_2$ flux tube $w$---the key concept introduced in \cref{sec:mod_T} (though not needed in the main text) to interpret the spectrum of the fermionic modular $T$ operator---is realized on the lattice. In \cite{Chen:2019mjw}, the coupling \cref{eqn:backgnd_U1} to the $U(1)$ background is introduced on the lattice as
\begin{align}
    \exp\left\{i\sum_l a_l \left( \frac{\tilde d\mathbf{A}}{2\pi}-\mathbf{s} \right)_l-i\sum_p s_p \mathbf{A}_p \right\}
\end{align}
where the $U(1)$ background $\mathbf{A}_p\in (-\pi, \pi]$ lives on the dual lattice link $p^\vee$, and $\mathbf{s}_l\in\mathbb{Z}$ (satisfying $\tilde d \mathbf{s}=0$) is its corresponding background Dirac string that lives on the dual lattice plaquette $l^\vee$; all desired gauge invariances are preserved (upon suitable treatments if spacetime has a boundary). Manifestly and crucially, $-\mathbf{s}_l$ is at the same time nothing but the charge of a Wilson loop insertion, therefore a narrow $2\pi$ background flux tube is indeed indistinguishable from creating an anyon worldline \cite{Chen:2019mjw}, as we said below \cref{eqn:backgnd_U1}. Now, to create a narrow $\pi$ flux tube, we simply set $\mathbf{A}_p=\pi \eta'_p$ with $\eta'_p=0, 1$, leading to the flux located at where $\tilde d\eta'_l=1 \mod 2$. If we define $\eta^{\mathrm{tot}}=\eta+\eta' \mod 2$, then in the path integral we have a factor of $(-1)^{\int s\eta^{\mathrm{tot}}}$, such that the fermion parity $\mathbb{Z}_2$ defect is defined to be the deviation
\begin{align}
    \tilde{d}\eta^{\mathrm{tot}}-{w_2}_{\text{rep}}=\tilde{d}\eta' \mod 2
    \label{eqn:fermion_parity_defect}
\end{align}
Along this defect, we have an additional coupling $e^{i\sum_l a_l \tilde{d}\eta'_l/2}$, so it is indeed like ``half of a Wilson loop'', as we said in \cref{sec:mod_T}.

In \cref{fig:cross_section} we consider two possible lattice realizations of the cross section of a solid torus \cref{fig:solid_torus}, demonstrating the key idea in \cref{sec:mod_T} that, the fermion being periodic or anti-periodic in the $x$-direction on the boundary only relies on whether a fermion parity $\mathbb{Z}_2$ defect \cref{eqn:fermion_parity_defect} is inserted in the cross section, regardless of how the cross section is realized as a lattice in detail, e.g. whether ${w_2}_{\text{rep}}$ and $\eta$ is trivial or not in the cross section.

\section{Algorithm for Evaluating $Z'_{\mathcal{T}}$ and $Z'_{\mathcal{T}_m}$}
% \subsection{Numerical Recursion}

Using the derivation in Appendix E of \cite{Xu:2024hyo}, we translated the Chern-Simons-Maxwell path integral to a Gaussian path integral along with some topological treatments, and the Gaussian integral is the $Z'$ we introduced in the main text, which we now evaluate. We will perform a complete gauge fixing of the $A'$ field on $\mathcal{T}$, but as we said in the main text, the $\mathrm{nlocFP}$ topological treatment of the $A'$ Gaussian integral in $Z'$ no only drops the zero modes from gauge invariance (exact forms), but also those from the artificially introduced 1-form $\mathbb{R}$ symmetries (closed non-exact forms). In practice, after the complete gauge fixing, we just drop any remaining zero eigenvalues in the determinant evaluation during the Gaussian integral. Now we introduce the algorithm which first constructs $e^{-\beta H'}$, and then performs the trace $Z'_{\mathcal{T}}=\Tr(T' e^{-\beta H'})$.

Assume we have a free field theory with degrees of freedom labeled by a vector $x\in\mathbb{R}^D$, where $D$ is the number of degrees of freedom. Since the path integral is Gaussian, the resulting matrix elements of $e^{-\beta H}$ take in the following form
\begin{equation}
    \langle x_F|e^{-\beta_1H}|x_I\rangle
    =\alpha_{\beta_1}\exp\left[-\frac{1}{2}\begin{pmatrix}
        x_F^t&x_I^t
    \end{pmatrix}
    \begin{pmatrix}
        M_{\beta_1} & N_{\beta_1}\\
        N^t_{\beta_1} & Q_{\beta_1}\\
    \end{pmatrix}
    \begin{pmatrix}
        x_F\\x_I
    \end{pmatrix}\right],
    \label{eqn:MNQ_def}
\end{equation}
where $x_I$, $x_F$ are the boundary conditions on the initial and final time slices, and $M, N, Q$ are matrices. We get a recursive relation by
\begin{equation}
    \begin{aligned}
        &\langle x_F|e^{-(\beta_1+\beta_2)H}|x_I\rangle=\int Dx_M\langle x_F|e^{-\beta_2 H}|x_M\rangle\langle x_M|e^{-\beta_1H}|x_I\rangle\\
        =&\alpha_{\beta_1}\alpha_{\beta_2}\int Dx_M\exp\left[-\frac{1}{2}\begin{pmatrix}
            x_F^t&x_M^t&x_I^t
        \end{pmatrix}
        \begin{pmatrix}
            M_{\beta_2} & N_{\beta_2}&\\
            N^t_{\beta_2} & Q_{\beta_2}+M_{\beta_1} &N_{\beta_1}\\
            &  N^t_{\beta_1}& Q_{\beta_1}\\
        \end{pmatrix}
        \begin{pmatrix}
            x_F\\x_M\\x_I
        \end{pmatrix}\right]\\
        =&\frac{\alpha_{\beta_1}\alpha_{\beta_2}}{\sqrt{\det (2\pi)^{-1}(Q_{\beta_2}+M_{\beta_1})}}\\
        &\exp\left[-\frac{1}{2}\begin{pmatrix}
            x_F^t&x_I^t
        \end{pmatrix}
        \begin{pmatrix}
            M_{\beta_2}-N_{\beta_2}(Q_{\beta_2}+M_{\beta_1})^{-1}N_{\beta_2}^t & -N_{\beta_1}(Q_{\beta_2}+M_{\beta_1})^{-1}N_{\beta_2}\\
            -N_{\beta_2}^t(Q_{\beta_2}+M_{\beta_1})^{-1}N_{\beta_1}^t&  Q_{\beta_1}-N_{\beta_1}^t(Q_{\beta_2}+M_{\beta_1})^{-1}N_{\beta_1}\\
        \end{pmatrix}
        \begin{pmatrix}
            x_F\\x_I
        \end{pmatrix}\right]
    \end{aligned}
\end{equation}
i.e.
\begin{equation}
\left\{
\begin{aligned}
    M_{\beta_1+\beta_2}&=M_{\beta_2}-N_{\beta_2}(Q_{\beta_2}+M_{\beta_1})^{-1}N_{\beta_2}^t\\
    N_{\beta_1+\beta_2}&=-N_{\beta_2}(Q_{\beta_2}+M_{\beta_1})^{-1}N_{\beta_1}\\
    Q_{\beta_1+\beta_2}&=Q_{\beta_1}-N_{\beta_1}^t(Q_{\beta_2}+M_{\beta_1})^{-1}N_{\beta_1}\\
    \alpha_{\beta_1+\beta_2}&=\frac{\alpha_{\beta_1}\alpha_{\beta_2}}{\sqrt{\det (2\pi)^{-1}(Q_{\beta_2}+M_{\beta_1})}}
\end{aligned}\right. \ \ \ \ .
\label{eqn:recursion_relations}
\end{equation}
Note that more generally the recursive relation for $\alpha_{\beta}$ may contain extra factors coming form the lattice path integral measure $\int Dx_M$. For lattice field theory, the $M$, $N$, $Q$ matrices at $\beta=1$ can be directly read-off from the action. Then by this recursive relation, an $O(\log_2\beta)$ algorithm for $e^{-\beta H}$ can be realized, see the pseudo-codes \cref{alg:single_layer_update,alg:iteration}. (We can also analytically find the exact ground state projector by solving for the $\beta\rightarrow\infty$ fixed point solution. We will not introduce the details here since the logarithmic algorithm for large $\beta$ is good enough for our practical purpose.)

\begin{algorithm}[H]
    \caption{Single layer update}
    \label{alg:single_layer_update}
    \begin{algorithmic}[1] % [1] enables line numbering
        \State \textbf{Input:} Matrices $M_{\text{new layer}},N_{\text{new layer}},N^t_{\text{new layer}},Q_{\text{new layer}}$;$M,N,N^t,Q$ 
        \State \textbf{Output:} Updated matrices $M',N',{N'}^t,Q'$; scalar $f$ ($\log$ of $\alpha$)
        \Function{single\_layer\_update}{$M_{\text{new layer}},N_{\text{new layer}},N^t_{\text{new layer}},Q_{\text{new layer}},M,N,N^t,Q$}
        \State $M' \gets M_{\text{new layer}}-N_{\text{new layer}}(Q_{\text{new layer}}+M)^{-1}N^t_{\text{new layer}}$ 
        \State $N' \gets -N_{\text{new layer}}(Q_{\text{new layer}}+M)^{-1}N^t$ 
        \State ${N'}^t \gets -N(Q_{\text{new layer}}+M)^{-1}N^t_{\text{new layer}}$ 
        \State $Q' \gets Q-N(Q_{\text{new layer}}+M)^{-1}N^t$ 
        \State $f \gets -0.5\log\det2\pi(Q_{\text{new layer}}+M)$ \Comment{the reason why we write $2\pi$ instead of $(2\pi)^{-1}$ here is that we have extra $2\pi$ factor in ``$DA'_M$''}
        \State \textbf{return} $M',N',{N'}^t,Q,f$
    \EndFunction
    \end{algorithmic}
\end{algorithm}
\begin{algorithm}[H]
    \caption{Iteration of $e^{-\beta H}$}
    \label{alg:iteration}
    \begin{algorithmic}[1] % [1] enables line numbering
        \State \textbf{Input:} Matrices of $\beta=1$ and some $k_x$  $M_1,N_1,N^t_1,Q_1$; $\beta$ size $\log_2\beta$ 
        \State \textbf{Output:} Matrices 
        $M_\beta,N_\beta,{N}^t_\beta,Q_\beta$; scalar $f_\beta$
        \Function{exp\_beta\_H\_matrix}{$M_1,N_1,N^t_1,Q_1,\log_2\beta$}
        \State $f_\beta\gets0$
        \State $M,N,N^t,Q\gets M_1,N_1,N^t_1,Q_1$
        \State $i \gets 0$
        \For{$i<\log_2\beta$}
            \State $M',N',{N'}^t,Q',f \gets \Call{single\_layer\_update}{M,N,N^t,Q,M,N,N^t,Q}$
            \State $f_\beta\gets 2f_\beta+f$
            \State $M,N,N^t,Q\gets M',N',{N'}^t,Q'$
            \State $i\gets i+1$
        \EndFor
        \State \textbf{return} $M,N,{N}^t,Q,f_\beta$
        \EndFunction
    \end{algorithmic}
\end{algorithm}

For the $e^{-\beta H'}$ in our $Z'$ path integral in particular, after gauge fixing $A'_\tau = 0$, on each time slice there are $2 \times L \times L$ remaining degrees of freedom $A'_l$ on the $x$- and $y$-direction links. Then $e^{-\beta H'}$ is then evaluated by the recursive algorithm above, leading to matrix elements of the form
\begin{equation}
    \langle A_F'|e^{-\beta H'}|A_I'\rangle=\alpha_\beta\exp\left[-\frac{1}{2}\begin{pmatrix}
        {A_F'}^t&{A_I'}^t
    \end{pmatrix}
    \begin{pmatrix}
        M_{\beta} & N_{\beta}\\
        N^t_{\beta} & Q_{\beta}\\
    \end{pmatrix}
    \begin{pmatrix}
        A_F'\\A_I'
    \end{pmatrix}\right],
\end{equation}
where $A_I',A_F'$ are vectors composed of all $A_l'$  corresponding to the links $l$ on each time slice. Note that Fourier transformation can be performed in both $x$ and $y$, so each of the $M, N, Q$ matrices is block diagonalized into one $2\times 2$ matrix for each Fourier mode; we can do so from the very beginning at $\beta=1$ before the recursive process in $\beta$, making the process much more efficient. (In practice, we only Fourier transformed the $x$-direction, because the main time consuming step is going to be the trace with $T'$ below, in which only the $x$-direction can be Fourier transformed anyways.)

As introduced in the main text, the $T'$ operator is realized as two special layers of cubes. Since the action is still quadratic in the special layers, we can also represent it by some matrices $M_{T'},N_{T'},Q_{T'}$ and a scalar $\alpha_{T'}$:
\begin{equation}
    \langle {\tilde A_F'}|T'|A_I'\rangle=\alpha_\beta\exp\left[-\frac{1}{2}\begin{pmatrix}
        {\tilde A_F'}{}^t&{A_I'}{}^t
    \end{pmatrix}
    \begin{pmatrix}
        M_{T'} & N_{T'}\\
        N^t_{T'} & Q_{T'}\\
    \end{pmatrix}
    \begin{pmatrix}
        {\tilde A_F'}\\A_I'
    \end{pmatrix}\right].
    \label{eqn:T'_matrices}
\end{equation}
Since the lattice used to compute $T'$ consists of two layers, we can first obtain the $M$, $N$, and $Q$ matrices corresponding to each of these two layers separately directly form the lattice action, and then combine them using  \cref{eqn:recursion_relations} to obtain $M_{T'}$, $N_{T'}$, and $Q_{T'}$. ote that Fourier transformation can be performed in the $x$-direction from the beginning, but the $y$-direction has no translation invariance in $T'$. So we will actually perform the computation above for each Fourier mode $k_x$.

\begin{figure}[!htbp]
    \centering
    \includegraphics[width=0.5\linewidth]{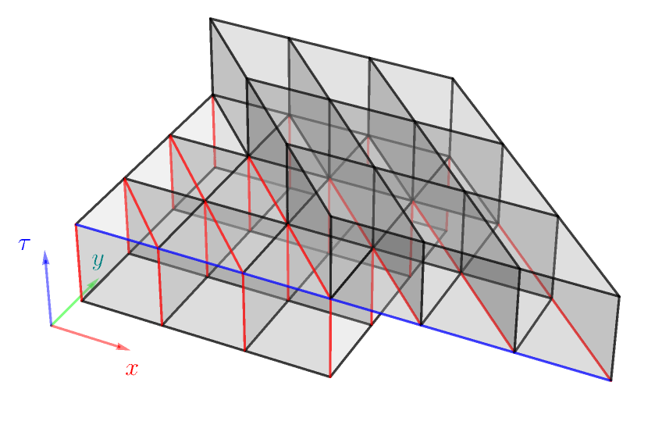}
    \caption{Gauge fixing on the special layer: $A'$ on red links are set to be $0$; the $k_x\neq0$ components of $A'$ on the blue links are also set to be $0$ (i.e., we only keep the holonomy along the loop represented by the blue line as a variable).}
    \label{fig:gauge_fixing}
\end{figure}

Notice that in \cref{eqn:T'_matrices} we have used a different notation $\tilde A'$ to denote the degrees of freedoms for $\tau=2$ time slice. Now we explain the reason. On the special layers, we use the gauge fixing condition illustrated in \cref{fig:gauge_fixing}. However, this implies that there are extra $L\times L$ degrees of freedom $A_\tau'$ on the $\tau$-direction links between $\tau=1$ and $\tau=2$ time slice (while extra degrees of freedom on the $\tau=1$ time slice does not matter since they have already been integrated out when combining the two layers for $T'$ operator). We combine them with the ordinary $2\times L\times L$ degrees of freedoms $A'$ to form $\tilde A'$. Correspondingly, the dimensions of the $M_{T'}$, $N_{T'}$ matrices are $(3\times L\times L)\times(3\times L\times L)$ and $(3\times L\times L)\times(2\times L\times L)$. We need to match the number of degrees of freedom in the $N_{\beta}$, $Q_{\beta}$ matrices in $e^{-\beta H'}$, and this simply done by ``padding'' them---i.e. setting the rows and columns corresponding to the additional degrees of freedom to $0$---to get $\tilde N_{\beta}$, $\tilde Q_{\beta}$. After that we can take the trace 
\begin{equation}
    Z'_{\mathcal{T}}=\left[\int D\tilde A_F'DA_I' \right]\alpha_\beta\alpha_{T'}\exp\left[-\frac{1}{2}\begin{pmatrix}
            {\tilde A_F'}{}^t&{A_I'}^t&{\tilde A_F'}{}^t
        \end{pmatrix}
        \begin{pmatrix}
            M_{T'} & N_{T'}&\\
            N^t_{T'} & Q_{T'}+M_{\beta} &\tilde N_{\beta}\\
            &  \tilde N^t_{\beta}& \tilde Q_{\beta}\\
        \end{pmatrix}
        \begin{pmatrix}
            \tilde A_F'\\A_I'\\\tilde A_F'
        \end{pmatrix}\right]
\end{equation}
which is just a Gaussian integral. Note that the Gaussian integral contains infinities arising from the $1$-form $\mathbb{R}$ symmetry. As we said, these are artifacts that are dictated to be dropped in the $\mathrm{nlocFP}$ prescription in the definition of $Z'$ (see Appendix E of \cite{Xu:2024hyo} for the rigorous derivaiton). This simply means, in the final result of the Gaussian integral, we simply replace the originally vanishing determinant by $\det'$, the product of all \emph{non-zero} eigenvalues.

Finally, recall that we have performed Fourier transform in the $x$-direction, so the computation above is actually applied to each $k_x$ Fourier mode, and the final result $Z'_\mathcal{T}$ seems to be simply the product of the result $Z'_\mathcal{T}(k_x)$ from each $k_x$ Fourier mode. However, there is a caveat: Due to the Fourier transformation and gauge fixing, after taking the said product we must be multiply an extra $\sqrt{3}L^2$ Jacobian factor to obtain the correct result for $Z'_\mathcal{T}$. Roughly speaking, the funny factor $3$ arises because the $y$-direction $1$-form $\mathbb{R}$ symmetry should be represented by a layer of links that form a non-contractible surface on the dual lattice of \cref{fig:gauge_fixing} with its top and bottom time slice identified, which consists of $3L$ original lattice links (it should not be represented by a layer of links that form a non-contractible surface on the dual lattice of $\mathcal{T}$ instead, because in the recursive process we have already integrated out all variables inside the lattice for $e^{-\beta H'}$). The pseudocode for computing $Z'_{\mathcal{T}}$, is presented in \cref{alg:evaluation_1} (In this practical algorithm, instead of first combining the two layers in $T'$ to get $T'$, we first combine the first layer of $T'$ with $e^{-\beta H'}$, and then take the trace of the second layer of $T'$ with the previously result.) 

Recall that we will evaluate $Z'_{\mathcal{T}}$ for different $L$'s, and perform a quadratic fit for the phase of $Z'_{\mathcal{T}}$ as a function of $L$. In this fitting, phase unwrapping is needed. 

The algorithm for computing $Z_{\mathcal{T}_m}'$ is similar. Note that in the gauge fixing, except for $A'_\tau$ on one layer of the $\tau$-direction links in the last $T'$ insertion, all other $A'_\tau$ are fixed to $0$. Therefore, we can first compute the matrix corresponding to $e^{-\beta H'/m}(T'e^{-\beta H'/m})^{m-1}$, and then obtain the final trace using the same method as in the calculation of $Z_{\mathcal{T}}'$. Moreover, by the same argument, we can show that the additional Jacobian in this case is still $\sqrt{3}L^2$.

\begin{algorithm}[H]
    \caption{Evaluation of $Z_\mathcal{T}$}
    \label{alg:evaluation_1}
    \begin{algorithmic}[1] % [1] enables line numbering
        \State \textbf{Input:} Lattice spacial size $L$; Maxwell coefficient $e^2$; Chern-Simons level $k$; $\beta$ size $\log_2\beta$ 
        \State \textbf{Output:} $\log Z_\mathcal{T}$
        \Function{log\_partition\_function}{$L,e^2,k,\log_2\beta$}
        \State $k_x=0$
        \For{$k_x<L$}\Comment{can be accelerated in parallel}
            \State Construct matrices for $e^{-\beta H}$ when $\beta=1$:  $M_1(k_x),N_1(k_x),N^t_1(k_x),Q_1(k_x)$
            \State\Comment{size: all $2L\times 2L$ since $A_\tau$ has been fixed to $0$ for these layers}
            \State Construct matrices for the first layer in the spacial layer:
            $M_{\text{sp1}}(k_x), N_{\text{sp1}}(k_x), N^t_{\text{sp1}}(k_x), Q_{\text{sp1}}(k_x)$
            \State\Comment{size: $3L\times 3L$, $3L\times 2L$, $2L\times 3L$, $2L\times 2L$, since in the middle of the special layer has extra diagonal $A$}
            \State Construct matrices for the second layer in the spacial layer:
            $M_{\text{sp2}}(k_x), N_{\text{sp2}}(k_x), N^t_{\text{sp2}}(k_x), Q_{\text{sp2}}(k_x)$
            \State\Comment{size:  $(2L+L)\times (2L+L)$, $(2L+L)\times 3L$, $3L\times (2L+L)$, $3L\times3L$, ``$+L$'' is for the unfixed $A_\tau$ in the special layer}
            \State $M_\beta,N_\beta,N^t_\beta,Q_\beta,f_\beta \gets \Call{exp\_beta\_H\_matrix}{M_1(k_x),N_1(k_x),N^t_1(k_x),Q_1(k_x)}$
            \State $M',N',{N'}^t,Q',f_1\gets \Call{single\_layer\_update}{M_{\text{sp1}}(k_x), N_{\text{sp1}}(k_x), N^t_{\text{sp1}}(k_x), Q_{\text{sp1}}(k_x),M_\beta,N_\beta,N^t_\beta,Q_\beta}$
            \State Padding $M',N',{N'}^t,Q'$ to size $3L\times 3L$, $3L\times (2L+L)$, $(2L+L)\times 3L$, $(2L+L)\times (2L+L)$
            \State Do gauge fixing on the middle layer of special layer: delete corresponding rows and columns of $M',N',{N'}^t,Q'$ and $M_{\text{sp2}}(k_x), N_{\text{sp2}}(k_x), N^t_{\text{sp2}}(k_x), Q_{\text{sp2}}(k_x)$
            \State $f_2\gets-0.5\log\det'2\pi\begin{pmatrix}
                M_{\text{sp2}}(k_x)+Q' &N_{\text{sp2}}(k_x)+{N'}^t\\
                N^t_{\text{sp2}}(k_x)+N' & Q_{\text{sp2}}(k_x)+M' 
            \end{pmatrix}$ 
            \Comment{$\det'$ means the product of all non-zero eigenvalues}
            \State $\log Z_\mathcal{T}(k_x)\gets f_\beta+f_1+f_2$
            \State $k_x\gets k_x+1$
        \EndFor
        \State $\log Z_\mathcal{T}\gets\sum_{k_x}\log Z_\mathcal{T}(k_x)$
        \State \textbf{return} $\log Z_\mathcal{T}$ \Comment{when processing the data, phase unwrapping must be performed on the imaginary part, and the real part needs to be added by $\log \sqrt{3}L^2$.}
        \EndFunction
    \end{algorithmic}
\end{algorithm}

The separately attached file \textattachfile[color=blue]{anc/chern_simons_calculation.jl}{chern\_simons\_calculation.jl} is the actual Julia code that we used for our calculations.

\bibliographystyle{apsrev4-2}
\bibliography{biblio.bib}